\documentclass[apj]{emulateapj}
\usepackage{natbib}

\newcommand{\um}{$\mu$m}
\newcommand{\hii}{\ion{H}{2}}
\newcommand{\degree}{^{\circ}}
\newcommand{\Spitzer}{{\it Spitzer}}
\newcommand{\MSX}{{\it MSX}}
\newcommand{\IRAS}{{\it IRAS}}

\newcommand{\Lsun}{L$_{\sun}$}
\newcommand{\kms}{km~s$^{-1}$}
\newcommand{\Msun}{M$_{\sun}$}
\newcommand{\peryr}{yr$^{-1}$}

\shorttitle{The Star Formation History of M17}
\shortauthors{Povich et al.}

\begin{document}


\title{The Extended Environment of M17: A Star Formation History}


\author{Matthew S. Povich\altaffilmark{1}, Ed
  Churchwell\altaffilmark{1}, John H. Bieging\altaffilmark{2}, Miju
  Kang\altaffilmark{2,3}, Barbara A. Whitney\altaffilmark{4}, \\ Crystal
  L. Brogan\altaffilmark{5}, Craig A. Kulesa\altaffilmark{2}, Martin Cohen\altaffilmark{6},
 Brian L. Babler\altaffilmark{1}, \\
  R\'emy Indebetouw\altaffilmark{7}, Marilyn R. Meade\altaffilmark{1},
  and Thomas P. Robitaille\altaffilmark{8}
}


\altaffiltext{1}{Department of Astronomy, University of Wisconsin at Madison,
    475 N. Charter Street, Madison, WI 53706; povich@astro.wisc.edu}
\altaffiltext{2}{Steward Observatory, University of Arizona, Tucson, AZ 85721}
\altaffiltext{3}{Korea Astronomy and Space Science Institute, Chungnam National University, Daejeon, Korea}
\altaffiltext{4}{University of Colorado, Space Science Institute, 1540
  30th Street, Suite 23, Boulder, CO 80303-1012}\altaffiltext{5}{National Radio Astronomy Observatory,
  Charlottesville, VA 22903}
\altaffiltext{6}{Radio Astronomy Laboratory, University of California, Berkeley, CA 94720}
\altaffiltext{7}{Department of Astronomy, University of Virginia,
  P.O. Box 3818, Charlottesville, VA 22903-0818}
\altaffiltext{8}{Spitzer Fellow, Harvard-Smithsonian Center for Astrophysics, 60 Garden Street, Cambridge, MA 02138}



\begin{abstract}
M17 is one of the youngest and most massive nearby star-formation
regions in the Galaxy. It features a bright \hii\ region erupting as a
blister from the side of a giant molecular cloud (GMC). 
Combining photometry from the \Spitzer\
Galactic Legacy Infrared Mid-Plane Survey Extraordinaire (GLIMPSE) with
complementary infrared (IR) surveys, we identify
candidate young stellar objects (YSOs) throughout a $1.5\degree \times
1\degree$ field that includes the M17 complex. The long sightline
through the Galaxy behind M17 creates significant contamination in
our YSO sample from unassociated sources with similar IR
colors. Removing contaminants, we produce a highly-reliable
catalog of 96 candidate YSOs with a high probability of association
with the M17 complex. We fit model spectral energy
distributions to these sources and constrain their physical properties.
Extrapolating the mass function of 62 
intermediate-mass YSOs ($M_{\star}>3$~\Msun), 
we estimate
that ${>}1000$ stars are in the process of 
forming in the extended outer regions of M17. The remaining 34
candidate YSOs are found in a 0.17 deg$^2$ field containing the well-studied M17 
\hii\ region and photodissociation region, where bright diffuse mid-IR
emission drastically reduces the sensitivity of the GLIMPSE
point-source detections.

By inspecting IR survey images from \IRAS\ and GLIMPSE, we find
that M17 lies on the rim of a large shell structure 
${\sim}0.5\degree$ in diameter (${\sim}20$ pc at 2.1 kpc).  We present
maps of $^{12}$CO and $^{13}$CO ($J=2\rightarrow 1$) emission observed
with the Heinrich Hertz Telescope. 
The CO emission shows that
the shell is a coherent, kinematic structure associated with M17,
centered at $v
= 19$ \kms.  
The shell is an extended bubble outlining the photodissociation
region of a faint, diffuse \hii\ region several Myr old. We identify
a group of candidate ionizing stars within the bubble.
YSOs in our catalog are concentrated around the bubble rim, providing
evidence that 
massive star formation has been triggered by the expansion of the bubble. 
The formation of the massive cluster
ionizing the M17 \hii\ region itself may have been similarly triggered.
We conclude that the star formation history in the extended
environment of M17 has been punctuated by successive waves of massive
star formation propagating through a GMC complex.  

\end{abstract}




\section{Introduction}
\begin{figure*}[ht]
\epsscale{1.0}
\plotone{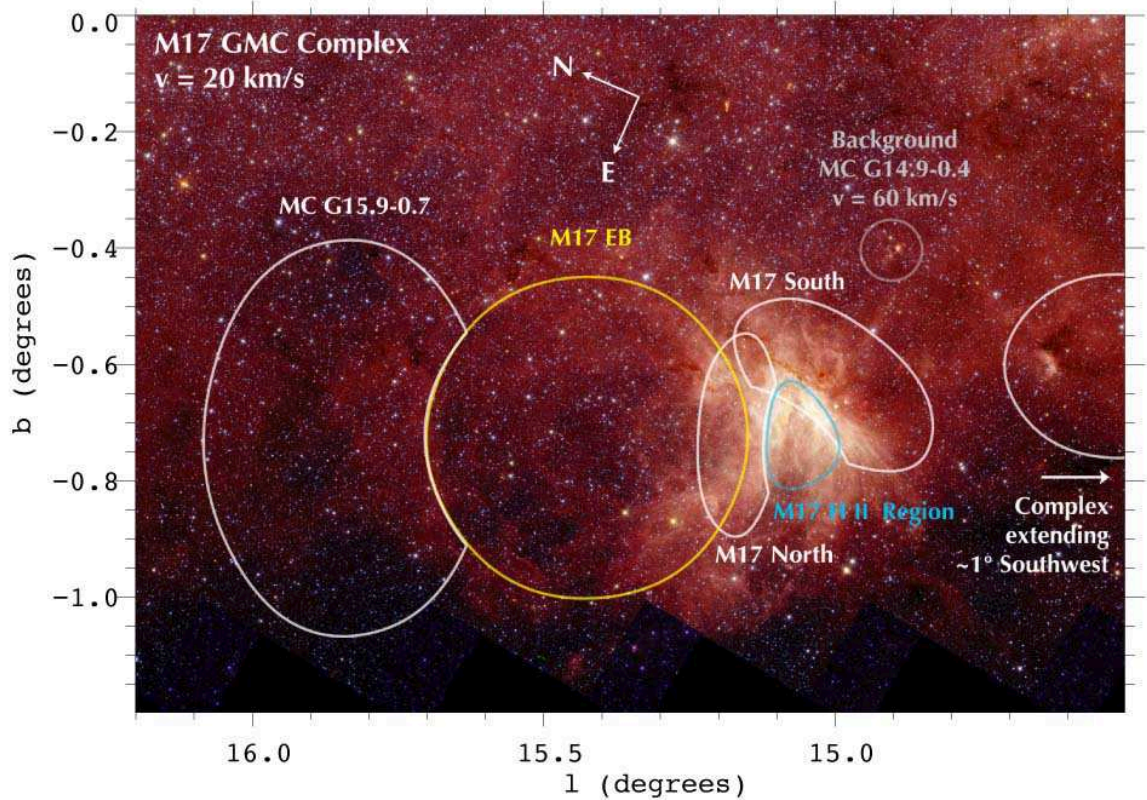}
\caption{\small GLIMPSE full-color image of the M17 field ({\it red:} 8.0
  \um, {\it orange:} 5.8 \um, {\it green:} 4.5 \um, and {\it blue:} 3.6
  \um). A schematic diagram showing the approximate boundaries of
  large-scale structures mentioned in the text is overlaid. The
  molecular clouds shown in {\it white} and {\it gray} have been
  identified from the CO survey data of \citet{S86}.
\label{fullcolor}}
\end{figure*}

Most star formation in the local universe is observed to occur in
dense, massive clusters of hundreds or thousands of stars that are
created by the gravitational 
collapse and fragmentation of massive cores within giant molecular 
clouds (GMCs). In the Galaxy, one of the best nearby
laboratories for the study of star formation in the environment of a
rich, massive 
cluster is M17, the Omega 
Nebula. 
Most of the
observational attention given to M17 has focused on the bright
``blister'' \hii\ region, ${\sim}5\arcmin$ in diameter, erupting
from the side of 
the M17 molecular cloud, a GMC at a velocity of 20~\kms\ 
with a mass ${>}3\times
10^4$~\Msun\ \citep{CL76}. The \hii\ region divides the M17 molecular
cloud into two components, and following the nomenclature of
\citet{W99} we will refer to the more massive, southern component as
``M17 South'' and the northern component as ``M17 North'' (see Fig.\ \ref{fullcolor}).
M17 has served both as an infrared (IR) spectral template for
photodissociation regions (PDRs) 
surrounding bright \hii\ regions \citep{DC96,LV96} and
as a prototype for \hii\ region structure
(Felli, Churchwell, \& Massi 1984; Brogan \& Troland 2001; Pellegrini
et al.\ 2007; Povich et al.\ 2007, hereafter PSC07).

The ionization of the M17
\hii\ region is provided by the rich cluster NGC 6618, which is deeply
embedded in the gas and 
dust of the M17 cloud. Optical and
near-IR photometric and 
spectroscopic studies have uncovered
at least 16 O stars and over 100 B
stars in NGC 6618 
(Ogura \& Ishida 1978; Chini, Els\"{a}sser, \& Neckel 1980; Lada et
al.\ 1991; Hanson, Howarth, \& Conti (1997); Hoffmeister et al.\ 2008,
hereafter H08).
Several water
masers have been observed around M17 \citep{CCLC78}, and a prominent
hypercompact \hii\ region, M17 UC-1, lies on the interface of the
\hii\ region 
with M17 South \citep{FCM84,MS04}. 
M17 has been searched repeatedly for massive
stars with circumstellar disks \citep{MN01,RC04b}, and thousands of
IR-excess sources have been 
reported in and around the \hii\ region 
(Lada et al.\ 1991; Jiang et al.\ 2002, hereafter J02; H08) 
High-resolution observations using the {\it Chandra X-ray
  Observatory} have provided a census of the
young stellar population of M17 (Broos et al.\ 2007; hereafter
BFT07). BFT07 found IR counterparts for 771 of the 886 X-ray sources
in their sample, and they report that only ${\sim}10$\% of these
sources exhibit IR emission 
in excess of a stellar photosphere due to the presence of dust in
circumstellar disks and/or 
infalling envelopes. Taken together,
the observations suggest that, while some fraction of the NGC 6618
stellar population has ceased
accreting, massive star formation is ongoing in and around the M17 \hii\ region.
Many authors have suggested
 that the birth of a second
generation of massive stars has been triggered on the periphery of
the \hii\ region by the expansion
of the ionized gas into the M17 molecular cloud
\citep{CCLC78,FCM84,ZJ02,VH08}.

The extended environment of M17 has often been overlooked. 
\citet{EL76} presented CO observations of a large
molecular cloud 
complex associated with M17 at $v=20$~\kms\ extending more than
$1\degree$ to the Southwest, parallel to the 
Galactic mid-plane (Fig.\ \ref{fullcolor}). 
They suggested that the entire complex is undergoing
sequential star formation, beginning with the OB stars of NGC 6618 at
the northeast tip. \citet{JF82} cast doubt on this
sequential star formation scenario by presenting 13 far-infrared
sources within the extended complex that did not show a progression of
decreasing age with increasing distance from M17.
Several high-resolution Galactic plane radio surveys in
molecular or atomic line emission have included M17 \citep{S86,NANTEN,SGPS}.
The Massachusetts--Stony Brook Galactic Plane CO 
Survey \citep{S86}, which provided $(b,v)$ maps with 45\arcsec\
resolution that include M17 and its vicinity.
The \citet{S86} CO
$J=1$--0 maps clearly show both the M17 molecular cloud and the complex identified by
\citet{EL76}, and they reveal an additional molecular cloud at $v=20$
\kms,
centered at $(l,b) = (15.9\degree,-0.7\degree)$. We
name this second cloud MC G15.9-0.7 (Fig.\ \ref{fullcolor}).

Unbiased surveys of the Galactic plane reveal large structures that
are missed by pointed observations with limited fields-of-view.
Visual inspection of images from the {\it Spitzer Space Telescope}
Galactic Infrared Mid-Plane Survey  
Extraordinaire (GLIMPSE) reveals a circular dust shell
${\sim}0.5\degree$ in diameter outlined by faint, diffuse 8 \um\
emission.
M17 marks the Southwest rim of the shell (see Fig.\
\ref{fullcolor}).
This structure is also apparent in {\it Infrared Astronomy Satellite}
(\IRAS) images at 60 \um\
and 100 \um. 
The dust shell
appears to extend from M17 to MC G15.9-0.7. 
Diffuse 8 \um\ emission is usually attributed
to polycyclic aromatic 
hydrocarbons (PAHs) that fluoresce when illuminated by ultraviolet
(UV) radiation. 
The shell morphology is
similar to that of the IR bubbles identified in GLIMPSE by
\citet{bubbles}. Such bubbles typically outline the
PDRs of \hii\ regions \citep{paper1,WP08}. 
The dust shell near M17 was not selected for the GLIMPSE bubbles
catalog because it 
is partially obscured by the bright diffuse emission
from M17 and lacks a 
well-defined inner rim. Nevertheless, we will demonstrate that the shell is
an extended 
bubble outlining the PDR of a large, faint \hii\ region associated with M17. We
call this structure the M17 extended bubble (M17 EB; Fig.\ \ref{fullcolor}).  

While the screen of cold, dense molecular gas in M17 South probably prevents the
stellar winds and ionizing radiation from the OB stars in NGC 6618
from influencing the extended molecular cloud complex
to the Southwest, on the Northeast side of
M17 there is evidence that M17
EB is interacting with both MC 
15.9-0.7 and M17 itself.
The M17 \hii\ region surrounds a central cavity that is filled by hot, diffuse
plasma observed by {\it Chandra} \citep{T03,paper1}. {\it ROSAT} observations
covering a larger area show soft X-ray emission spilling out of the
\hii\ region and apparently filling part of M17 EB; this X-ray 
emission is
spatially coincident with faint H$\alpha$ emission \citep{D03}.
A search for SIMBAD objects in the interior of M17 EB reveals several
stars with OB spectral types. We will show that these stars 
represent the most massive members of the cluster ionizing M17 EB
and that this cluster is both distinct from and older than NGC 6618. We
refer to this progenitor cluster as NGC 6618PG.

The distance to M17 has been somewhat disputed in recent
years. Most estimates have been based upon optical/near-IR
photometry of the most 
luminous stars. \citet{HHC97} reported a distance of
$1.3^{+0.4}_{-0.2}$ kpc, a significant 
departure from the kinematic distance of 2.3 kpc. \citet{MN01}
subsequently employed a revised extinction law \citep{CW98} and
derived a distance of $1.6\pm 0.3$ kpc. The value of 1.6 kpc has since been widely
adopted. PSC07 balanced the flux in the integrated IR
SED of M17 with the bolometric luminosity of all known ionizing stars
in NGC 6618 
and reported a distance of $1.6^{+0.3}_{-0.2}$ kpc. This luminosity
distance assumed both that all of the major ionizing stars in the region
were known and, like the previous spectrophotometric distance
estimates, that the majority of the known ionizing stars were {\it not} in
unresolved binary or otherwise multiple systems. Recently, \citet{VH08}
showed that the most luminous stellar system in M17, the O4+O4 binary
star called Kleinmann's Anonymous star or CEN 1 \citep{CEN}, is itself
composed of two spectroscopic binary systems of nearly equal-mass
components. This 
discovery, along with new distance moduli for other massive stars
in NGC 6618, lead \citet{VH08} to revise the M17 distance once again,
to $2.1\pm 0.2$ kpc, in agreement with the kinematic distance. In this
work, we will generally adopt the 2.1 kpc distance, but we will also
consider 1.6 kpc where the distance uncertainty could
significantly impact our results.

We take advantage of the large area surrounding M17 observed by the
{\it Spitzer} Galactic plane surveys to investigate
star formation in the extended environment of M17, emphasizing the
newly-discovered structures to the Northeast.
We have combined GLIMPSE with longer-wavelength IR survey data from
both the Multiband Imaging Photometer for {\it Spitzer} (MIPS) Galactic plane
survey (MIPSGAL) and the {\it Midcourse Space 
  Experiment} (\MSX). We use these datasets to investigate the
population of young stellar objects 
(YSOs), identified by their IR excess emission, over a large (${\sim
  1.5\degree \times 1\degree}$) area 
surrounding M17. Our investigation includes the full spatial extent of
the M17 molecular cloud, M17 EB, 
and MC G15.9-0.7.
We have followed up the infrared survey data with high-resolution
$^{12}$CO and $^{13}$CO 
$J=2$--1 observations.
In \S2 we summarize our observations and data reduction. We describe
our methods for selecting and characterizing YSO candidates in \S3,
and we discuss the extended young stellar population of M17 in relation
to the molecular cloud structures in \S4. In
\S5, we synthesize these results into a picture of propagating massive star
formation in M17, and we summarize our conclusions in \S6.

\section{Observations and Data Processing}

\subsection{IR Galactic Plane Survey Data} 

\subsubsection{GLIMPSE}
As part of the {\it Spitzer} Legacy Science Program, GLIMPSE
\citep{GLIMPSE} imaged the Galactic plane from $|l| =
0\degree$--$65\degree$, $|b|\le 
1\degree$ using the four mid-IR bands  (3.6, 4.5, 5.8, and 8.0 \um) of the
Infrared Array Camera (IRAC) on the {\it 
  Spitzer Space Telescope} \citep{IRAC}. 
High-resolution (1.2\arcsec\ pixels) image mosaics
were created by the GLIMPSE pipeline\footnote{Details of the data
  processing can be found at
  http://www.astro.wisc.edu/glimpse/docs.html} from Basic Calibrated
Data (BCD) image frames
processed by the {\it Spitzer} Science Center (SSC). 
The GLIMPSE point-source extractor, a modified version of Daophot
\citep{Daophot} 
was used to extract point sources from the individual image
frames. The $5\sigma$ point-source detection limits of GLIMPSE are
nominally 0.2, 0.2, 0.4, and 0.4 mJy for the [3.6],
[4.5], [5.8], and [8.0] bands, respectively, and
higher in regions of bright diffuse emission.
 Flux densities of sources detected
at greater than 5$\sigma$ 
at least twice in one or more of the 4 IRAC bands were included in the
GLIMPSE Point Source Archive.
The highly-reliable GLIMPSE Point Source
Catalog is a subset of the Archive. Both the Archive and the Catalog
incorporated $JHK_S$ flux 
densities from the 
2MASS point source catalog \citep{2MASS} 
to produce final sourcelists containing flux densities from 1--8 \um\
in 7 near- to mid-IR bands. 
For this study, we use the GLIMPSE v2.0 Point
Source Archive and the GLIMPSE v3.0 image mosaics. 
The majority of the Archive sources used in our analysis will also
appear in the highly-reliable Catalog because of our additional
requirement that
each source must be detected in at least 4 bands (see \S3.2). We use
the Archive instead of the catalog because the Archive identifies
sources with near-saturated fluxes that can later be used as lower
limits and contains more sources in crowded regions.
Mosaics produced from 2MASS data provide
near-IR images that complement GLIMPSE.

\subsubsection{MIPSGAL}

We downloaded 24 \um\ and 70 \um\ post-BCD mosaics of the MIPSGAL
survey \citep{MIPSGAL} from
SSC. The 24 \um\ mosaics have 2.4\arcsec\ pixels 
and the 70 \um\
mosaics have 4.8\arcsec\ pixels. 
MIPSGAL point-source catalogs were not yet available at the time
of this study, so we extracted
24 \um\ point-source 
fluxes using the GLIMPSE extractor. Because the GLIMPSE pipeline
software is not optimized for MIPSGAL data, this procedure generates
the following types of false source detections: noise peaks, small
resolved structures, and artifacts from the point-spread function
wings of bright sources. As a first cut to reduce the number of false
sources in our analysis, we removed all 24 \um\ sources with
larger than 25\% error in measured flux density ($F/\delta F < 4$)
from the sourcelist. We then spatially correlated (bandmerged) the 24 \um\
sources with the 
GLIMPSE Archive sources. If an
 Archive source was located within 2.4\arcsec\ of the central position
 of a 24 \um\ source, the source was considered to have a 24
 \um\ detection and the [24] flux was appended to the Archive
 fluxes. If more than one Archive source was located within 2.4\arcsec\
 of a 24 \um\ source, the [24] flux was appended as an upper limit
 for each correlated Archive source. The
 correlation procedure favors ``true'' 24 \um\ point sources because
 these are more likely to have a GLIMPSE counterpart, but
 occasionally a false source or false correlation makes it into the
 bandmerged list. The final 8-band (combined 2MASS, GLIMPSE, and MIPSGAL [24])
 sourcelists were inspected by eye to reject 
 any remaining
 suspicious sources (see \S3.2).

\subsubsection{Aperture Photometry}

We did not attempt to automatically extract MIPSGAL [70] point source
fluxes. Only ${\sim}10$ obvious 70 \um\ unresolved sources are
visible in the BCD mosaics within our analysis region, so we extracted
the [70] fluxes of 
interesting sources manually, using
aperture photometry. Visual inspection of the GLIMPSE mosaics revealed
${\sim}10$ compact red sources that are resolved by IRAC and hence not
in the Archive. Such sources may be luminous YSOs 
that heat ambient dust in molecular clouds out to a few pc distance.
One pc corresponds to 78 1.2\arcsec\
GLIMPSE mosaic pixels at 2.1 kpc. 
Our aperture photometry procedure used the Funtools\footnote{See http://hea-www.harvard.edu/RD/funtools/ds9.html.} package with the
SAOImage DS9\footnote{See http://hea-www.harvard.edu/RD/ds9/ref.}
image display program to extract fluxes from apertures defined by
eye. For each source extracted, we used the same target apertures
for the 7 2MASS+IRAC bands and 
increased the aperture size as needed for the 2 MIPSGAL bands. We varied
the position of the background sampling a minimum of 3 times to estimate the
contribution of the background subtraction to the flux
uncertainty. Although the background subtraction generally dominated
the uncertainty on the extracted fluxes, our uncertainty estimates
also included photon counting 
statistics (prior to the correction for Zodiacal emission) and the
uncertainty introduced by applying the IRAC extended aperture
corrections\footnote{See
  http://ssc.spitzer.caltech.edu/irac/calib/extcal/.} to sources with
effective radii $>9$\arcsec. 

\subsubsection{\MSX}
The Spirit III instrument on board the \MSX\ satellite surveyed
the Galactic plane in four IR bands \citep{MSX}: A (isophotal wavelength 8.28~\um), C (12.13~\um), D
(14.65~\um), and E (21.3~\um). 
The $5\sigma$ point-source detection limits for \MSX\ are 100, 1100,
900, and 200 mJy for A, C, D, and E bands, respectively. 
The sensitivity of the \MSX\ A and E bands is just below the saturation
limits of IRAC and MIPS, so these bands can be used to replace the fluxes of
point sources saturated at IRAC [8.0] and MIPS [24], respectively. 
The
spatial resolution of the \MSX\ images is $\sim$18.3\arcsec, so
confusion is an issue when correlating \MSX\ Catalog\footnote{See
  http://irsa.ipac.caltech.edu/applications/Gator/.} sources with
the GLIMPSE Archive. 
After selecting YSO candidates based on the
combined GLIMPSE and MIPS [24] photometry (see \S3.1), we correlated
only the brightest saturated and near-saturated YSO candidates with
the \MSX\ catalog, using a 
6\arcsec\ correlation radius.
We did not use the \IRAS\ Point-Source Catalog for this work
because its low resolution causes too much confusion with multiple
GLIMPSE sources. 

\subsection{CO Observations with the Heinrich Hertz Submillimeter Telescope}

We mapped the M17 region in the $J=2\rightarrow 1$ transitions of
$^{12}$CO and $^{13}$CO with the 10-m Heinrich Hertz
Telescope (HHT) on Mt.\ Graham,
Arizona between 2008 February and 2008 June. The receiver was a dual
polarization superconductor-insulator-superconductor (SIS) mixer
system employing single-sideband (SSB) mixers with 
outputs for both upper and lower sidebands, each with a 4--6 GHz IF
band. The $^{12}$CO $J=2\rightarrow 1$ line at 230.538 GHz was placed in the upper
sideband and the $^{13}$CO $J=2\rightarrow 1$ line at 220.399 GHz in the lower
sideband, with a small offset in frequency to ensure that the two
lines were adequately separated in the IF band, since the sideband
rejection was typically 17--20 dB.  The spectrometers, one for each
of the two polarizations and the two sidebands, were filter banks with
256 channels of 1 MHz width and separation.  At the observing
frequencies, the spectral resolution was 1.3 km s$^{-1}$ and the
angular resolution of the telescope was 32$\arcsec$ (FWHM). A set of
$10\arcmin \times 10\arcmin$ fields were mapped with on-the-fly (OTF)
scanning in RA at 10$\arcsec$ per sec with row spacing of 10$\arcsec$
in declination, over a total of 60 rows. Each field required about 2
hours of elapsed time. System temperatures were calibrated by the
standard ambient temperature load method \citep{KU81} 
after
every other row of the map grid. Atmospheric conditions were generally
clear and stable, and the system temperatures were usually nearly
constant at $T_{\rm sys} \sim 200$ K (SSB) for each $10\arcmin \times
10\arcmin$ field.

Data for each polarization and CO isotopomer were processed with the
{\it CLASS} reduction package (from the University of Grenoble
Astrophysics Group), by removing a linear baseline and convolving the
data to a square grid with 10$\arcsec$ grid spacing (equal to
approximately one-third the telescope beamwidth).  The intensity
scales for the two polarizations were determined from observations of
W51D (W51-IRS2) made just before the OTF maps. The gridded spectral data cubes
were processed with the {\it Miriad} software package \citep{RS95} 
for further analysis.  The two polarizations were averaged
for each sideband, yielding images with RMS noise per pixel and per
velocity channel of 0.15 K-T$_A^*$ for both the $^{12}$CO and
$^{13}$CO transitions. 

We mapped a total of 26 $10\arcmin \times 10\arcmin$ fields, arranged
in a semi-regular pattern to cover the main M17 \hii\ region and 
M17 EB.
These 26 fields were combined onto a common
spatial grid and converted to Galactic coordinates for comparison with
the GLIMPSE images.


\section{YSO Selection and Characterization}

An overview of the target field of the GLIMPSE survey searched for
YSOs is presented in Figure \ref{fullcolor}, with the approximate
boundaries of the M17 \hii\
region, the bubble M17 EB, and the major molecular
cloud structures from \citet{S86} overlaid. 
The top of the image is the Galactic
mid-plane ($b=0\degree$), and the jagged bottom edge is the boundary of the
GLIMPSE survey between $b=-1\degree$ and $-1.2\degree$. Most of this
image is occupied by the 
molecular cloud complex at $v=20$ \kms\ that includes M17. A smaller
cloud, MC G14.9-0.4 associated with compact, bright diffuse mid-IR
emission, lies in 
the background at $v=60$ \kms. 

We identify candidate YSOs in this image based on
their IR excess emission, with the goal of
determining the YSO population associated with M17.
The location of M17 near the Galactic mid-plane and only $15\degree$ from the
Galactic center produces a long sightline through the Galaxy in the
mid-IR that passes through multiple spiral arms; hence we expect that
the GLIMPSE Archive contains many YSOs or other objects with similar
IR colors that lie at different distances, unassociated with M17.
The average extinction produced by the molecular clouds in
the target field is
too low to significantly reduce the number of background
sources detected in the mid-IR (see \S4.1.1).
Because M17 is bounded in longitude by
potentially star-forming 
molecular clouds and in latitude by the 
Galactic mid-plane and the survey edge, there is no portion of the
GLIMPSE image in Figure \ref{fullcolor} that could be used as an
off-source region to estimate the 
contamination in our sample from foreground or background sources (see \S3.3). 
Fortunately, the region symmetric to M17, reflected above the Galactic
mid-plane, lacks obvious CO emission marking molecular clouds \citep{S86,NANTEN} and is
correspondingly free of the bright diffuse 8.0 \um\ emission in
GLIMPSE that usually identifies \hii\ regions. 
We selected a circular control field of
$0.5\degree$ radius centered at $(l,b)=(15.08\degree,0.61\degree)$.
Since this region has the same Galactic longitude and absolute latitude
occupied by M17 but exhibits no obvious signs of massive star
formation, it can be used to address the question of what the Galactic
plane would look like if M17 were not 
present. 
 We analyzed the control field in parallel with the M17 target
field, using the same procedure to select candidate YSOs from both fields.

\subsection{Fitting YSO Model SEDs to Broadband IR Fluxes}

Our principal tools for identifying and characterizing candidate YSOs
from this rich dataset were the grid of YSO models from
\citet[][hereafter RW06]{grid} and the spectral energy distribution (SED) fitting tool of
\citet[][hereafter RW07]{fitter}.
The model grid consists of 20,000 2-dimensional Monte Carlo radiation
transfer models 
\citep{BW03a,BW03b,BW04} 
spanning a complete range of stellar mass and evolutionary stage and
output at 10 viewing angles (inclinations), so the model fitting tool
actually has 200,000 SEDs to choose from.
The model fitting tool uses a fast $\chi^2$-minimization linear regression
algorithm (RW07). 
We can robustly distinguish between YSOs and reddened
photospheres of main-sequence and giant stars because YSOs require a
thermal emission component from circumstellar dust to reproduce the
shapes of their mid-IR excesses. 
The Monte Carlo radiative transfer models
upon which the
fitting tool relies have been tested extensively by
successfully fitting the SEDs of numerous well-characterized YSOs in Taurus
(RW07). The models and fitting tool have also been employed to analyze the
YSO populations of the 
Eagle Nebula \citep{I07}, several other Galactic massive star
formation regions observed as part of GLIMPSE \citep{SP07,WP08}, and
the Large Magellanic Cloud \citep[][]{BW08}.

The radiation transfer
technique employed in the RW06 models
propagates ``photons'' from the central source through the
circumstellar environment. The models solve for the temperature
structure of the circumstellar material and include absorption and
re-emission by dust along with photons produced by disk accretion and
backwarming of the stellar photosphere \citep{BW03a,BW03b,BW04}. 
When we fit models to observed YSOs, we essentially
use Monte Carlo radiation transfer to look through the veil of circumstellar
dust, placing the central star on the H-R diagram by interpolating between
pre-main-sequence (PMS) evolutionary tracks \citep{BM96,SDF00}. 
Like all complex models, the YSO models have built-in  assumptions,
degeneracies, and a multi-dimensional parameter space to
explore. In general, fitting these models to real data results
in many different sets of parameters that can describe a YSO
almost equally 
well. Rather than attempt (futilely) to find a unique solution, we
accept a range of well-fit models to investigate how well we can {\it
  constrain} the physical properties of each YSO. 
The distance to the modeled sources places an important external
constraint on the allowable luminosities of the well-fit YSO
models. For M17, we used the distance range of 1.6--2.3 kpc, 
effectively building the distance uncertainty into our constraints on
the YSO properties from the model parameters.
The number of acceptable models also decreases rapidly with increasing
range of IR 
wavelengths from available data included in the SED.

\begin{deluxetable*}{lcccc}
\tabletypesize{\scriptsize} 
\tablecaption{Source Counts in the YSO Search\label{counts}}
\tablewidth{0pt}
\tablehead{
  \colhead{} & \multicolumn{2}{c}{M17 Target Field} &
  \multicolumn{2}{c}{Control Field} \\
  \colhead{} & \multicolumn{2}{c}{(1.67 deg$^{2}$)} &
  \multicolumn{2}{c}{(0.78 deg$^{2}$)} \\
  \colhead{Sources} & \colhead{Number} & \colhead{Density (deg$^{-2}$)} &
  \colhead{Number} & \colhead{Density (deg$^{-2}$)}
}
\startdata
In GLIMPSE Archive & 352,225 & 211,000 & 191,010 & 245,000 \\
Fit with SED models ($N_{\rm data} \ge 4$) & 126,385 & 75,700 & 71,391 &
91,500 \\
Well-fit by stellar photosphere SEDs ($\chi^2/N_{\rm data} \le 5$) &
124,768 & 74,700 & 70,628 & 90,500 \\
Poorly-fit by stellar photosphere SEDs ($\chi^2/N_{\rm data} > 5$) & 
1,617 & 970 & 763 & 980 \\
Well-fit by YSO SEDs ($\chi^2/N_{\rm data} \le 5$) & 979 & 590 & 370 &
470 \\
In final sample of YSO candidates  & 406 & 240 & 106 & 140 \\
\tableline
In M17 YSO candidate subsample\tablenotemark{a}/control
subsample\tablenotemark{b} & 195 & 205 & 25 & 100 \\
\enddata

\tablenotetext{a}{Irregularly shaped region with area 0.95
  deg$^{2}$ (See \S4).}
\tablenotetext{b}{Square box centered at
  $(l,b)=(15.1\degree,0.7\degree)$ with area 0.25 deg$^{2}$.}
\end{deluxetable*}
The steps of our process for selecting candidate YSOs are outlined in
Table \ref{counts}, along with the number and density of sources in
both the target and control fields at each step.
We began by
removing sources with SEDs that are consistent with stellar
photospheres. We fit all sources that are detected in $N_{\rm
  data}\ge 4$ of the 8 near- through mid-IR
bands in the GLIMPSE Archive combined with MIPSGAL 24 \um\ fluxes
(126,385/352,225 sources in the target field and 71,391/191,010
sources in the control field; Table \ref{counts})
with SEDs from 7853 model stellar
atmospheres \citep{Gaia} included in the RW07 fitting tool. 
Interstellar extinction based upon the mid-IR extinction law of
\citet{I05} was included in the model fits, so even highly
reddened stars (we allow a maximum $A_V$ of 30 mag when fitting stellar
photosphere SEDs) returned good fits. We considered all sources with
a best-fit $\chi^2$, normalized by the number of flux
datapoints used 
in the fit, satisfying $\chi^2/N_{\rm data} \le 5$ to be well-fit by stellar
photospheres. We determined this threshold value by visual inspection
of the best-fit SEDs for a variety of sources.  To avoid biasing
the fits for sources where the flux uncertainties have been
underestimated, before fitting any models we conservatively reset the
uncertainties to a floor 
value of $\delta
F/F = 10\%$ for all
sources in the GLIMPSE Archive with $\delta F/F < 10\%$. For the MIPS
[24] fluxes, we used an uncertainty floor of $15\%$.

We then took the 1,617 sources in the target field and 763
sources in the control field that were poorly fit by stellar
photospheres and fit them a second time using the RW06 YSO
models. Again, we considered sources with a best-fit $\chi^2/N_{\rm
  data} \le 5$ to be well-fit, and these formed the basis of our YSO
sample. The vast majority of sources that were well-fit neither by
stellar photospheres nor YSO models are actually stars with saturated
fluxes or poor
photometry (often in the [24] band where our photometry is
less robust) or signs of variability (a mismatch between 2MASS and
GLIMPSE photometry due to the different epochs of the
observations). Eleven sources in the target field appeared to be good 
YSO candidates with marginal fits. Several of these had saturated
fluxes in one or more bands, and one (G015.6653-00.4989) exhibited
excess emission in the [4.5] band similar to previously-studied
sources associated with molecular outflows \citep{S06,D07,SP07,egos}.
These were moved to the ``well-fit'' sample after the
questionable flux measurements were replaced with upper or lower
limits.

\subsection{A Highly Reliable Sample of Candidate YSOs}
Our primary goal was to produce a highly reliable sample of
candidates YSO. As shown in Table \ref{counts}, only 979/126,385 or
0.77\% of sources fit with SED models in the M17 target field were fit
well by YSO   
models but could not be fit well by stellar photospheres. This fraction
was significantly lower  
for the control field, 
with 370/71,391 
or 0.5\% of sources (the
 counting uncertainty is 0.02\%). While the model fitting
process performed the service of discarding the ${>}99\%$ of sources
that we were not interested in studying, simply being well-fit by YSO
model SEDs was a necessary but insufficient criterion for a source to
be included in the final sample. As a last step, we
culled
sources from the sample by inspecting all of the SED fits by eye, 
in many cases returning to the GLIMPSE and MIPSGAL images to confirm
visually the existence of a valid source in each band with a reported
detection. Two common reasons to reject a well-fit YSO candidate
necessitated this final inspection:
\begin{enumerate}
\item The IR excess emission that prevented the source from being fit
  well by stellar photosphere models occurs only in the IRAC [8.0]
  band, with no detection at MIPS [24]. Such an SED can be produced by
  poor source extraction in regions with highly structured diffuse
  emission or
  a Malmquist bias affecting faint sources in the GLIMPSE sourcelists. 
  The Malmquist bias occurs because when a source is selected for the
  Archive on the 
  basis of a $5\sigma$ detection in any combination of bands, the
  fluxes of the remaining bands can be entered in the Archive even if
  they are at lower confidence levels. Stars are typically faint at 8
  \um, and often a noise peak or a diffuse background feature can 
  be extracted from the position of a star observed in the other
  bands, causing an artificially high measured [8.0] flux. While a few
  of the sources exhibiting this SED may actually be YSOs, we make
  the conservative decision to discard all questionable candidates.
\item A suspicious IR excess is observed in the MIPS [24] band only. 
  The difference in resolution between MIPS and
  IRAC increases the chance of a false match when correlating sources
  between MIPSGAL and GLIMPSE. 
  This often happens near
  bright, saturated 24 \um\ sources where the extended wings of the
  MIPS point-spread function can be extracted as individual point
  sources and matched with an overlapping GLIMPSE source. This creates
  a spurious 24 \um\ excess attributed to the GLIMPSE source.
\end{enumerate}
Both of the above issues constitute important caveats to consider if searching
for YSOs possessing ``transition disks'' with large inner holes that
might have been cleared out by planets, since
such sources exhibit IR excess emission only at longer wavelengths.

\begin{figure*}[ht]
\epsscale{0.9}
\plotone{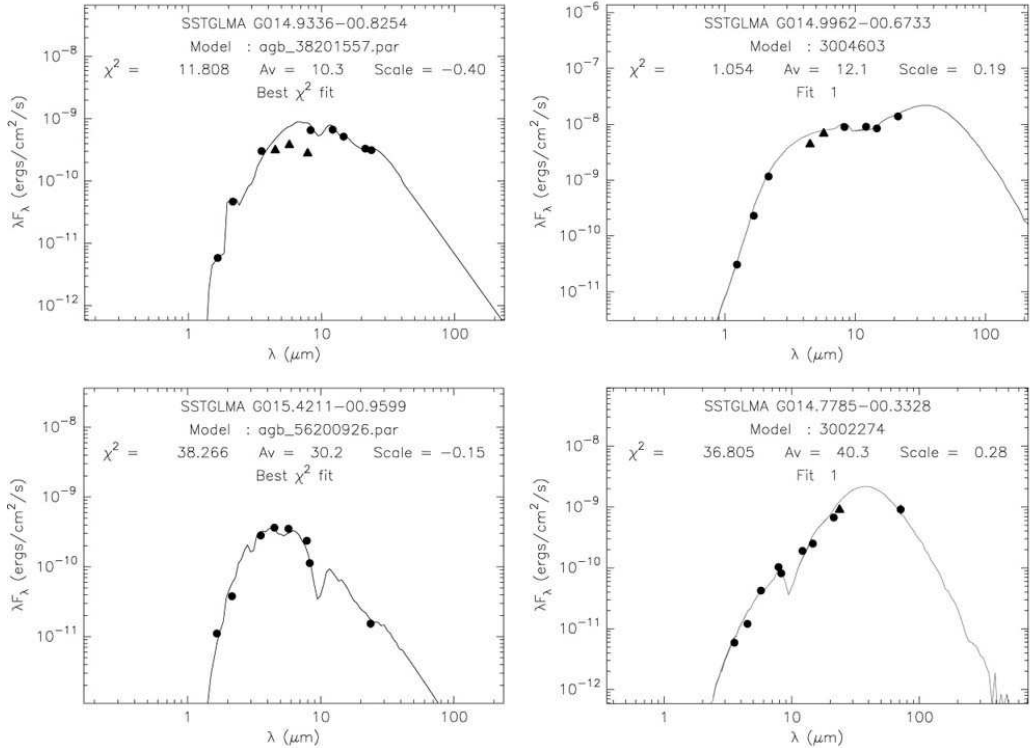}
\caption{\small Example plots of best-fit model SEDs to available near- and mid-IR
  photometry. Two sources well-fit by AGB star spectral templates ({\it
    left} panels) are compared to two sources well-fit by YSO models
  ({\it right} panels) with similar brightnesses. Triangles denote
  datapoints with saturated fluxes that were used only as lower limits
  for the model fitting. The best-fit SED to the KW Object is plotted
  in the {\it upper-right} panel.
\label{AGBvsYSO}}
\end{figure*}
The brightest sources in our sample also appear in the \MSX\ point
source catalog, so we used available \MSX\ fluxes to fill in the SED
between IRAC and MIPS and replace saturated IRAC [8.0] and MIPS [24]
fluxes, where possible. Bright sources with very red SEDs from
$K$--[8.0] that are isolated from other IR-excess sources are candidate
evolved stars on the asymptotic giant branch (AGB). Carbon 
stars, for example, have mean absolute $K$-band magnitudes of
${\sim}6.8$ \citep{WK98} and are often enshrouded by dusty envelopes
generated by strong winds \citep{RG06}; such stars have similar
SEDs to YSOs from $K$ to [8.0] and, even when located on the far side
of the Galaxy, 
will be among the brightest sources observed in GLIMPSE. The RW07
fitting tool incorporates 75 AGB star SED templates derived from
{\it Infrared Space Observatory} ({\it ISO}) spectroscopy
\citep{AGBO,AGBC}. 
After fitting bright sources with the AGB star templates, we removed
30 and 15 likely AGB stars 
from the target and control samples, respectively. 

Example SEDs of two probable AGB stars and two bright candidate YSOs
from the target field are plotted in Figure \ref{AGBvsYSO}. The most
distinctive features of the AGB star SEDs are the precipitously red 
$K$--[8.0] spectral index and the turnover to a
Rayleigh-Jeans spectrum at wavelengths longward of ${\sim}24$
\um. The YSO models that fit the SEDs of these bright sources tend to be
massive stars with optically thick disks but without substantial infalling
envelopes. Such objects are relatively
rare, the best example in M17 being the Kleinmann-Wright (KW) Object
\citep{KW73}, a suspected Herbig Be star \citep{RC04a}. The SED of
the KW Object is plotted in the 
upper-right panel of Figure \ref{AGBvsYSO}.
Several candidate YSOs, including the one shown in the
lower-right panel of Figure \ref{AGBvsYSO}, are detected at 70 \um\ by
MIPSGAL, but no candidate AGB stars are observed at 70 \um. Generally,
YSO disks and envelopes contain cool ($T_d\sim 30$ K) 
dust that becomes optically thin at relatively large radii from
the central star,
while the 
dust distribution in AGB stellar winds has a much smaller effective
radius, giving 
higher effective temperatures ($T_d \ga 200$ K).
It is therefore reasonable to expect that the SEDs of AGB stars
will peak at shorter wavelengths than those of YSOs, but this
difference cannot be observed without data at wavelengths of 24 \um\ or longer,
and there may be exceptions to this general rule. 

\begin{deluxetable*}{ccccccccccc}
\tablewidth{0pt}
\tabletypesize{\scriptsize}
\tablecaption{Candidate YSOs in the Extended Environment of M17\label{M17e}
}
\tablehead{
  \colhead{Index} & \colhead{$l$} & \colhead{$b$} &
  \colhead{[4.5]} & \colhead{[8.0]} & \colhead{[24]} &
  \multicolumn{2}{c}{$M_{\star}$ (\Msun)} &
  \multicolumn{2}{c}{$\log(L_{\rm TOT}/{\rm L_{\odot}})$} & \colhead{Evolutionary} \\
  \colhead{No.} & \colhead{(deg)} & \colhead{(deg)} &
  \colhead{(mag)} &   \colhead{(mag)} & \colhead{(mag)}
  & \colhead{Min} & \colhead{Max}
  & \colhead{Min} & \colhead{Max}
  & \colhead{Stage\tablenotemark{a}}
}
\startdata
\multicolumn{11}{l}{The Bridge} \\
\tableline
E1 &  14.7207 &  -0.4859 &   6.4 &   6.0 &   3.3  &    6 &    6 & 3.1 & 3.2 & II \\
E2 &  14.7392 &  -0.4736 &  12.0 &  10.4 &   5.7  &  0.3 &    4 & 0.1 & 2.2 & Amb. \\
E3 &  14.7762 &  -0.4657 &   7.4 &   6.8 &   3.9  &    6 &    6 & 3.1 & 3.2 & II \\
E4 &  14.7776 &  -0.4900 &   9.7 &   9.5 &   5.8  &    1 &    4 & 1.3 & 2.1 & Amb. \\
E5 &  14.7963 &  -0.4565 &  12.3 &  11.2 &   6.9  &    1 &    4 & 1.3 & 2.5 & II \\
E6 &  14.8039 &  -0.5135 &   7.4 &   7.0 &   4.3  &    3 &    6 & 2.1 & 2.9 & II \\
E7 &  14.8214 &  -0.5136 &   7.5 &   6.5 &   2.8  &    4 &    8 & 2.4 & 3.4 & II \\
E8 &  14.8217 &  -0.4782 &  11.3 &  11.1 &   6.8  &  0.6 &    4 & 0.6 & 1.7 & 0/I \\
E9 &  14.8291 &  -0.5279 &   9.2 &   8.4 &   4.8  &    3 &    5 & 1.8 & 2.6 & II \\
E10 &  14.8462 &  -0.4890 &  12.7 &  11.9 &   7.0  &  0.2 &    4 & 0.1 & 1.7 & Amb. \\
\tableline
\multicolumn{11}{l}{The Outer Regions of the M17 GMC} \\
\tableline
E11 &  14.8338 &  -0.6974 &   9.6 &   8.8 &   6.0  &    3 &    5 & 1.7 & 2.6 & II \\
E12 &  14.9743 &  -0.5079 &  11.7 &   9.8 & $>$  3.3  &  0.3 &    4 & 0.4 & 2.3 & Amb. \\
E13 &  15.1289 &  -0.4958 &  11.5 &   9.0 &   3.2  &  0.2 &    5 & 1.0 & 2.7 & 0/I \\
E14 &  15.1426 &  -0.4883 &  11.4 &   9.8 & $>$  4.1  &  0.2 &    5 & 0.5 & 2.8 & II \\
E15 &  15.1473 &  -0.5299 &   9.2 &   8.3 &   3.5  &    1 &    5 & 1.5 & 2.5 & Amb. \\
E16 &  15.1511 &  -0.4837 &  10.2 &   9.2 &   5.8  &    2 &    4 & 1.3 & 2.3 & II \\
E17 &  15.1971 &  -0.6266 &   7.0 &   5.7 & \nodata  &    5 &   10 & 2.6 & 3.5 & Amb. \\
E18 &  15.2039 &  -0.6325 &   8.2 &   7.3 & \nodata  &    3 &    6 & 2.1 & 2.7 & II \\
E19 &  15.2052 &  -0.6451 &   9.2 & \nodata & \nodata  &    3 &    9 & 2.1 & 3.7 & Amb. \\
E20 &  15.2336 &  -0.6279 &   9.2 &   7.8 &   3.2  &  0.2 &    5 & 0.9 & 2.5 & Amb. \\
\tableline
\multicolumn{11}{l}{M17 EB} \\
\tableline
E21 &  15.3512 &  -0.3525 &  10.7 &  10.1 &   6.8  &    2 &    4 & 1.0 & 2.0 & II \\
E22 &  15.3555 &  -0.2921 &   9.8 &   8.7 &   2.5  &  0.9 &    6 & 1.3 & 2.4 & 0/I \\
E23 &  15.3572 &  -0.3770 &  10.3 &   9.4 &   4.7  &  0.9 &    4 & 1.0 & 2.4 & Amb. \\
E24 &  15.3599 &  -0.2973 &  11.4 &  10.5 &   5.2  &  0.2 &    4 & 0.2 & 2.1 & 0/I \\
E25\tablenotemark{b} &  15.3649 &  -0.4164 &   7.2 &   3.4 & $<$ -0.4  &    6 &   10 & 2.9 & 3.4 & Amb. \\
E26 &  15.3708 &  -0.3350 &   7.1 &   6.4 &   2.4  &    5 &    6 & 2.6 & 3.0 & II \\
E27 &  15.3733 &  -0.3714 &  12.9 &  12.5 &   7.7  &  0.2 &    3 &-0.2 & 1.6 & Amb. \\
E28 &  15.3886 &  -0.3073 &  10.9 &  10.1 &   5.9  &  0.8 &    4 & 0.8 & 2.0 & Amb. \\
E29 &  15.3990 &  -0.3633 &  12.9 &  10.7 &   7.1  &    1 &    5 & 0.5 & 2.7 & Amb. \\
E30 &  15.4620 &  -0.7198 &  10.3 &   9.5 &   6.2  &    2 &    4 & 1.3 & 2.2 & II \\
E31 &  15.4854 &  -0.7175 &   4.6 &   3.7 &   1.7  &    9 &   10 & 3.6 & 3.7 & II \\
E32 &  15.4878 &  -0.4327 &  12.8 &  11.8 &   6.9  &  0.1 &    4 &-0.1 & 1.8 & Amb. \\
E33 &  15.4932 &  -0.4162 &  11.7 &  10.1 &   5.9  &  0.4 &    4 & 0.6 & 2.4 & II \\
E34 &  15.5115 &  -0.9519 &   9.9 &   8.8 &   6.7  &    2 &    7 & 1.3 & 3.3 & II \\
E35 &  15.5202 &  -0.4190 &  10.9 &  10.0 & $>$  5.4  &  0.7 &    4 & 0.8 & 1.7 & II \\
E36 &  15.5221 &  -0.4880 &   8.2 &   7.0 &   2.4  &  0.2 &    1 & 1.3 & 1.6 & 0/I \\
E37 &  15.5268 &  -0.4078 &  11.1 &  11.3 & $>$  6.1  &  0.9 &    1 & 1.0 & 1.1 & 0/I \\
E38 &  15.5281 &  -0.4207 &  12.5 &  11.9 & $>$  4.6  &  0.2 &    4 & 0.3 & 1.7 & 0/I \\
E39 &  15.5326 &  -0.3903 &  12.5 &  11.2 &   8.0  &  0.4 &    4 & 0.1 & 2.2 & II \\
E40 &  15.5330 &  -0.4066 &  10.6 &   9.4 & $>$  5.3  &    3 &    6 & 1.8 & 2.9 & II \\
E41 &  15.5437 &  -0.9338 &   9.0 &   7.3 &   4.0  &    3 &    8 & 2.1 & 3.4 & Amb. \\
E42 &  15.5471 &  -0.3938 &  11.9 &  11.5 &   7.5  &  0.3 &    4 & 0.1 & 1.7 & II \\
E43 &  15.5489 &  -1.0066 &   8.3 & $>$  5.6 & $<$  0.2  &    2 &    7 & 2.2 & 2.7 & 0/I \\
E44 &  15.5502 &  -1.0063 &  10.1 &   8.0 & $>$  0.2  &    1 &    5 & 1.4 & 2.0 & 0/I \\
E45 &  15.5553 &  -0.4603 &   7.1 &   6.2 &   2.5  &    5 &    8 & 2.9 & 3.4 & II \\
E46 &  15.5583 &  -0.4646 &  12.6 &  11.0 & $>$  1.1  &  0.2 &    5 & 0.4 & 2.2 & 0/I \\
E47 &  15.5584 &  -0.4622 &   6.5 & $<$  5.0 &   1.1  &    5 &   10 & 2.8 & 3.8 & II \\
E48 &  15.5625 &  -0.3870 &  13.3 &  11.0 &   4.9  &  0.1 &    5 & 0.4 & 2.0 & 0/I \\
E49 &  15.5691 &  -0.5076 &  12.0 &  11.2 &   7.4  &  0.9 &    3 & 0.4 & 1.4 & II \\
E50 &  15.6050 &  -0.4788 &  11.4 &  10.4 &   6.7  &  0.6 &    4 & 0.5 & 2.1 & II \\
E51 &  15.6092 &  -0.4878 &  12.0 &  10.5 &   5.4  &  0.1 &    4 & 0.0 & 2.3 & Amb. \\
E52 &  15.6385 &  -0.5225 &  10.1 &   9.0 &   6.0  &    2 &    3 & 1.2 & 1.7 & II \\
E53 &  15.6448 &  -0.4847 &  13.0 & \nodata &   6.1  &  0.1 &    4 &-0.0 & 2.4 & Amb. \\
E54 &  15.6588 &  -0.5279 &  10.2 &   9.4 &   5.8  &    3 &    4 & 1.7 & 2.5 & II \\
E55\tablenotemark{c} &  15.6653 &  -0.4989 &   9.2 &   9.5 &   0.7  &    7 &    8 & 2.6 & 3.0 & 0/I \\
E56 &  15.6918 &  -0.8014 &  11.2 &  10.7 &   8.3  &    2 &    3 & 0.8 & 1.5 & III \\
E57 &  15.7296 &  -0.7243 &  13.5 &  11.4 &   7.0  &  0.2 &    4 & 0.0 & 2.0 & II \\
E58 &  15.7408 &  -0.6349 &  12.0 &  11.2 &   8.1  &  0.5 &    3 & 0.3 & 1.7 & II \\
E59 &  15.7478 &  -0.6914 &  13.1 &  10.7 &   5.5  &  0.1 &    4 & 0.2 & 2.1 & Amb. \\
E60 &  15.7677 &  -0.7310 &  10.6 &   9.4 &   5.4  &    1 &    4 & 1.1 & 2.4 & II \\
\tableline
\multicolumn{11}{l}{MC G15.9-0.7 (Likely Unassociated)} \\
\tableline
E61 &  15.8856 &  -0.6053 &   9.0 &   8.7 &   6.0  &    3 &    4 & 1.7 & 2.2 & II \\
E62 &  15.9210 &  -0.6134 &  12.6 &  11.9 &   8.4  &  0.3 &    3 &-0.1 & 1.6 & II \\
\enddata
\tablecomments{The sources presented in this table are
  found within the {\it white} boundary
  lines but outside the rectangle in Fig.\ \ref{clusters}. While the
  association of any given source with 
  M17 is uncertain, the groupings of candidate YSOs are
  significant. The ``Bridge'' is a molecular gas structure connecting
  the M17 molecular cloud with the GMC complex extending 85 pc to the
  southwest. 
  Although our cluster-finding algorithm
  selects 2 candidate YSOs apparently located in MC 15.9-0.7, these
  sources are likely unassociated.}
\tablenotetext{a}{``Amb.'' means that the well-fit models are divided
   between different evolutionary Stages}
\tablenotetext{b}{Source E25 is resolved by IRAC and is also an {\it
    MSX} point source. Its SED exhibits PAH emission.}
\tablenotetext{c}{Source E55 is the candidate protostar
  shown in Fig.\ \ref{IRDC}.}
\end{deluxetable*}
\begin{deluxetable}{ccccccc}
\tablewidth{0pt}
\tabletypesize{\scriptsize}
\tablecaption{Other Candidate Clustered YSOs in the M17 Target Field\label{field}
}
\tablehead{
  \colhead{Index} & \colhead{$l$} & \colhead{$b$} &
  \colhead{[4.5]} & \colhead{[8.0]} & \colhead{[24]} 
 & \colhead{} \\
  \colhead{No.} & \colhead{(deg)} & \colhead{(deg)} &
  \colhead{(mag)} &   \colhead{(mag)} & \colhead{(mag)}
  & \colhead{Group\tablenotemark{a}}
}
\startdata
F1 &  14.5865 &  -0.5708 &  10.6 &   9.1 &   4.9  & SW \\
F2 &  14.5898 &  -0.5822 &  12.8 &  11.6 &   7.3  & SW \\
F3 &  14.5999 &  -0.5422 &  10.2 &   9.7 &   6.3  & SW \\
F4 &  14.6005 &  -0.6248 &   9.9 &   9.9 &   5.6  & SW \\
F5 &  14.6017 &  -0.6177 &  10.9 &  10.1 &   5.3  & SW \\
F6 &  14.6043 &  -0.6229 &  10.3 &   9.7 &   6.4  & SW \\
F7 &  14.6116 &  -0.5520 &  11.2 &   9.9 &   5.7  & SW \\
F8 &  14.6117 &  -0.6259 &   9.7 &   9.3 &   6.4  & SW \\
F9 &  14.6148 &  -0.5843 &  11.4 &  10.4 & $>$  5.7  & SW \\
F10 &  14.6235 &  -0.5775 &  10.7 &   9.9 & $>$  5.5  & SW \\
F11 &  14.6285 &  -0.5781 &   9.5 &   8.6 & $>$  4.6  & SW \\
F12 &  14.6290 &  -0.7614 &  12.5 & \nodata &   5.3  & SW \\
F13 &  14.6297 &  -0.5724 &  11.1 &  10.1 & $>$  5.4  & SW \\
F14 &  14.6311 &  -0.5784 &  10.6 &   9.3 & $>$  2.1  & SW \\
F15 &  14.6314 &  -0.5774 &  11.7 &  10.2 &   2.1  & SW \\
F16 &  14.6327 &  -0.5719 &  10.9 &  10.2 & $>$  6.4  & SW \\
F17 &  14.6348 &  -0.8730 &  13.1 &  12.1 &   8.1  & SW \\
F18 &  14.6395 &  -0.7286 &  11.6 &  11.3 &   8.3  & SW \\
F19 &  14.6416 &  -0.8928 &  13.1 &  11.8 &   6.1  & SW \\
F20 &  14.6426 &  -0.7639 &  11.2 &   9.7 &   5.3  & SW \\
F21 &  14.6428 &  -0.8833 &  12.3 &  11.5 &   7.2  & SW \\
F22 &  14.6437 &  -0.9011 &  10.9 &  10.3 &   6.8  & SW \\
F23 &  14.6533 &  -0.7411 &  12.6 &  11.5 &   8.2  & SW \\
F24 &  14.6660 &  -0.8838 &  12.6 &  11.9 &   8.0  & SW \\
F25 &  14.6745 &  -0.8863 &  10.9 &  10.6 &   7.0  & SW \\
F26 &  14.6771 &  -0.7672 &   9.7 &   9.0 &   5.1  & SW \\
F27 &  14.6802 &  -0.8755 &  11.8 &  10.6 &   6.0  & SW \\
F28 &  14.6990 &  -0.8916 &  12.3 &  11.9 &   6.9  & SW \\
F29 &  14.7720 &  -0.3570 & \nodata &   8.6 &   3.1  &  \\
F30 &  14.7868 &  -0.2176 &  11.9 &  10.2 & $>$  6.4  &  \\
F31 &  14.7878 &  -0.2353 &  10.9 &  10.5 &   7.3  &  \\
F32 &  14.7952 &  -0.2062 &  12.0 &  11.7 &   7.7  &  \\
F33 &  14.7999 &  -0.2072 &  12.6 &  11.1 &   7.7  &  \\
F34 &  14.8122 &  -0.2346 &   9.9 &   9.3 &   6.4  &  \\
F35 &  14.8209 &  -1.0206 &   8.0 &   7.4 &   4.7  & SW \\
F36 &  14.8247 &  -0.2663 &  11.7 &  10.2 &   5.9  &  \\
F37 &  14.8302 &  -0.1774 &  11.1 &  10.7 &   2.2  &  \\
F38 &  14.8443 &  -1.0049 &   6.9 &   6.4 &   3.3  & SW \\
F39 &  14.8516 &  -0.9890 & $>$ 10.7 &  10.1 &   3.2  & SW \\
F40 &  14.8519 &  -0.9927 & $>$ 10.8 &  10.0 &   3.2  & SW \\
F41 &  14.8521 &  -1.0007 &  11.4 &  10.7 &   7.7  & SW \\
F42 &  14.8522 &  -0.3314 &  11.5 &  11.0 &   7.8  &  \\
F43 &  14.8566 &  -0.1674 &  13.0 &  12.1 & $>$  7.0  &  \\
F44 &  14.8607 &  -0.4358 &  10.1 &   9.4 &   6.8  & MC G14.9 \\
F45 &  14.8621 &  -0.1594 &  13.8 &  12.1 &   7.6  &  \\
F46 &  14.8624 &  -0.9974 &  12.8 &  11.8 &   8.0  & SW \\
F47 &  14.8637 &  -0.1750 &  11.0 &  10.7 &   7.0  &  \\
F48 &  14.8695 &  -0.3952 &  11.3 &   9.9 &   5.9  & MC G14.9 \\
F49 &  14.8709 &  -0.3886 &  11.9 & \nodata &   5.5  & MC G14.9 \\
F50 &  14.8747 &  -0.4154 &  10.8 &   9.2 &   5.7  & MC G14.9 \\
F51 &  14.8840 &  -0.3558 &  12.4 & \nodata &   7.0  &  \\
F52 &  14.8844 &  -0.3114 &  11.7 & \nodata &   6.3  &  \\
F53\tablenotemark{b} &  14.8891 &  -0.4018 &   6.1 &   2.5 & $<$ -1.0  & MC G14.9 \\
F54 &  14.8892 &  -0.1977 &  11.8 &  11.1 &   7.4  &  \\
F55 &  14.8951 &  -0.4004 &  11.2 &   7.6 & $>$  2.9  & MC G14.9 \\
F56 &  14.8993 &  -0.2249 &  10.0 &   9.7 &   5.2  &  \\
F57 &  14.9224 &  -0.2113 &  12.9 &  12.0 &   8.1  &  \\
F58 &  14.9234 &  -0.2147 &  12.3 & \nodata &   5.4  &  \\
F59 &  14.9251 &  -0.4058 &   9.2 &   8.9 &   5.8  & MC G14.9 \\
F60 &  14.9261 &  -0.1578 &  12.4 & \nodata &   7.9  &  \\
F61 &  14.9314 &  -0.2239 &  12.4 & \nodata &   8.0  &  \\
F62 &  14.9340 &  -0.1358 &  11.1 &  10.3 &   7.0  &  \\
F63 &  14.9346 &  -0.2295 &  12.0 &  11.7 &   7.7  &  \\
F64 &  14.9372 &  -0.4296 &  10.5 &   9.8 &   6.2  & MC G14.9 \\
F65 &  14.9378 &  -0.4229 &  11.8 & \nodata &   7.9  & MC G14.9 \\
F66 &  14.9402 &  -0.1474 &  11.7 &  11.1 &   7.8  &  \\
F67 &  14.9607 &  -0.1868 &   6.8 &   6.1 &   3.3  &  \\
F68 &  14.9622 &  -0.1400 &   9.7 &   9.4 &   6.6  &  \\
F69 &  14.9685 &  -0.1384 &  11.6 &  10.8 &   7.8  &  \\
F70 &  14.9764 &  -0.2147 &  12.0 &  10.5 &   6.3  &  \\
F71 &  14.9804 &  -0.4319 &  10.0 &   9.2 &   6.2  &  \\
F72 &  14.9960 &  -0.3090 &  11.2 &  10.8 &   7.8  &  \\
F73 &  15.0626 &  -0.2450 &  10.7 &  10.2 &   7.2  &  \\
\enddata
\end{deluxetable}
\setcounter{table}{2}
\begin{deluxetable}{ccccccc}
\tablewidth{0pt}
\tabletypesize{\scriptsize}
\tablecaption{---Continued}
\tablehead{
  \colhead{Index} & \colhead{$l$} & \colhead{$b$} &
  \colhead{[4.5]} & \colhead{[8.0]} & \colhead{[24]} 
 & \colhead{} \\
  \colhead{No.} & \colhead{(deg)} & \colhead{(deg)} &
  \colhead{(mag)} &   \colhead{(mag)} & \colhead{(mag)}
  & \colhead{Group\tablenotemark{a}}
}
\startdata
F74 &  15.0634 &  -0.3109 &  10.8 &  10.2 &   7.4  &  \\
F75 &  15.0768 &  -0.2528 &  11.4 &  10.7 &   6.7  &  \\
F76 &  15.0773 &  -0.2511 &  11.1 &  10.2 &   7.1  &  \\
F77 &  15.0784 &  -0.2678 &  12.4 &  11.6 &   7.1  &  \\
F78 &  15.0795 &  -0.2609 &   9.2 &   8.3 &   4.2  &  \\
F79 &  15.0817 &  -0.2699 &  13.0 &  11.7 & $>$  5.2  &  \\
F80 &  15.0832 &  -0.2686 &  10.1 &   9.2 &   5.2  &  \\
F81 &  15.0841 &  -0.1548 &  10.8 &  10.0 &   6.5  &  \\
F82 &  15.1135 &  -0.1485 &   9.6 &   7.8 &   4.7  &  \\
F83 &  15.1512 &  -0.1216 &  11.9 &  11.2 &   6.5  &  \\
F84 &  15.1530 &  -0.1909 &  12.4 &   9.7 &   4.9  &  \\
F85 &  15.1684 &  -0.1267 &  11.2 &  11.4 &   4.1  &  \\
F86\tablenotemark{b} &  15.1761 &  -0.1585 &  12.0 &   9.4 &   3.7  &  \\
F87\tablenotemark{b} &  15.1831 &  -0.1625 &   7.8 &   4.7 & $<$ -0.5  &  \\
F88 &  15.1841 &  -0.1589 &   8.9 &   6.6 & $>$  1.2  &  \\
F89\tablenotemark{b} &  15.1850 &  -0.1561 &   9.7 &   6.1 & $>$  2.3  &  \\
F90 &  15.1898 &  -0.1400 &  12.2 &  11.6 &   7.1  &  \\
F91 &  15.1904 &  -0.1692 &  12.6 &   9.9 & $>$  5.9  &  \\
F92 &  15.2199 &  -0.1451 &  10.8 &   9.7 & $>$  4.1  &  \\
F93 &  15.2354 &  -0.1750 &   9.1 &   8.3 &   5.7  &  \\
F94 &  15.5750 &  -0.3101 &  11.0 &  10.4 & $>$  6.5  & Bu G15.7 \\
F95 &  15.5754 &  -0.3218 &  12.3 &  11.1 &   7.2  & Bu G15.7 \\
F96 &  15.5815 &  -0.2963 &  11.1 &   9.9 & $>$  6.3  & Bu G15.7 \\
F97 &  15.5869 &  -0.3003 &   9.1 &   8.3 &   3.7  & Bu G15.7 \\
F98 &  15.5920 &  -0.2923 &  13.4 &  10.9 &   6.6  & Bu G15.7 \\
F99 &  15.5998 &  -0.2861 &  10.8 &  10.3 &   6.6  & Bu G15.7 \\
F100 &  15.6028 &  -0.2040 &   9.5 &   8.6 &   5.0  & Bu G15.7 \\
F101 &  15.6051 &  -0.2537 &  11.3 &  10.6 &   6.6  & Bu G15.7 \\
F102 &  15.6337 &  -0.2282 &  10.3 &   9.7 & $>$  6.0  & Bu G15.7 \\
F103 &  15.6408 &  -0.2189 &   7.9 & $>$  6.2 &   1.0  & Bu G15.7 \\
F104 &  15.6541 &  -0.2249 &  11.6 &   7.9 &   2.1  & Bu G15.7 \\
F105 &  15.6603 &  -0.1189 &  11.4 &   9.1 &   5.6  &  \\
F106 &  15.6638 &  -0.1231 &   9.9 &   9.1 &   5.7  &  \\
F107 &  15.6747 &  -0.1476 &  10.5 & \nodata &   5.4  &  \\
F108 &  15.6768 &  -0.1465 &  11.9 &   8.2 &   4.6  &  \\
F109 &  15.6781 &  -0.2651 &  10.3 &   9.9 &   6.2  & Bu G15.7 \\
F110 &  15.6807 &  -0.2613 &  10.7 &   8.9 &   5.2  & Bu G15.7 \\
F111 &  15.6829 &  -0.1424 &  10.3 &   9.3 &   5.6  &  \\
F112 &  15.6929 &  -0.2576 &  11.4 &  11.1 &   7.6  & Bu G15.7 \\
F113 &  15.6929 &  -0.1355 &  10.5 &   7.0 &   2.2  &  \\
F114 &  15.6956 &  -0.1367 &  10.8 &  10.5 & $>$  2.2  &  \\
F115 &  15.7181 &  -0.1648 &  10.6 &   9.3 & $>$  3.0  &  \\
F116 &  15.7237 &  -0.2617 &  11.5 &  10.5 &   7.3  & Bu G15.7 \\
F117 &  15.7250 &  -0.2480 &  11.9 &  11.5 &   5.0  & Bu G15.7 \\
F118 &  15.7574 &  -0.2234 &  12.2 &   8.8 &   4.7  &  \\
F119 &  15.7580 &  -0.1887 &   8.9 &   8.2 &   5.8  &  \\
F120 &  15.7600 &  -0.2195 &   9.3 &   8.6 &   6.1  &  \\
F121 &  15.7826 &  -0.3897 &  12.1 &  10.5 &   6.5  &  \\
F122 &  15.7848 &  -0.3864 &  12.8 & \nodata &   7.4  &  \\
F123 &  15.7911 &  -0.2355 &  11.3 &  10.3 &   6.0  &  \\
F124 &  15.7951 &  -0.4058 & $>$ 11.8 &  12.1 &   7.0  &  \\
F125 &  15.7991 &  -0.3927 &   9.9 &   8.8 &   5.5  &  \\
F126 &  15.8012 &  -0.3760 &   8.8 &   8.4 &   5.6  &  \\
F127 &  15.8067 &  -0.3818 &  12.0 &  11.2 &   7.5  &  \\
F128 &  15.8108 &  -0.2328 &  11.7 &  10.5 &   8.0  &  \\
F129 &  15.8113 &  -0.2394 &  12.6 &  11.1 &   6.8  &  \\
F130 &  15.8367 &  -0.2653 &  10.9 &   9.9 &   6.3  &  \\
F131 &  15.8612 &  -0.2350 &  11.1 & \nodata &   5.7  &  \\
F132 &  15.9580 &  -0.1866 &  11.8 &  10.7 &   6.3  &  \\
F133 &  15.9750 &  -0.2097 &   9.5 &   8.3 &   3.9  &  \\
F134 &  15.9920 &  -0.2097 &  12.1 &  10.6 &   4.6  &  \\
F135 &  15.9930 &  -0.1868 &  10.9 &  10.7 &   6.4  &  \\
\enddata
\tablecomments{The sources presented in this table are found outside
  the {\it white} boundary lines in Fig.\ \ref{clusters}. The majority
of these sources are at unknown distances.}
\tablenotetext{a}{Groups of these candidate YSOs are apparently
  associated with prominent molecular cloud structures: ``SW''
  refers to the near end of the large GMC complex at $v=20$ \kms\
  extending southwest 
  from M17; ``MC G14.9'' refers to the molecular cloud G14.9-0.4 at
  $v=60$ \kms\ (see Fig.\ \ref{fullcolor}); and ``Bu G15.7'' refers to
  an IR/CO shell structure, 4\arcmin\ in diameter, centered at
  $(l,b)=(15.67\degree,-0.29\degree)$ and $v \approx 45$ \kms.}
\tablenotetext{b}{These 4 sources are resolved by IRAC. F53, F87,
  and F89 are detected by {\it MSX} and may be compact H II regions
  ionized by young clusters 
  rather than single YSOs.} 
\end{deluxetable}
Visual inspection of the M17 target field revealed 6 bright, compact
resolved IRAC sources 
that have SEDs well-fit by YSO models. These sources are noted in Tables
\ref{M17e}--\ref{x} (see \S4.1.2). Four of the 6 are \MSX\ point
sources, and their SEDs are suggestive of a PAH spectrum. The model
fits to these sources were improved dramatically by assuming
PAHs contributed significantly to the emission in the 4 bands with the
strongest PAH 
features, IRAC [5.8] \& [8.0] and \MSX\ A \& C. 
The RW06 YSO models did not include emission from PAHs; hence to fit
sources with potentially PAH-dominated spectra we set these 4 bands as 
upper limits. Using the fluxes of the well-fit models to estimate
 the continuum levels in the bands affected by PAHs, we find that PAHs
 contribute anywhere from 30\% to 90\% of the broadband fluxes. The
YSO models for 3 of these 4 sources indicate central stars with
photospheric temperatures of $T_{\rm eff} \sim 2\times 10^4$ K, hot
enough to produce the UV radiation necessary to excite PAH
transitions \citep[][and references therein]{paper1}. The fourth
source, F53 in Table \ref{field}, was fit by few high-$T_{\rm eff}$
models. This object appears to 
be associated with the molecular cloud MC G14.9-0.4 (see Fig.\
\ref{fullcolor}), giving it a near kinematic distance of 4.8 kpc at
the cloud velocity of 60 \kms. This is far beyond the maximum distance
of 2.3 kpc given to the model fitting tool, and at the
larger distance this source
is likely to be a compact \hii\ region. This interpretation is
supported by the fact that the 
source also appears to have associated radio emission from the 11-cm continuum
survey of \citet{11cm}.
For consistency, we leave this source in the final
sample of candidate YSOs, but we exclude it, along with other
sources listed in Table \ref{field} that are found outside the M17
subregion defined in \S4.1, from the 
analysis of
YSOs associated with M17.

\subsection{YSO Evolutionary Stages}
RW06 divided the YSO models into evolutionary Stages according to the
envelope accretion rate $\dot{M}_{\rm env}$ and the 
circumstellar disk mass $M_{\rm disk}$, both normalized by the mass of the
central star $M_{\star}$: 
\begin{center}
\begin{tabular}{lcc}
  Stage 0/I: & $\dot{M}_{\rm env}/M_{\star} > 10^{-6}$ yr$^{-1}$ & \\
  Stage II: &  $\dot{M}_{\rm env}/M_{\star} < 10^{-6}$ yr$^{-1}$; & $M_{\rm
    disk}/M_{\star} > 10^{-6}$ \\
  Stage III: &  $\dot{M}_{\rm env}/M_{\star} < 10^{-6}$ yr$^{-1}$; & $M_{\rm
    disk}/M_{\star} < 10^{-6}$. \\
\end{tabular}
\end{center}
\begin{figure*}[ht]
\epsscale{1.0}
\plotone{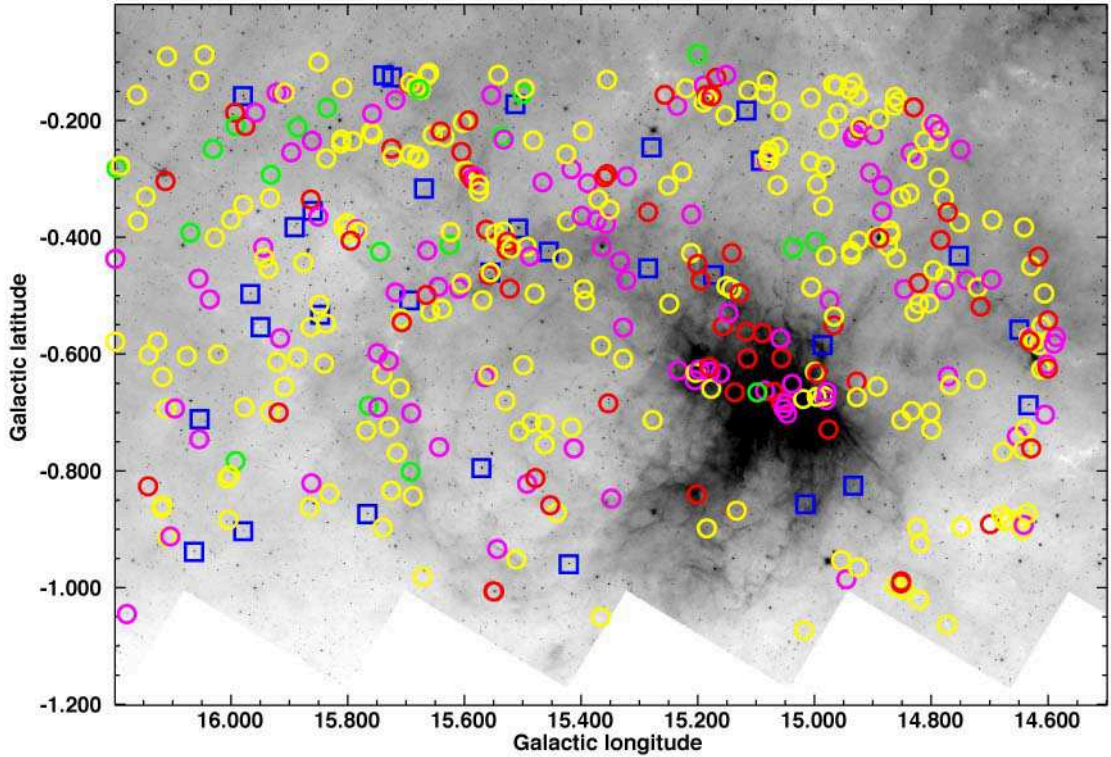}
\caption{\small GLIMPSE 8.0 \um\ image of the M17 analysis target field (inverted
  logarithmic grayscale). Positions
  of the 406
  candidate YSOs in the final sample are marked by circles
  color-coded according to the most probable evolutionary Stage of
  each source: {\it red} circles are Stage 0/I candidates (deeply embedded in
  infalling envelopes); {\it yellow} are Stage II 
  (optically thick circumstellar disks); {\it green} are Stage III
  (optically thin circumstellar disks); and {\it magenta}
  are candidate YSOs for which the Stage is more ambiguous. {\it
    Blue} squares denote candidate AGB stars removed from the YSO
  sample. The
  apparent lack of candidate YSOs in the extreme top and right portions of the
  image reflects the boundaries of the search
  area. 
\label{YSOverview}}
\end{figure*}
These YSO classifications are based upon 
model parameters corresponding to physical properties, but they
parallel the popular observational T Tauri 
classification system that uses IR spectral indices
\citep{CL87}. Stage 0/I YSOs are protostars 
heavily embedded in the infalling gas and dust of their natal
envelopes. Stage II YSOs, like Class II T Tauri stars, have optically
thick circumstellar disks that dominate their near- to mid-IR
SEDs. Stage III YSOs have optically-thin remnant disks and their SEDs
are dominated by photospheric emission; these objects are difficult to
identify by IR excess alone (given our error bars
of $\ge 10$\%--15\%), hence very few examples are
found in our conservative YSO sample. Because each candidate YSO 
was fit with multiple YSO models, we determined the evolutionary Stage
of each source statistically. We defined the set of ``well-fit'' models
to each source as those models with fits meeting the
criterion of
\begin{equation}\label{best}
  \frac{\chi^2}{N_{\rm data}} - \frac{\chi^2_{\rm min}}{N_{\rm data}} \le 2,
\end{equation}
where $\chi^2_{\rm min}$ is the goodness-of-fit parameter for the
best-fit model. We estimated the {\it relative} probability of each
model in the set according to
\begin{equation}\label{prob}
  P(\chi^2)=e^{-(\chi^2-\chi^2_{\rm min})/2}
\end{equation}
and normalized such that $\Sigma P = 1$. This allowed us to construct a
probability distribution for the evolutionary Stage of each source
from the Stages of all the well-fit models. The most probable Stage of
each source was defined by $\Sigma P$(Stage)$\ge 0.67$; if this
condition could not be met, then the Stage of the source was
considered to be ambiguous.

The 406 candidate YSOs in the final sample of the M17 target field are
overplotted as circles on a GLIMPSE [8.0] image in Figure \ref{YSOverview}. 
The color-coding of the circles corresponds to the most probable
evolutionary Stage of each source.
Candidate YSOs are distributed
throughout the image (the absence of sources from the top
and right margins of the image is due simply to the boundaries of our
search area). Numerous small ``clusters'' of 5--10 YSOs are apparent, and
larger groups are associated with the M17 molecular cloud and the
rim of M17 EB. Candidate YSOs associated with the M17 molecular cloud would
dominate the sample if not for the strong masking effect of the bright diffuse
mid-IR  emission from the \hii\ region and PDR.
The point-source detection limit rises dramatically
near the M17 \hii\ region, and only the most luminous YSOs are
detectable within the area where the 8.0 \um\ image in Figure
\ref{YSOverview} appears black. The MIPSGAL 24 \um\ image is
completely saturated within the \hii\ region, and a relatively high
fraction of candidate YSOs near 
the \hii\ region have ambiguous Stage ({\it magenta} circles) because 24 \um\
fluxes are an important discriminator between disk- and
envelope-dominated sources (RW06).

\subsection{Mid-IR Color-Color Plots}
\begin{figure}
\epsscale{0.85}
\plotone{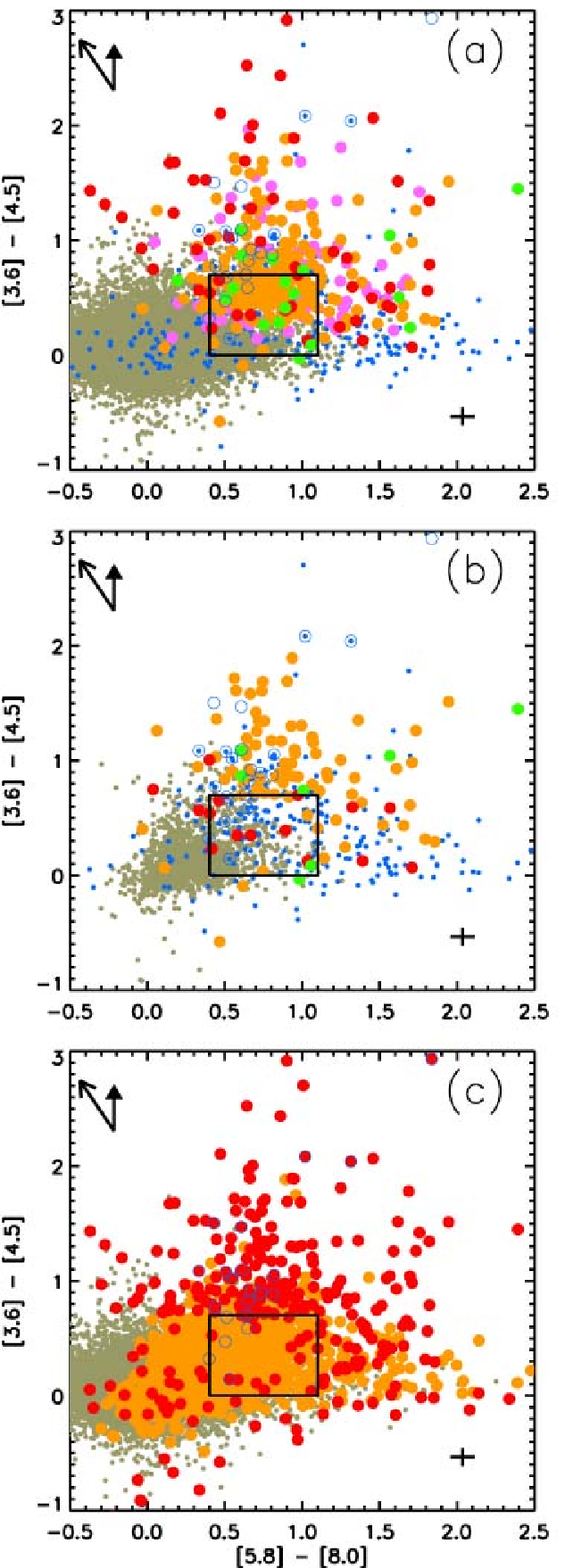}
\caption{\small Mid-IR color-color space showing sources in the
  target field that were detected in all 4 IRAC bands. 
   The ``disk domain'' of A04 is
  contained within the black box. ({\it a}) Candidate YSO selection and
  classification from model 
  fitting. Small {\it gray} dots are sources well-fit by
  stellar photosphere SEDs.   Large dots are sources
  in the final YSO sample ({\it red:} Stage 0/I; {\it orange:}
  Stage II; {\it 
    green:} Stage III; and {\it lavender:} ambiguous). 
 ({\it b}) Sources for which our model-based classification
  differs from the IRAC color-color classification of
  G08. Sources discarded from the final YSO sample (small {\it
    blue} dots) remain in this plot if they have IRAC colors consistent
  with YSOs according to G08.  ({\it c}) G08 YSO selection and classification
 ({\it gray:} stellar photospheres; {\it red:}
  protostars; and {\it orange:} Class II). 
  In all plots, 
  {\it blue} circles are candidate AGB stars, reddening vectors
  for $A_V=30$~mag based on the extinction laws 
  of \citet{WD01} and \citet{I05} are
  shown as open and filled arrows, respectively, and black crosses give
  typical photometric errors.
\label{IRACcc}}
\end{figure}

\begin{figure}
\epsscale{0.85}
\plotone{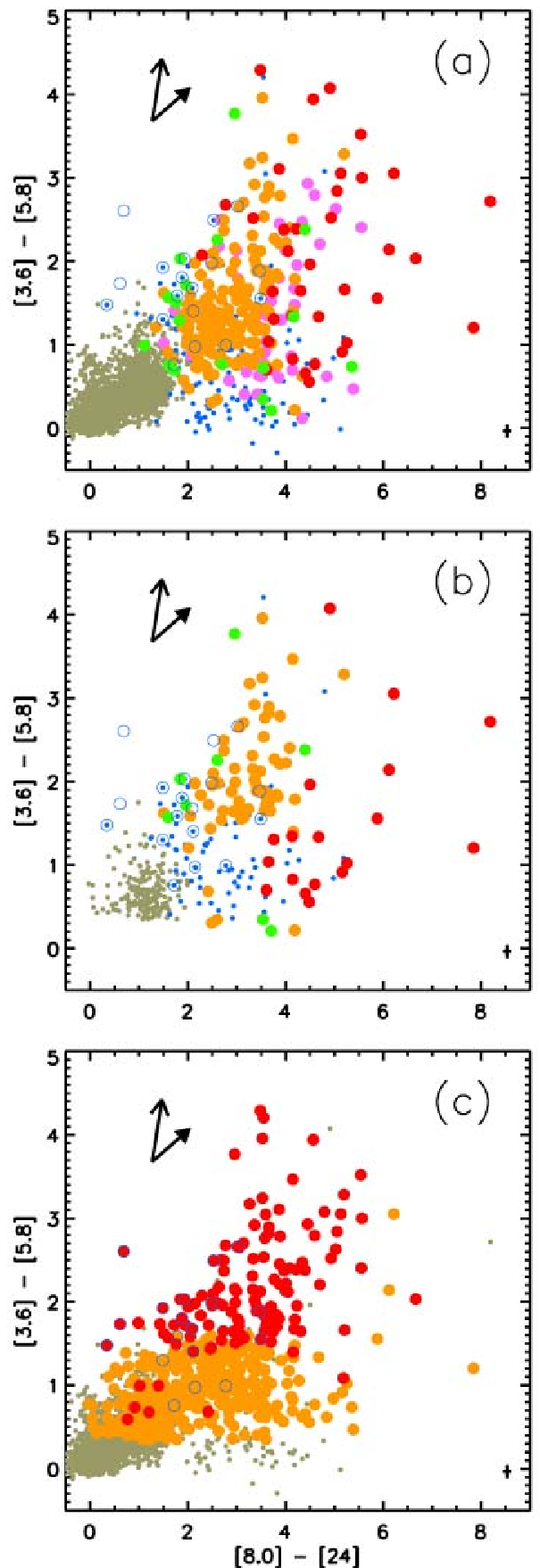}
\caption{\small Same as Fig.\ \ref{IRACcc}, except a color space combining
  IRAC with MIPS [24] photometry is shown. 
\label{IRACMIPScc}}
\end{figure}
In Figures \ref{IRACcc} and \ref{IRACMIPScc} we compare our model-based
classifications of all stars and YSOs in the M17 target field
with classification systems derived from IRAC
mid-IR colors (Allen et al.\ 2004, hereafter A04; Gutermuth et al.\
2008, hereafter G08).
A04 showed,
through a study of several nearby young clusters,
that a box-shaped
region in the IRAC $[3.6]-[4.5]$ versus $[5.8]-[8.0]$ color-color diagram
(shown in Figs. \ref{IRACcc}) contains primarily Class II YSOs with
disks. A04 found that protostars were
located in the redward 
extension of this ``disk domain'' along both color axes.
G08 built upon the work of A04 to devise a more
sophisticated system for the 
classification of YSOs in the young embedded cluster NGC 1333, utilizing
multiple IRAC color-color planes. 
In Figures
\ref{IRACcc} and \ref{IRACMIPScc} the YSO selection and classification criteria of A04 and G08 are compared to our
YSO selection and Stage classification using the RW06 models.
Sources plotted in Figures \ref{IRACcc}{\it a,b} and
\ref{IRACMIPScc}{\it a,b} are color-coded according to their most
probable evolutionary Stage from the RW06 models: {\it red, orange, green},
and {\it lavender} for Stages 0/I, II, III, and ambiguous,
respectively, while {\it gray} dots represent sources best fit by
stellar photospheres, and 
{\it blue} dots show sources that were well-fit by YSO models but were
excluded from the final sample (see \S3.2 above). 
Figures \ref{IRACcc}{\it b} and
\ref{IRACMIPScc}{\it b} show only sources for which the classification
using model fitting differs from the classification using the G08
color-color criteria (sources with ambiguous Stage are not plotted).
In Figures \ref{IRACcc}{\it c} and
\ref{IRACMIPScc}{\it c}, sources 
are plotted according to their G08 color classification: protostars
are {\it red} while Class II sources are {\it orange}. Like virtually
all classification systems based upon colors, neither the A04
nor the G08 criteria allow ambiguity between
protostars and Class II sources, but we stress that {\it ambiguity is an
inherent property of any YSO taxonomy, whether based upon colors,
spectral indices, or SEDs.}

Although the
A04 disk domain contains many disk 
sources, it does not
contain {\it all} candidate disk sources and
selects a significant number of diskless stars with reddened
photospheres. The locus of Stage II 
sources in Figure \ref{IRACcc}{\it a} is redder than the disk
domain by $[3.6]-[4.5] \sim 0.3$~mag . The same
displacement was found by \citet{I07} 
for YSOs in the Eagle Nebula (M16), another massive star formation region
near the Galactic mid-plane, located ${\sim}2.5\degree$ from M17. The discrepancy
could be largely resolved by reddening the disk domain by an amount equivalent
to $A_V\approx 20$ mag, a reasonable modification considering that the
regions studied by A04 were not
highly-reddened massive star
formation regions 
in the 
Galactic plane.
The G08 color classification selects 1882 
Class II sources and 57 protostars 
that we 
find to be well-fit by reddened stellar photospheres ({\it gray} dots
in Figures
\ref{IRACcc}{\it b} and \ref{IRACMIPScc}{\it b}). These sources are a
striking feature 
of Figure \ref{IRACcc}{\it c}, where they are found predominantly to
the left of the disk domain, in the locus of photospheres (A08).
Largely on account of these sources, the total
number of YSOs selected by the A04 criteria or especially the G08 criteria is
far greater than the number of candidate YSOs in our conservative sample.
Again, the fact that the
G08 color criteria were derived through the study of a
nearby star formation region located away from the Galactic plane
would explain the large number of reddened stellar photospheres
selected as YSOs. In addition, both the A04 and G08 criteria were derived from
studies of regions that, unlike M17 and M16, lack many massive YSOs,
which should contribute to the disparities between the color-color
classifications and the model-based classifications \citep{BW04}. 

When compared to our evolutionary Stage classifications, the G08
classification system confuses a significant number of protostars for
more evolved YSOs, and vice versa.
This disagreement is evident in Figure \ref{IRACcc}, but it can be more readily
explained by examination of Figure \ref{IRACMIPScc}, in which sources
are plotted 
using a color-color diagram that combines
IRAC with MIPS [24] photometry. As shown by RW06, this color space
illustrates the value of 24 \um\ data for distinguishing
between protostars (Stage 0/I) and disk-dominated (Stage II) YSOs. 
Figure
\ref{IRACMIPScc}{\it a} shows a progression toward less evolved
sources with increasing $[8.0]-[24]$ color, and many sources with
ambiguous Stage determinations ({\it lavender}) lie along the
boundary between the Stage 0/I and II populations. 
In
Figure \ref{IRACMIPScc}{\it b}, all Stage II sources ({\it orange}
dots) plotted for $[3.6]-[5.8]>1.5$ are protostars according to the
G08 classification system (these sources are {\it red} dots in Fig.\
\ref{IRACMIPScc}{\it c}), which relies upon photometry at
wavelengths ${\le}8.0$ \um.

Two main classes of ``false'' YSOs are illustrated in the IRAC
color-color plots of Figures \ref{IRACcc}. 
(1) Candidate AGB stars (open {\it blue} circles) are intermingled with
YSOs and appear  
very difficult if not impossible to distinguish by IRAC colors
alone. (2) The lower-right part of Figure \ref{IRACcc}{\it a}, centered on
$[3.6]-[4.5]=0$ for $1<[5.8]-[8.0]<2.5$, contains numerous sources
well-fit by YSO models ({\it blue} dots) that were discarded from the
final sample 
because the IR excess emission appears at [8.0] only (see
\S3.1 above).

Spurious correlations of 24 \um\ sources with IRAC
sources can also create artificial IR excesses (see \S3.1), and
consequently many sources discarded from our final YSO sample appear
in Figure \ref{IRACMIPScc}{\it a} 
as {\it blue} dots 
 near $[3.6]-[5.8]=0$ for
$2<[8.0]-[24]<4$. Candidate AGB stars are better distinguished from
YSOs in Figures 
\ref{IRACMIPScc} than in Figures \ref{IRACcc}. 
Three candidate AGB stars (see \S3.2 above) with
$[3.6]-[5.8] \ge 1.5$ but $[8.0]-[24] < 1$ are located far
from YSOs and stellar photospheres. Several other candidate AGB stars
have $[8.0]-[24] \sim 2$, placing them on
the blue edge of the Stage II YSO locus. This
supports the interpretation that AGB stars are 
typically surrounded by dust shells with smaller effective radii than the disks
and envelopes of YSOs \citep[see][for example]{R08}.


We have experimented with the A04 and G08 color-color
criteria for selecting and classifying candidate YSOs because these
are among the 
most well-developed analysis techniques available for IRAC. It is not 
appropriate to apply these precise color cuts directly to the interpretation
of sources in 
the M17 region, because they were developed through the study of
relatively nearby ($d<1$ kpc) regions that, unlike M17, are
located in relatively low-density molecular clouds away from the
Galactic plane. This illustrates one of the
principal advantages of 
the model-fitting approach. {\it All methods for selecting populations of
objects based on broadband photometric colors are based
upon models.} The RW07 model fitter, with its integrated capacity to apply
interstellar extinction to SED models before matching them with
observed fluxes, can
be employed consistently for the study of star formation in various
environments without the need to empirically redraw the selection
criteria. A second major advantage is the ability to use
all available photometric information simultaneously to constrain the
physical parameters of large 
numbers of sources. For more on the advantages of the
model-fitting approach, see \citet{I07}.

\subsection{Contaminants in the YSO Sample}

Even the most carefully selected sample of candidate YSOs will
identify other objects that have similar IR colors. External, dusty
galaxies are the most common
contaminant in YSO searches using
deep IRAC observations of nearby star formation regions 
away from the Galactic plane, such as the c2d
Survey \citep{AP07}. GLIMPSE is 
a shallow survey, and employing the \citet{AP07} criterion for galaxy selection
to our final M17 target sample of YSOs yields no 
galaxy candidates (the control sample contains a single source that
may be a galaxy). In the
Galactic plane, the objects most likely to masquerade as YSOs in
significant numbers are evolved stars, specifically AGB stars with
dusty winds and unresolved planetary nebulae (PNe).
In addition, because we are
primarily interested in analyzing the population of YSOs associated
with M17, YSOs in the foreground or background must also be treated as
contaminants to our sample.

\begin{figure}
\epsscale{1.0}
\plotone{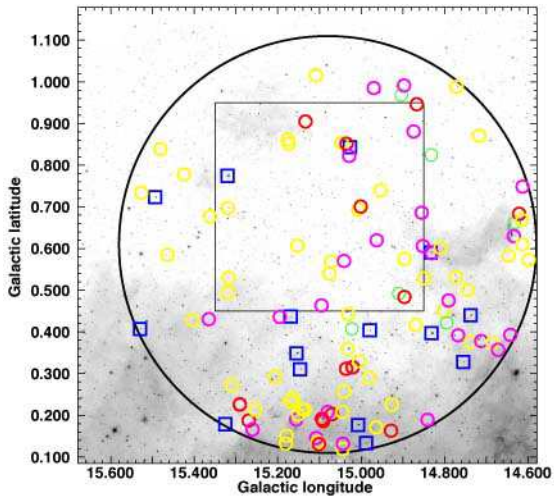}
\caption{\small GLIMPSE 8.0 \um\ image of the control field, with the same
  color overlays as in Fig.\ \ref{YSOverview}. The
  area searched for YSOs is shown by the black
  circle. The control subregion chosen to mirror the position of M17 about
  the Galactic mid-plane is the boxed area, $0.5\degree$
  on a side. 
\label{control}}
\end{figure}
A GLIMPSE 8.0 \um\ image of the control field is presented in Figure
\ref{control}, with the final sample of 106 candidate YSOs and AGB stars
overplotted 
as in Figure \ref{YSOverview}. Candidate
YSOs are distributed throughout the 
search area with an average density of 140 sources deg$^{-2}$, or nearly 60\%
of the average source density in the target sample (see Table
\ref{counts}). The source density in the control sample is not uniform,
decreasing with increasing Galactic latitude across the
1-degree span of the field, which implies a low Galactic scaleheight for
the source population. Because of this variation, we define a box
$0.5\degree$ on a side  
centered at $(l,b)=(15.1,0.7)$ in the control field, to sample directly the density of
contaminants at Galactic
latitudes mirroring the location of M17. The source density in
this control subfield is 100 deg$^{-2}$ (Table \ref{counts}).

The control sample represents the contamination expected to be present
in the sample
of candidate YSOs in the M17 target field. To estimate the
numbers of evolved stars expected in the control
and target samples, we turned to a detailed model of the IR point source sky
\citep{Sky1,Sky2}. This Sky model incorporates 87 different source
populations, including PNe and several types of AGB stars, into a
geometric representation of the Galactic bulge, halo, and thin and
thick disks. The model has been shown to agree well with \IRAS\
observations \citep{Sky1}.

\begin{figure*}[ht]
\epsscale{0.9}
\plotone{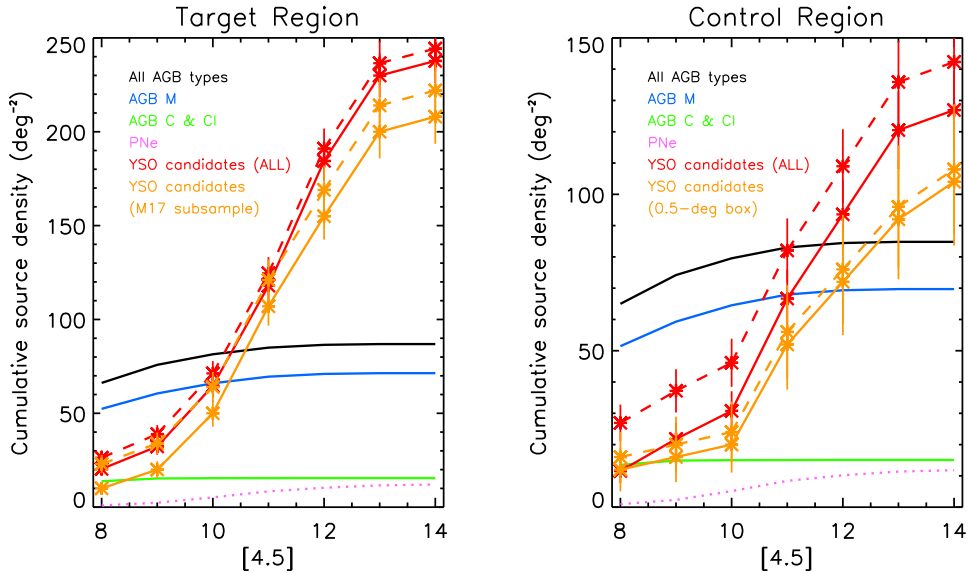}
\caption{\small Cumulative densities of different types of red sources as a
  function of [4.5] magnitude. The curves for AGB stars (AGB M are O-rich
  and AGB C \& CI are optically visible \& obscured C-rich AGB stars, respectively)  and PNe were
  generated by the Sky model \citep{Sky1,Sky2}. The solid YSO curves represent
  the target and control samples and subsamples, while the
  dashed curves show the solid curves before bright candidate AGB stars
  were removed.
\label{AGBcontrol}}
\end{figure*}
Cumulative source densities as functions of [4.5] magnitude are
presented in Figure 
\ref{AGBcontrol}. Curves generated by the Sky model for oxygen-rich AGB
stars (AGB M), optically visible/obscured carbon stars (AGB C/CI), and
PNe are plotted along with YSO candidates detected in 
the target and control samples and subsamples (see \S4). One striking
feature of these plots is the high brightness of AGB stars in the
IR. The brightest magnitude plotted in Figure \ref{AGBcontrol} is
magnitude 8, because
nonlinearity begins to affect the [4.5] band for brighter sources.
The source density of all AGB types brighter than magnitude 8 is more
than twice the density of comparably bright sources selected for our
YSO sample, even before suspected AGB candidates are removed (dashed
curves in Fig.\ \ref{AGBcontrol}). While the brightest AGB stars will
saturate the IRAC detectors and hence not
be extracted into the point-source Archive, there are
too few such sources visible in the target and control fields to
account for a significant number of the AGB stars predicted by the Sky
model. While most AGB stars in the inner Galaxy will be in
the GLIMPSE Archive, evidently not all types of AGB
star have large IR excesses that allow them to be confused with YSOs. 

Removing suspected AGB stars from the control field sample cut the
cumulative source density at magnitude 8 by more than half (the
difference between the dashed and solid {\it red} curves in the {\it
  right} panel of Figure \ref{AGBcontrol}). As the cumulative source
density with magnitude plotted in Figure
\ref{AGBcontrol} shows, the Sky model predicts that ${<}30\% =
(70~{\rm deg}^{-2}-50~{\rm deg}^{-2})/70~{\rm deg}^{-2}$ of
O-rich AGB stars ({\it blue} 
curves) and
virtually no carbon stars ({\it green} curves) are
fainter than $[4.5]=8$~mag. Suppose that {\it all} the IR excess
sources remaining in the control 
sample brighter than $[4.5]=8$~mag are O-rich AGB stars, with {\it no}
fainter AGB stars having been removed.
Scaling down the {\it black} curve to match the solid {\it red} curve
at 8 mag in
the control field plot of Figure \ref{AGBcontrol} shows that the
final control sample could still contain up to 15 AGB stars
deg$^{-2}$, or ${\sim}12$\% of the overall source density
(${\sim}125$~deg$^{-2}$; value of the 
solid {\it red} curve at $[4.5]=14$~mag). 
This represents
 the {\it maximum} possible fraction of AGB stars remaining in the
 control sample of candidate YSOs.
The fraction of AGB
stars remaining in the 
M17 target sample will be even lower (${\sim}7$\%), because the overall
source density 
is 1.7 times higher (Table \ref{counts}). 

There remains some potential contamination from unresolved
PNe. According to the predictions of the Sky model (Fig.\
\ref{AGBcontrol}), these represent $<10\%$ of sources in the control
sample and $<5\%$ of sources in the target sample.

We conclude that the majority (${>}80\%$) of sources selected as
candidate YSOs in
the control field are actual YSOs, albeit at unknown
distances. This is higher than the 60\% fraction of YSOs found in the recent
catalog of dusty red sources in GLIMPSE compiled by \citet{R08}. 
This catalog contains only sources with $[8.0]> 10$~mJy. 
Virtually all Galactic AGB stars meet this brightness criterion, but
only ${\sim}50\%$ of the sources in our control sample do. Therefore,
if we apply the [8.0] cut of \citet{R08}
to our control sample, the fraction of candidate evolved stars {\it doubles},
from 20\% to 40\%, in good agreement with the new red source catalog
despite the small size of our control sample. 
 
The density of IR excess sources drops noticeably across
the $1\degree$ increase in Galactic latitude on either side of the
mid-plane (Figs.\ \ref{YSOverview} and \ref{control}).  
This suggests that star formation, 
in the form of isolated YSOs or
small clusters,
is distributed throughout
the Galactic plane with a low scaleheight. The number of candidate
YSOs in the M17 vicinity represents a significant enhancement 
over this Galactic background YSO population, as we discuss in the
following section.

\section{The Extended YSO Population and Molecular Cloud Structure of M17}

\begin{figure*}
\epsscale{0.95}
\plotone{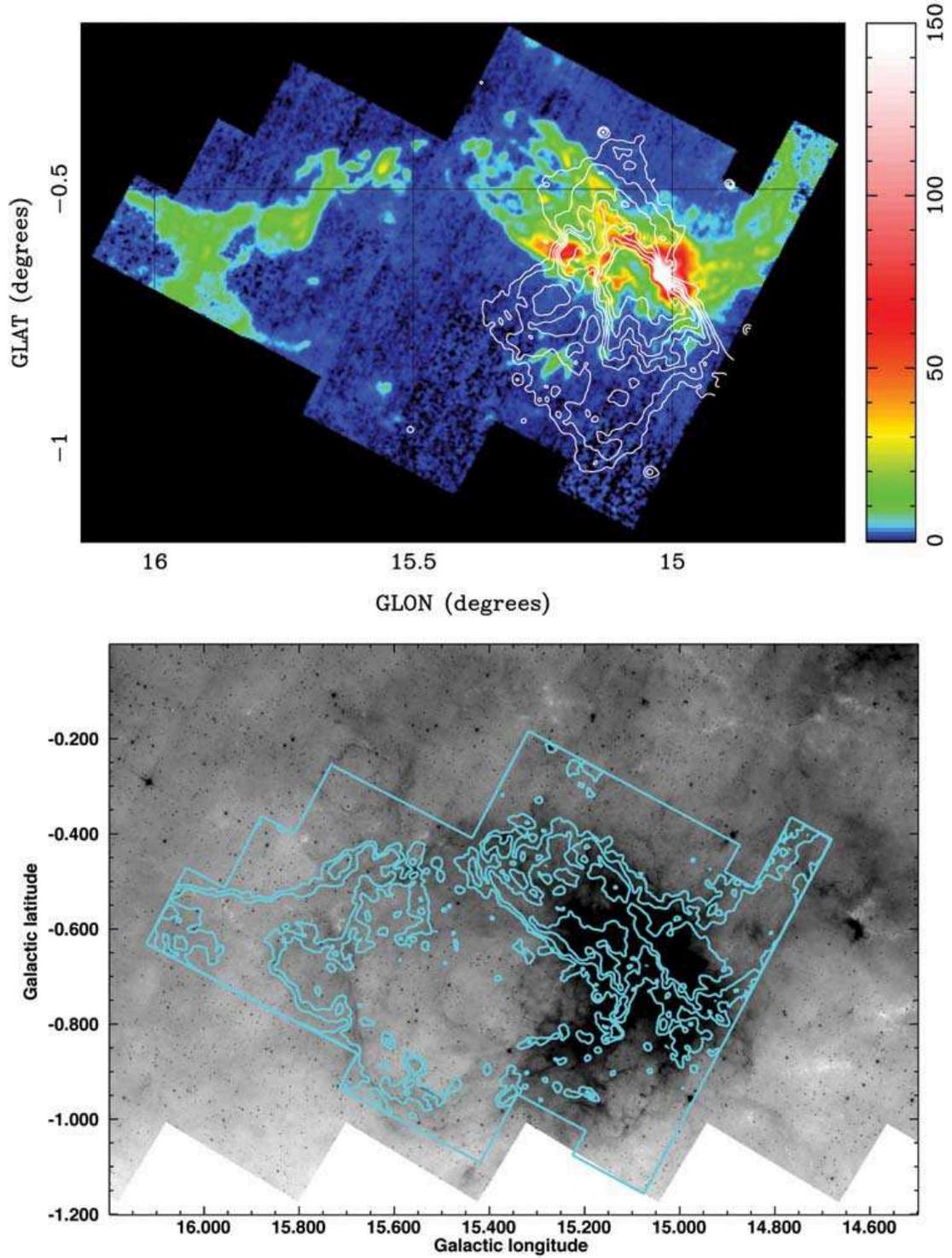}
\caption{\small CO maps of extended molecular cloud
  structures associated with M17. ({\it Top}) 
  HHT $^{13}$CO moment 0 map for $v= 12$--26 \kms. The image colorbar
  is in units of K$(T_{A}^{\star})$ \kms. Contours show 90 cm
  continuum observations from the VLA (at intervals of 5, 10, 20, 30,
  40, 60, and 80\% of the map peak) that trace thermal emission from
  ionized gas associated with the M17 \hii\ region. ({\it Bottom})
  GLIMPSE 8 
  \um\ image. Contours show HHT $^{12}$CO emission at $v=19$ \kms. \label{CO}}
\end{figure*}

A $^{13}$CO zero moment map for $v=12$--26
\kms\ and a $^{12}$CO channel ¯map at $v=19$~\kms\
are compared to the 8~\um\ image of the M17 target
field in Figure \ref{CO}. The $^{12}$CO and $^{13}$CO maps from the HHT
reveal the underlying structure of the molecular gas outlining the
extended regions of M17 along with M17 EB
and portions of MC G15.9-0.7. M17 EB is a
coherent, albeit clumpy, kinematic structure in molecular gas
at $v=19$~\kms. The CO data thus confirm that the entire dust
shell observed in the IR has a common velocity with M17, supporting
our conclusion that M17, M17 EB, and MC
G15.9-0.7 are all part of the same dynamical complex.
M17 EB, as a bubble, should exhibit a
3-dimensional, roughly spherical structure. Many astrophysical shell
structures exhibit a ring-like morphology in optically thin emission
due to limb-brightening, and emission from the interior of the
ring may be difficult to detect. Integrating the $^{13}$CO data cube spatially
over a $17\arcmin \times 
11\arcmin$ box centered at $(l,b)=(15.6\degree,-0.7\degree)$,
we definitively detect an emission line near $v=20$ \kms\ with a peak
at $\langle T_B\rangle = 0.05$ K. The line does not exhibit the
double-peaked profile of an expansion signature, which means that if
the bubble is still expanding, the expansion velocity is low, less
than a few \kms, and the line profile is dominated by internal motions
within the shell.

\subsection{Correction for Contamination from Unassociated Sources}
We have claimed that the control field sample of candidate YSOs mirrors the
population of unassociated sources contaminating the M17 target field
sample. We can use the control sample to remove contaminants
statistically from the target sample, but 
first we must verify this claim.
The molecular cloud complex associated with M17
contributes a significant enhancement to the gas column density in the
target field with respect to the control field. This additional source of
interstellar extinction could bias the 
source selection between our flux-limited samples. 

\subsubsection{Extinction Map}
To estimate the extinction through the molecular cloud, we employ a
model for CO line excitation based on an escape probability formalism
for the CO line radiation \citep{CK05}. 
The observed $^{12}$CO and
$^{13}$CO $(J=2\rightarrow 1)$ line intensities, velocity widths, and peak
brightness temperatures constrain the model at each map pixel to give
the total CO column density, $N({\rm CO})$, and the mean gas kinetic
temperature, $\langle T_{kin}\rangle$.  We estimate the mean H$_2$ density,
$\langle n({\rm H_2})\rangle$ from the derived $N(^{13}{\rm CO})$
column density. $\langle n({\rm H_2})\rangle$ varies from
$10^3$~cm$^{-3}$ in the extended cloud structures to $10^5$~cm$^{-3}$ in the M17
South core near the H II region. 
The derived $N({\rm CO})$, $\langle n({\rm H_2})\rangle$, and $\langle
T_{\rm kin}\rangle$ are then applied as
constraints on a CO PDR model based on 
\citet{vDB88} 
to estimate the far UV radiation field,
$G_o$, incident on the molecular gas, and the total hydrogen column
density $N({\rm H})=N({\rm H^0}) + 2N({\rm H_2})$ (H-atoms cm$^{-2}$).  The
corresponding extinction $A_V$, is calculated from $N({\rm H})$ using
the value $N({\rm H})/A_V = 1.9 \times 10^{21}$~cm$^{-2}$ mag$^{-1}$
\citep{BSD78}. 
Our analysis thus attempts to
incorporate the variations in both gas density and the far UV
radiation throughout the cloud complex.  The resulting  distribution
of extinction is shown in Figure 9, as a map of $A_V$ over the region
mapped in $^{12}$CO and $^{13}$CO.

The PDR model outputs the total hydrogen column density used in the
extinction calculation. Abundances measured from the M17 \hii\
region give $X=0.698$ \citep{CP08}, and including unobserved He and
metals the entire region mapped by HHT contains $1.35\times 10^5$
\Msun\ total gas mass in the velocity range occupied by M17.

\begin{figure*}[ht]
\epsscale{0.95}
\plotone{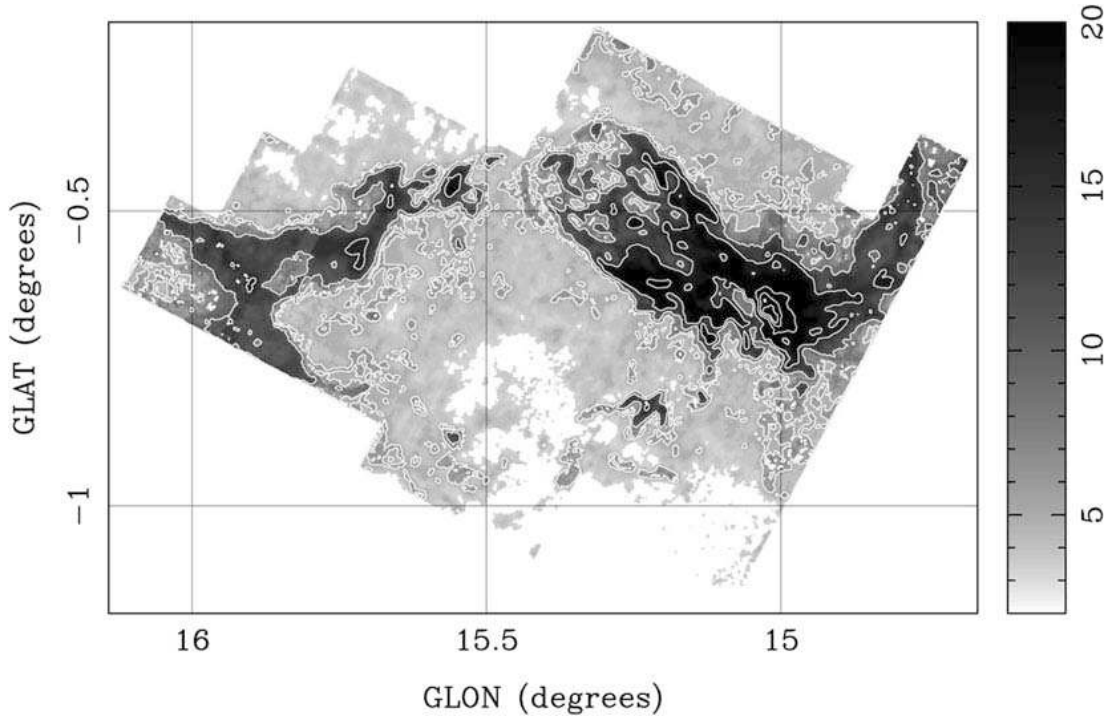}
\caption{\small Extinction map produced from CO at
  $v=12$--26~\kms. Grayscale is labeled in magnitudes of $A_V$.
  Contour levels are 
  $A_V$ of 5, 10, 15, 30, 60, and 90 mag. 
\label{extinction}
}
\end{figure*}
The extinction map in Figure \ref{extinction} represents the {\it
  differential} extinction 
between the target and control fields due to the molecular and atomic gas
associated with M17. 
The molecular gas occupies less than half of the
image area, while candidate YSOs are found throughout (Fig.\
\ref{YSOverview}). Much of the M17 molecular cloud has $A_V \ga 15$~mag,
reaching $A_V > 90$~mag in the dense core of M17 South.
The extended
outer regions of M17 and MC G15.9-0.7, however, produce $A_V<
15$~mag. 
This corresponds to $A_K < 2$ mag and $A_{[4.5]} < 1$
mag \citep{I05} generally, and much less through the interior of
M17 EB and other areas lacking high-density molecular gas. 
The map in Figure
\ref{extinction} represents the spatial average of $A_V$ over
each beam. 
Hence the extinction could be higher along  
localized sightlines through
compact, dense clumps of molecular gas that are smaller than the
32\arcsec\ (FWHM) resolution of our map (0.33 pc at 2.1 kpc), but
such clumps occupy an insignificant fraction of the mapped area.
We thus do not expect the
differential extinction between 
the target and control fields 
to produce a significant disparity in mid-IR point source detections,
except in the M17 molecular cloud itself.

The M17 molecular cloud and PDR do create significant,
spatially varying extinction and, more importantly, produce bright
diffuse mid-IR emission that saturated the MIPSGAL 24 \um\ images and
drastically reduced the point-source 
sensitivity of GLIMPSE. The surface density of GLIMPSE
Archive sources is 14\% higher in the control field than in the target
field. Much of this discrepancy is due to the sharp drop in source
detections near the M17 \hii\ region and molecular cloud, which occupy
${\sim}10\%$ of the area in the target field. We therefore consider
the control field ill-suited for comparison with the central regions
of M17, but for the remaining 90\% of our extended target field the control
sample gives an accurate picture of the contamination to the
target YSO sample.

\subsubsection{Selection of YSOs Associated With M17}
\begin{figure*}
\epsscale{1.0}
\plotone{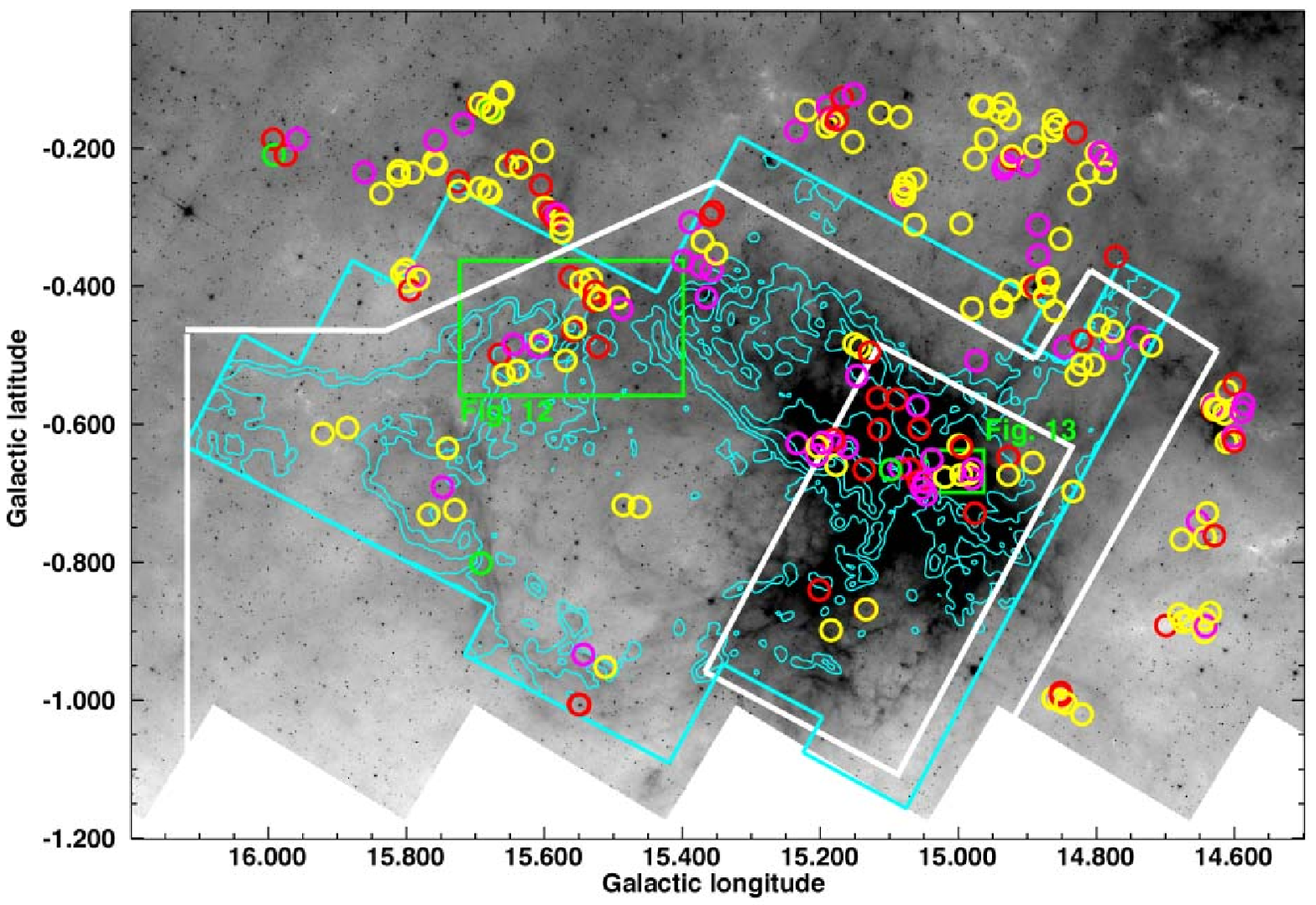}
\caption{\small Candidate YSOs exhibiting significant clustering (see
  text). Circles mark YSO positions and are color-coded according to
  evolutionary Stage as in 
  Fig.\ \ref{YSOverview}. The 8~\um\ image and CO contours are the same as in Fig.\
  \ref{CO} ({\it bottom}). The thick {\it white} boundary lines
 enclose the CO emission and pass
  through breaks in the spatial distribution of YSOs. The thick {\it
    white} rectangle shows the field that has been observed by {\it
    Chandra}, within which bright diffuse mid-IR background emission severely
  affects the IRAC point-source sensitivity. {\it Green} rectangles
  show the fields enlarged in Figs.\ \ref{IRDC} \& \ref{seqSF}.
\label{clusters}
}
\end{figure*}

The prevalence of candidate YSOs distributed throughout the M17 target
field requires that a distinction be drawn between areas that are
likely dominated by sources associated with M17 
and external star formation regions.
Using the distribution of 8 \um\ emission along with the CO velocity
structures revealed by our HHT data, we
defined a subregion of the target field that is dominated by molecular
gas associated with M17 (area bounded by {\it white} in Fig.\
\ref{clusters}). The 195 candidate YSOs within this region make up the
M17 target subsample (Table \ref{counts}). This subsample is directly
comparable to the control subsample found within the boxed
region in Figure \ref{control}.

In Figure \ref{clusters}, the background contamination to
the YSO sample has been statistically subtracted such that only
candidate YSOs exhibiting a
significant degree of clustering are plotted. This was accomplished
using a simple ``friends-of-friends'' nearest-neighbors
analysis. We first computed the median angular separation $\theta_{\rm con}$
between each source in the 
control field sample and its $N$ nearest neighbors, where $3 \le N\le
10$. 
We then calculated $\theta_{\rm targ}$, the
corresponding nearest-neighbors separation for each of the
candidate YSOs in the target sample.
The median value of $\theta_{\rm con}$ for all sources in
the control sample is $\Theta_{\rm con}$. 
Finally, we selected the subset of sources in the target field
for which $\theta_{\rm targ} < f \Theta_{\rm con}$.
The factor $f$ comes from the ratio of surface densities between the target
subsample and the control subsample $(f \equiv [\Sigma_{\rm
  con}/\Sigma_{\rm targ}]^{0.5})$, with $f= 0.7
\pm 0.1$ using the values from the bottom row of Table \ref{counts} and assuming
counting statistics.
We found that the subset of candidate YSOs selected by
this procedure 
was generally insensitive to the choice of $N$ for $N > 3$, while the
uncertainty in $f$ creates a ${\sim}20\%$ uncertainty on the total number of
sources selected. 
The final sample of candidate YSOs shown outside the {\it white}
rectangle in  Figure \ref{clusters} was selected using $N = 5$, for which
  $\Theta_{\rm 
    con}=4.28\arcmin$. For the central parts of M17 
we did not correct for contaminants because our
  control field is unsuitable, as noted above. The 34 candidate YSOs
  found within
 the {\it
    white} rectangle in Figure \ref{clusters} are presented in Table \ref{x}.
Excluding these 34 sources, we have discarded $197/372 = 0.47$ of the
sources from the original target sample. Given that we expect a
fraction $f^2 \sim 100/205 = 0.49$ of the sources in the target sample to be
contaminants (Table \ref{counts}), our cluster-finding procedure has
succeeded in removing the expected contamination.
\begin{deluxetable*}{cccccccccc}
\tablewidth{0pt}
\tabletypesize{\scriptsize}
\tablecaption{Candidate YSOs in the Region Observed By
    Chandra\label{x}
}
\tablehead{
  \colhead{Index} & \colhead{$l$} & \colhead{$b$} &
  \colhead{[4.5]} & \colhead{[8.0]} &
  \multicolumn{2}{c}{$M_{\star}$ (\Msun)} &
  \multicolumn{2}{c}{$\log(L_{\rm TOT}/{\rm L_{\odot}})$} & \colhead{Evolutionary} \\
  \colhead{No.} & \colhead{(deg)} & \colhead{(deg)} &
  \colhead{(mag)} &   \colhead{(mag)} 
  & \colhead{Min} & \colhead{Max}
  & \colhead{Min} & \colhead{Max}
  & \colhead{Stage\tablenotemark{a}}
}
\startdata
\multicolumn{5}{l}{With X-ray Counterparts} \\
\tableline
X1 & 14.9846 & -0.6718 &   9.9 &   8.7  &    2 &    3 & 1.2 & 1.6 & II \\
X2\tablenotemark{b} & 14.9855 & 0.6716 & 10.6 &   9.6  &    1 &    5 & 1.0 & 2.5 & II \\
X3 & 14.9954 & -0.6332 &   8.4 &   5.7  &    1 &    7 & 2.1 & 2.8 & 0/I \\
X4\tablenotemark{c} & 14.9962 & -0.6733 & $<$  3.6 & \nodata  &    9 &   10 & 3.7 & 3.9 & II \\
X5 & 14.9994 & -0.6304 &   7.3 &   6.9  &    5 &    6 & 2.6 & 3.2 & II \\
X6 & 15.0198 & -0.6768 &   8.6 & \nodata  &    4 &   10 & 2.3 & 3.8 & II \\
X7\tablenotemark{b} & 15.0386 & -0.6513 &   6.2 & \nodata  &    6 &   15 & 3.0 & 4.4 & Amb. \\
X8 & 15.0470 & -0.7023 &   9.9 &   8.3  &    1 &    5 & 1.1 & 2.7 & Amb. \\
X9 & 15.0530 & -0.6934 &   7.1 &   6.0  &    5 &    9 & 2.6 & 3.6 & Amb. \\
X10 & 15.0568 & -0.6068 &   9.9 &   8.1  &    2 &    5 & 1.5 & 2.0 & 0/I \\
X11 & 15.0699 & -0.6646 &   9.4 &   6.8  &    1 &    5 & 1.3 & 2.4 & 0/I \\
X12 & 15.1146 & -0.6079 &  11.4 & \nodata  &    1 &    7 & 1.1 & 2.6 & 0/I \\
X13 & 15.1613 & -0.6335 &  10.4 &   9.5  &  0.5 &    4 & 0.8 & 1.9 & Amb. \\
X14\tablenotemark{b} & 15.1772 & -0.6593 &   9.0 &   8.2  &    2 &    5 & 1.5 & 2.3 & II \\
\tableline
\multicolumn{5}{l}{Without X-ray Counterparts} \\
\tableline
NX1\tablenotemark{d} & 14.8923 & -0.6556 &  11.0 &  10.1  &    2 &    4 & 1.1 & 2.0 & II \\
NX2\tablenotemark{d} & 14.9277 & -0.6740 &   8.0 &   7.5  &    4 &    5 & 2.3 & 2.7 & II \\
NX3\tablenotemark{d} & 14.9278 & -0.6473 &  11.5 &  10.3  &  0.4 &    7 & 0.7 & 2.7 & 0/I \\
NX4 & 14.9761 & -0.7292 &  11.3 & \nodata  &  0.4 &    7 & 0.8 & 3.0 & 0/I \\
NX5\tablenotemark{e} & 14.9791 & -0.6651 &   6.6 &   3.3  &    8 &   11 & 3.5 & 3.9 & Amb. \\
NX6 & 14.9806 & -0.6787 &  11.0 &   9.9  &  0.5 &    5 & 1.0 & 2.4 & Amb. \\
NX7 & 15.0494 & -0.6838 &   7.6 & \nodata  &    4 &   10 & 2.2 & 3.8 & Amb. \\
NX8 & 15.0576 & -0.6830 &  10.8 &   8.7  &  0.6 &    7 & 1.0 & 2.9 & 0/I \\
NX9 & 15.0589 & -0.5735 &  10.8 &   9.9  &  0.5 &    4 & 0.7 & 1.5 & Amb. \\
NX10\tablenotemark{f} & 15.0848 & -0.6631 &   7.7 &   6.5  &    2 &   10 & 2.1 & 3.6 & Amb. \\
NX11 & 15.0895 & -0.5641 &  11.0 &   9.2  &  0.7 &    5 & 1.0 & 2.6 & 0/I \\
NX12 & 15.0984 & -0.6658 &   6.3 &   6.0  &    5 &   14 & 2.9 & 4.2 & III \\
NX13 & 15.1176 & -0.5618 &   9.2 &   8.3  &    3 &    7 & 2.0 & 2.7 & 0/I \\
NX14\tablenotemark{d} & 15.1337 & -0.8681 &  10.3 &   8.5  &  0.4 &    6 & 1.0 & 3.0 & II \\
NX15\tablenotemark{d} & 15.1348 & -0.8683 &  13.0 &  10.7  &  0.2 &    6 & 0.3 & 2.3 & 0/I \\
NX16 & 15.1366 & -0.6657 & \nodata &   5.8  &    7 &   14 & 3.2 & 4.2 & 0/I \\
NX17 & 15.1784 & -0.6299 &  10.5 &   9.6  &  0.3 &    3 & 0.7 & 1.6 & Amb. \\
NX18 & 15.1808 & -0.6222 &   8.2 &   6.7  &  0.5 &    8 & 1.5 & 2.9 & 0/I \\
NX19\tablenotemark{d} & 15.1844 & -0.8984 & \nodata &   5.6  &    5 &    8 & 2.9 & 3.3 & II \\
NX20 & 15.2017 & -0.8404 &   8.5 &   7.5  &    2 &    9 & 2.0 & 3.2 & 0/I \\
\enddata
 \tablenotetext{a}{The ``Amb.'' Stage designation is the same as in Table \ref{M17e}.
}
 \tablenotetext{b}{Selected as X-ray emitting candidate ``protostars'' in
   Table 7 of BFT07.}
\tablenotetext{c}{X4 is the KW Object (see Fig.\ \ref{seqSF}).}
\tablenotetext{d}{These 6 sources are located away from the bright 8
  \um\ emission of the M17 PDR, increasing the odds that they are
  unassociated contaminants.}
\tablenotetext{e}{NX5 is resolved by IRAC (see Fig.\ \ref{seqSF}).}
\tablenotetext{f}{NX10 is located at the tip a prominent pillar
  structure in the M17 H II region. \citet{ZJ02} resolve this source
  into two YSOs of comparable brightness in the near-IR, a Class I and
  a Class II/III candidate. The Class
  I source likely dominates the mid-IR emission.
}
\end{deluxetable*}

After
correcting for contaminants the final M17 YSO subsample contains 96 sources
(Tables \ref{M17e} \& \ref{x}).
Two main concentrations of candidate YSOs within the final subsample
are apparently associated 
with M17 EB and the M17 \hii\ region, respectively
(Fig.\ \ref{clusters}).
The final subsample excludes 135 candidate YSOs (Table \ref{field}) that 
either are unassociated with M17 or are associated
with the extended molecular cloud complex to the southwest
\citep{EL76,S86}. YSOs in the latter group are at the same distance as M17
and are associated with the M17 complex, but they
belong to an additional extended YSO population that is beyond the scope
of this work.

\subsection{Star Formation in M17 EB and MC G15.9-0.7}


\subsubsection{YSO Mass Function}
\begin{figure}
\epsscale{1.1}
\plotone{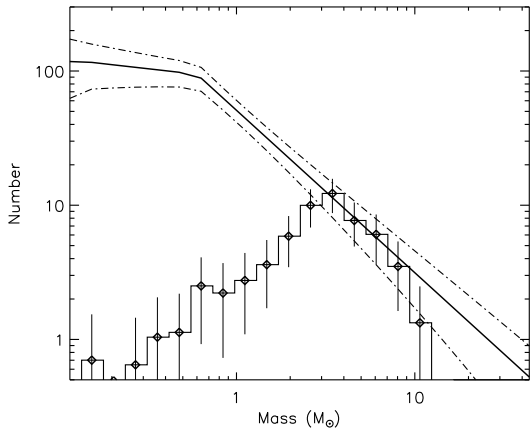}
\caption{\small Mass function of
   the 62 candidate YSOs found inside the {\it white} boundary lines but
  outside the {\it white} rectangle in Fig.\ \ref{clusters}  (histogram with error
  bars).
  The Orion Trapezium cluster IMF \citep{Muench} 
  has been
  scaled to match the mass function for ${\sim}3$~\Msun\ $< M_{\star} < 8$~\Msun\ (heavy curve with dash-dotted
  curves showing uncertainty 
  envelopes).
  \label{IMF}}
\end{figure}

We constructed a mass function for the observed YSO population following
a procedure  
employed previously by \citet{SP07} and \citet{BW08}.
Using Equation 2, we assigned a $\chi^2$-weighted
probability for the stellar mass $M_{\star}$ of each well-fit model
to a given candidate YSO. By summing over the probabilities of
all well-fit models (defined by Equation 1) and normalizing to unity,
we derived a probability distribution of $M_{\star}$ for each
candidate YSO. This procedure is similar to that employed to find the
probability distributions of 
evolutionary Stage (\S3.1), but a finer binning was used. Results of
fitting models to 62 sources in the extended M17 region (within the
{\it white} boundary lines in Figure \ref{clusters} but excluding the {\it
  white} rectangle) are presented in Table \ref{M17e}.
Ranges giving 95\% confidence intervals for
$M_{\star}$ and
$L_{\rm TOT}$ (total bolometric luminosity, including luminosity
produced by accretion) for each candidate YSO are listed in Table \ref{M17e}.
Summing
the probability distributions of $M_{\star}$ over the 62 sources in
Table \ref{M17e} produces the YSO mass function (YMF) plotted
in Figure \ref{IMF}. 


The YMF in Figure \ref{IMF} exhibits, for intermediate-mass
sources, the ubiquitous \citet{Salpeter} power-law 
slope. The turnover
for $M_{\star} \la 3$ \Msun\ is not a real break in the YMF but instead
reflects the incompleteness of our YSO sample with respect to
the detection of 
lower-mass sources. 
A completeness limit of ${\sim}3$ \Msun\ is reasonable for the
detection of disk-dominated sources. The [8.0] flux density of an A0 V photosphere with
$M_{\star}=3.2$ \Msun, $T_{\rm eff}=10,800$ K, and $R_{\star}=2.5$
R$_{\sun}$ \citep{allen} at 2.1 kpc is 3.9 mJy, or ${\sim}1$ dex above
the nominal detection limit of the point-source Archive. The extended
M17 field from which the YMF has been sampled exhibits neither
high source confusion nor very bright diffuse 8~\um\ emission, hence
our YMF represents a nearly complete sample of intermediate-mass
IR excess sources that will become main-sequence types A0 or earlier.

We estimated the size of the larger YSO population represented by our
intermediate-mass 
YSO sample
by scaling the IMF of \citet{Muench} to fit the part of our YMF ($3\la
M_{\star} \la 8$ \Msun) where it is most complete (solid curves with
dash-dotted uncertainty envelopes in Fig.\ \ref{IMF}). This procedure
generated a model IMF,
$\xi(M_{\star})$, for the YSO population.
$\xi(M_{\star})$ has a power-law slope
of $\Gamma = -1.2\pm 0.2$, as defined by
\begin{displaymath}
  M_{\star}^{\Gamma} \propto \frac{d\xi(M_{\star})}{d(\log M_{\star})},
\end{displaymath}
for $M_{\star}>0.6$ \Msun.
After the break point, $\xi(M_{\star})$ flattens to
$\Gamma = -0.2\pm 0.2$. 
Although the full \citet{Muench} IMF is a 4-part power law extending
to brown dwarf masses, we use only $\xi(M_{\star}\ge 0.1$ \Msun) for
consistency with the BFT07 
results\footnote{
Extrapolating down to the hydrogen-burning limit at
  0.08 \Msun\ increases neither the total number of YSOs nor the total
  stellar mass by a significant amount.} (see \S4.3 below).
Integrating $\xi(M_{\star} \ge 0.1$ \Msun), we predict
that the extended YSO population of M17 numbers $950\pm 90$ and
represents a total stellar mass of $750\pm 50$ \Msun. {\it These values
are lower limits.} We 
cannot be certain that our final YMF is 100\% complete even for
intermediate-mass sources. We treat any unresolved multiple sources as
single YSOs, which 
means that we are generally sensitive only to the most luminous component of a
multiple system, given that $L_{\rm TOT} \propto M_{\star}^\alpha$,
with $\alpha > 2.5$ for PMS stars \citep{BM96}.  The mass determined from the YSO fits will be
slightly higher than the most massive source in the beam but
substantially lower than the combined mass of the sources \citep{BW08}.


The YMF in Figure \ref{IMF} contains candidate YSOs distributed
throughout a large volume of space that cannot be regarded as members
of any single cluster. The YMF is, however, drawn from a coeval stellar
population comprising the most recent generation of star formation
in the extended M17 target field  (excluding YSOs associated with the central
NGC 6618 cluster). Because these sources were selected 
on the basis of 
mid-IR excess emission, and a negligible fraction were fit with
Stage III YSO models, the YMF population is characterized by optically-thick
circumstellar disks and/or infalling envelopes. The disk destruction
timescale is a decreasing function of the mass of
the central star, with
solar-mass YSOs losing their disks on timescales of ${\la}2$ Myr
\citep{HLL01}. 
The intermediate-mass YSOs dominating our sample are expected to evolve more
quickly, making them even
younger.


\subsubsection{Timescales of Evolutionary Stages and Lower Limit on
  Star Formation Rate}
The duration of the Class I phase for YSOs in low-mass star formation regions
is ${\sim}0.1$ Myr \citep{KH95,HLL00}.
We can make an independent estimate of the duration of the Stage 0/I
phase from the accretion ages of the well-fit models for each
source. We define ``accretion age'' as $t_A=10^{-6}M_{\star}/\dot{M}_{\rm
  env}$ [Myr]. The accretion age is a proxy for the
evolutionary age $t$ of a YSO, assuming both that $\dot{M}_{\rm env}$ is
constant, on average, over $t_A$ and that all of the accreting gas eventually
reaches the central star. In reality, average accretion rates decrease with
time, causing $t_A$ to overestimate $t$, but accretion-driven
outflows prevent 3--30\% of the 
infalling gas from accreting \citep{RS00}, driving $t_A$
in the opposite direction.
Combining the probability distribution of $t_A$ for
the 25 candidate Stage 0/I YSOs in Tables \ref{M17e} \& \ref{x}, we
find that $t_A$ has a median value of $0.075$ Myr, in good
agreement with other estimates of Class I lifetimes. 

Assuming a steady rate of star formation, the relative
number of Stage II versus Stage 0/I YSOs gives the relative
lifetime of each Stage. But we must take care to compare
populations with similar mass distributions. 
Most YSOs fit by Stage I models have steep, Class I spectral
indices, and even low-mass Stage 0/I YSOs can be bright in the
mid-IR. Consequently, Stage 0/I sources disproportionately populate the
YMF for $M_{\star}<3$ \Msun\ (Fig.\ \ref{IMF}), with the 12 Stage 0/I
candidates in the YMF (Table \ref{M17e})
overrepresented at 1~\Msun $\la M_{\star}\la 2$~\Msun\
compared to Stage II candidates. For $M_\star\ga 3$ \Msun,
the YMF contains 5.6 times as many Stage II as Stage 0/I candidates,
while this ratio is 2.6 for the entire mass range.
The median Stage II disk lifetime in our sample is therefore
$t_D=5.6t_A\sim 0.4$ Myr.
The above calculations exclude the single Stage III candidate and 18
candidate YSOs with ambiguous Stage determinations in Table \ref{M17e}. Nearly all of
the ambiguous candidates have model fits that are divided between
Stage 0/I and Stage II models. In the limiting case where all
ambiguous sources are treated as Stage II candidates, we have $t_D=
8.2t_A\la 0.6$ Myr. Intermediate-mass YSOs therefore lose their
disks in less than half of the corresponding ${\sim}1$ Myr timescale
observed for their solar-mass counterparts. This age differential
tends to steepen the slope of the YMF compared to a disk-unbiased IMF,
hence {\it the intermediate-mass YMF probably underestimates the true
number of low-mass YSOs.}

The characteristic age of the source population in the YMF is
$t_A+t_D\sim 0.5$ Myr.
Using the stellar mass estimated from the IMF model, we arrive at a
star formation rate (SFR) of 0.0016 \Msun~\peryr\ in the
extended target field, which applies to
M17 EB and the extended outer regions of the M17 molecular
cloud. Due to the potential incompleteness and the bias against
diskless sources inherent in the
intermediate-mass YMF, this value 
must be regarded as a   
lower limit. 

\begin{figure*}
\epsscale{1.0}
\plotone{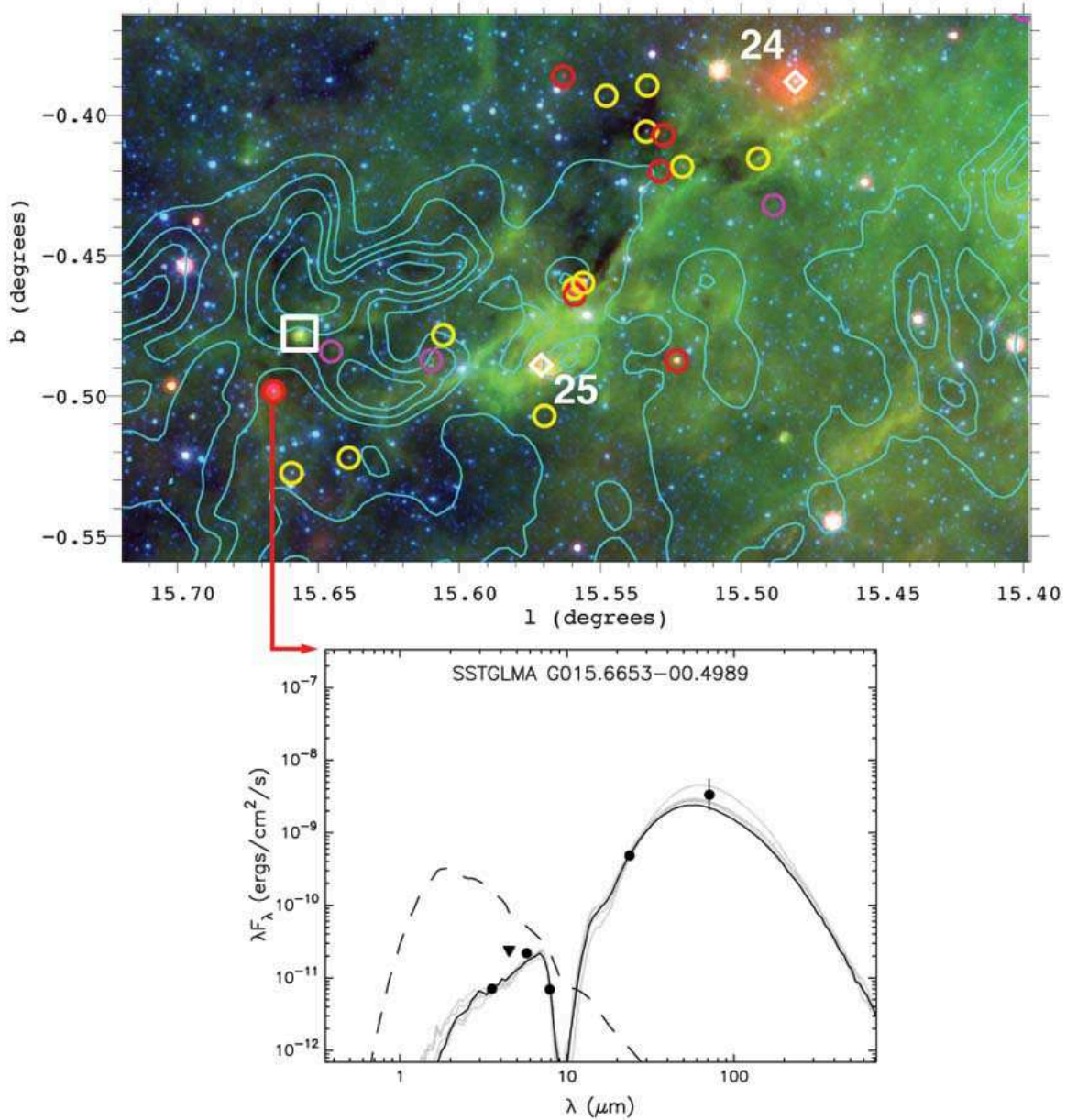}
\caption{\small Image showing part of the
  interface between M17 EB and MC G15.9-0.7 (IRAC [4.5]={\it
    blue}, IRAC [8.0]={\it green}, MIPS [24]={\it red}). The {\it
    cyan} contours are $^{12}$CO 
  emission at $v=19$ \kms. The rim of the bubble M17 EB
  is delineated by the bright {\it green} filament of 8~\um\ emission
  crossing from 
  the lower-left to the upper-right corner of the image.
  Candidate
  YSOs 
  are overplotted as in Fig.\ \ref{clusters}. Two candidate OB stars
  associated with diffuse 24 \um\ emission are marked by {\it white}
  diamonds and numbered as in Table \ref{OBstars}. The
  candidate Stage 0/I YSO G015.6653-00.4989 (E55 in
  Table \ref{M17e}) is
  highlighted, and the model fits to its SED are plotted. The [4.5]
  flux was used as an upper limit to the fitting and is plotted as a
  triangle. 
  The {\it dashed} curve shows the reddened photosphere of the central
  star as it would appear in the absence of circumstellar material.
  \label{IRDC}}
\end{figure*}
\subsubsection{Triggered Star Formation in MC G15.9-0.7?}
A concentration of star formation activity is observed in
the vicinity of an IR dark cloud (IRDC) complex centered at
$(l,b)=(15.55\degree,-0.45\degree)$. This region is shown in a
composite GLIMPSE and MIPSGAL image in Figure \ref{IRDC}. The bright,
filamentary 8 \um\ emission slicing diagonally through this image,
part of which makes up a previously identified optical reflection nebula,
outlines the rim of M17 EB where it intersects MC
G15.9-0.7 (Fig.\ \ref{fullcolor}). Because significant
concentrations of candidate YSOs are observed on the bubble rim while no
group of candidate YSOs is reliably detected in the adjacent regions
of MC G15.9-0.7 (Fig.\ \ref{clusters}), the situation is strongly suggestive of star
formation triggered by the expansion of M17 EB. 
The morphology of the IRDCs, bright 8 \um\ emission, and CO emission
provides
evidence for increased 
gas density resulting from the compression of the cloud. 

The SED of G015.6653-00.4989, 
a bright 24 \um\ source in MC G15.9-0.7
(Fig.\ \ref{IRDC}), is very similar to the SEDs of several massive
protostars associated with molecular outflows identified by
\citet{SP07}. It is characterized by excess emission at 4.5 \um\ (an outflow
signature) and a very
red IR spectral index peaking near 70 \um\ (this object is the
brightest 70 \um\ point source detected in the M17 target field). The
models fit to this source (E55 in Table \ref{M17e}) indicate a
luminous candidate protostar with mass 7 \Msun\ $\le M_{\star}\le 8$
\Msun\ and luminosity 400 \Lsun\ $\le L_{\rm TOT}\le 10^3$ \Lsun\
accreting from its 
circumstellar envelope at a high rate of $2\times 10^{-4}$~\Msun~\peryr\
$ < \dot{M}_{\rm env} < 8\times 10^{-4}$ \Msun\ \peryr. 
The accretion age of this candidate protostar is $t_A<0.04$~Myr. 
A nearby resolved, compact, red IR source (in the {\it white} box
in Fig.\ \ref{IRDC}) is associated with a second 70 \um\ point
source. The SED of this source cannot be fit well by any YSO model,
and we speculate that it may be either a luminous binary YSO or a compact
cluster. 
Both sources are apparently located in the cloud region
compressed by M17 EB.

Also shown in Figure \ref{IRDC} are 2 compact, resolved sources of
24 \um\ emission (30\arcsec--60\arcsec\ in size) that are not associated with any candidate YSO. This
emission suggests a localized region of heated dust, and the most
likely heat source is a massive young star or cluster. Five such
24 \um\ sources are observed in the M17 target field, and none 
are found in the control field. At the
center of every 24 \um\ source lies a GLIMPSE Archive source with an
SED consistent with a hot stellar photosphere. Following the method
pioneered by \citet{WP08}, we estimated the spectral 
types of these stars by employing the RW07 model fitting tool to fit
\citet{Kurucz} stellar atmospheres (including hot photospheres up to
$T_{\rm eff}= 50,000$~K) to available $BVJHK_S+$IRAC photometry. We
found that all 
of the stars  
associated with compact 24 \um\ emission have luminosities consistent
with B or late O stars, assuming they are associated with
M17 at either a distance of 1.6 kpc or 2.1 kpc
(see Table \ref{OBstars}).
Because these stars exhibit little or no IR excess emission in the 2MASS/GLIMPSE
bands, they either have already cleared out the inner regions of their
circumstellar disks or have dissipated their disks entirely. The observed
24 \um\ emission could arise from remnant disk material or heated dust
in the surrounding molecular cloud at 
distances of ${\sim}2$--$5\times 10^4$ AU from the central star. This
interpretation suggests that some of these stars may ionize
low-luminosity compact 
\hii\ regions that have not been detected by previous surveys
\citep[e.g.][]{WC89}. None of these sources is detected by the
high-resolution CORNISH survey at 6 cm \citep{CORNISH}, indicating
that any gas ionized by these B stars is diffuse enough to be resolved
out by the VLA in the B configuration. This rules out the presence of
hypercompact or ultracompact \hii\ regions around the stars.
The most extended of these  24 \um\ sources,
associated with star 24 (Fig.\ \ref{IRDC}, Table \ref{OBstars}), is
associated with 11 cm 
radio emission \citep{11cm} and enclosed by a partial ring of 8 \um\
emission at a projected distance of 0.5 pc from the star, suggesting a
small \hii\ region and PDR. 

Whatever the interpretation of the physical origin of the 24 \um\
sources, the circumstellar environments of the associated OB stars are
significantly more evolved than those of
the candidate YSOs, which exhibit IR excess emission at
wavelengths ${<}10$ \um. 
The comparison of star 25 with YSO E55, the candidate
protostar in 
Figure \ref{IRDC}, is particularly interesting. These two sources have
comparable masses (${\sim}10$ \Msun; E55 is still
gaining mass through accretion) and are apparently located within
the same molecular cloud, but they occupy strikingly different phases
of early massive stellar evolution, suggesting a difference in age 
comparable to the disk destruction timescale derived above,
$t_D\sim 0.4$ Myr. This difference in age could be 
explained naturally by triggered star formation.
The more evolved B star lies in the midst of the 8
\um\ emission defining the inner rim of M17 EB, while the young
candidate protostar
is located in a region of the molecular cloud that has only recently been
overtaken by the expanding bubble. The expansion velocity of the
bubble implied by this scenario is slow, a few \kms\ at most, which is
consistent with the non-detection of an expansion signature in the CO
line (\S4 above).

\subsection{The M17 Molecular Cloud and \hii\ Region}

The contours in the top panel of Figure \ref{CO} show a 90-cm Very
Large Array (VLA) radio image \citep{CB06} delineating the
extent of the ionized gas associated with the main M17 \hii\
region. The densest part of the M17 molecular cloud is located
in M17 South (see Fig.\ \ref{fullcolor}) along
the interface between the \hii\ region and the bright $^{13}$CO
($J=2\rightarrow 1$)
emission (colored {\it red-white} in Fig.\ \ref{CO}). The total
gas mass (molecular and atomic) contained in the main M17 molecular
cloud is ${\sim}6\times 10^4$ \Msun,
found by integrating the column density traced
by the CO emission.
The \hii\
region occupies a cavity cleared of molecular 
gas between M17 North  
and M17 South. The morphology of the
CO emission from M17 closely resembles the maps of the
region obtained by \citet{W99}, but the older maps have slightly lower
resolution and do not include M17 EB.
The cavity in the molecular cloud has been sculpted by the winds and
radiation of the O stars in NGC 6618 \citep{paper1}. 

\begin{deluxetable*}{clcccccc}
\tabletypesize{\scriptsize}
\tablecaption{Selected OB Stars Associated with M17\label{OBstars}}
\tablewidth{0pt}
\tablehead{\colhead{Index} & \colhead{} & \colhead{$l$} & \colhead{$b$} &
  \colhead{Cataloged} & \multicolumn{2}{c}{Spectral Type from SED} & $A_V$ \\
  \colhead{No.\tablenotemark{a}} & \colhead{Name\tablenotemark{b}} & \colhead{(deg)} & \colhead{(deg)} &
  \colhead{Sp.\ type\tablenotemark{c}} & \colhead{$d=1.6$ kpc} & \colhead{$d=2.1$ kpc} & \colhead{(mag)}
}  
\startdata
\multicolumn{8}{l}{O Stars in NGC 6618} \\
\tableline
1,2\tablenotemark{d} & CEN 1a,b  & 15.0562 & -0.6884 & O4+O4 V  & \nodata  & \nodata & \nodata \\
3 & CEN 43 & 15.0533 & -0.7045 & O3--O5 V & O4--O5 V & O5+O5 V & 12.3 \\
4 & CEN 2  & 15.0731 & -0.7004 & O5 V     & O5.5 V & O5.5+O7 V & 5.2 \\
5 & CEN 37 & 15.0559 & -0.6883 & O3--O6 V & \nodata & \nodata  & \nodata \\
6 & OI 345 & 15.0110 & -0.7020 & O6 V     & \nodata & \nodata  & \nodata \\
7 & CEN 18 & 15.0812 & -0.6569 & O6--O8 V & O6 V    & O6+O7.5 V  & 7.6 \\
8 & M17-S3 & 15.1032 & -0.6487 & \nodata  & O7 V    & O5 V     & 11.2 \\
9 & CEN 25 & 15.0673 & -0.6879 & O7--O8 V & O9 V    & O6.5 V   & 8.0 \\
10 & OI 352 & 14.9945 & -0.7486 & O8 V    & O5.5+O5.5 V & multiple\tablenotemark{e} & ${\sim}7$ \\
11 & OI 174 & 15.1325 & -0.5257 & O9 V    & O6 V    & O4 V or O6+O7 V  & 7.0 \\
12 & CEN 3  & 15.0658 & -0.7084 & O9 V    & O7 V    & O5 V or O7+O7 V & 3.7 \\
13 & CEN 16 & 15.0748 & -0.6460 & O9--B2 V & O9.5--B0.5 V & O8 V or B0+B0 V & 5.9 \\
14 & CEN 61 & 15.0594 & -0.6884 & O9--B2 V & O9 V  & O6.5 V or O9.5+O9.5 V & 9.9 \\
15 & CEN 27 & 15.0433 & -0.6950 & O9 V     & B1 V  & O9 V & ${\sim}9.5$ \\
16 & CEN 31 & 15.0759 & -0.6534 & O9.5 V   & \nodata  & \nodata & \nodata \\
\tableline
\multicolumn{7}{l}{Candidate OB Stars in NGC 6618PG} \\
\tableline
17 & BD-16 4831 & 15.3213 & -0.7758 & O+ & O5.5 V or O9 III & O4 V or O7 III & 4.4 \\
18\tablenotemark{f} & BD-16 4826 & 15.2604 & -0.7263 & O5 & O5.5 V or O9 III & O5 V or O6.5 III & 3.9 \\
19 & BD-16 4822 & 15.2244 & -0.6796 & B...  & O8.5 V & O6.5 V & 2.9 \\
20 & TYC 6265-1174-1 & 15.1882 & -0.7598 & B0 & B1 V & O9 V & 3 \\
21 & HD 168585  & 15.2855 & -0.7498 & B7--B8 II & B1 V & O9 V & 1.6 \\
\tableline
\multicolumn{7}{l}{Candidate OB Stars Associated with 24 \um\ Emission in M17 EB} \\
\tableline
22\tablenotemark{g} & ISOGAL YSO & 15.3320 & -0.7174 & \nodata & late B V & B1.5 V & 2.1 \\
23\tablenotemark{h} & TYC 6265-347-1 & 15.3583 & -0.6543 & O... & B2 V & B1 V & 2.6 \\
24 & BD-15 4928   & 15.4801 & -0.3889 & B & B0.5 V & O8 V & 2.8 \\
25 &  & 15.5702 & -0.4895 & \nodata & B1 V & B0.5 V & ${\sim}4.8$ \\
26 & TYC 6265-2079-1 & 15.5900 & -0.8268 & O... & B4 V & B2 V & 1.7 \\
\enddata
\tablecomments{All O stars in NGC 6618 are {\it
    Chandra} point sources (BFT07). The only star in NGC 6618PG that
  has been observed by {\it Chandra}, No.\ 20, is also an
  X-ray point source.}
\tablenotetext{a}{OB stars in each of the first two groups are
  listed in order of decreasing luminosity, while the candidate B
  stars in the final 
group are listed in order of increasing Galactic longitude.}
\tablenotetext{b}{CEN is \citet{CEN}, and OI is
  \citet{OI}. SWB M17-3 is a recently-discovered O star driving a
  stellar wind bow shock \citep{bowshock}. Other identifiers are from
SIMBAD. 
}
\tablenotetext{c}{Spectral types for NGC 6618 are from OI, CEN, \citet{HHC97},
  and \citet{VH08}; others are from SIMBAD.}
\tablenotetext{d}{CEN 1a,b refer to the main components of an apparent O4+O4
  visual binary that are saturated and marginally resolved by
  IRAC. CEN 1a,b each are spectroscopic binary systems with nearly
  equal-luminosity components \citep{VH08}.}
\tablenotetext{e}{At 2.1 kpc, the SED fits to OI 352 return a radius
  larger than that of an O supergiant. It may
  be an unresolved multiple O system (see text).
}
\tablenotetext{f}{
  BD-16 4826 is a {\it
    ROSAT} point source \citep{D03}.}
\tablenotetext{g}{This source, ISOGAL-P J182108.8-155658 from the
  catalog of \citet{ISOGAL}, exhibits a marginal IR excess at 5.8 \um\
and is not detected as a point source at 8.0 \um.}
\tablenotetext{h}{This source exhibits an IR excess at 8 \um.}
\end{deluxetable*}
\subsubsection{Prominent Massive Stars}
The spectral
classifications derived from 
SED fits to 27 prominent OB stars in the M17 target field are
presented in Table \ref{OBstars}. 
We fit \citet{Kurucz} stellar atmospheres to the broadband fluxes
of 11 of the 16 known O
stars in NGC 6618 that have reliable, unsaturated fluxes in the GLIMPSE Archive
to estimate their spectral types. Although all known O stars in NGC 6618 except
No.\ 8 have previously reported spectral types from optical and
near-IR spectroscopy and photometry \citep{OI,CEN,HHC97,VH08}, mid-IR
photometry offers the advantage of a relatively extinction-independent
determination of the luminosity in the Rayleigh-Jeans regime of the
stellar SED. We
incorporated available $B,V$ photometry \citep{OI,VH08} into each
SED to obtain more reliable estimates of $A_V$ toward each
star than would be possible from the 2MASS photometry alone.
Because the optical-IR SED shape does not vary significantly for
stars earlier than B3, the $A_V$ derived from our SED fitting
is independent of adopted distance.
As shown in Table \ref{OBstars},
our SED-based luminosity classification at 1.6 kpc agrees with previously
reported spectral types to within 2 subclasses for 8 of the O stars
analyzed in NGC 6618. 
This might seem surprising, given that the most recent distance
determination found that M17 cannot be significantly nearer than 2 kpc
\citep{VH08}. However, previous photometric determinations of spectral type
assumed that each source was a single star, and  
our method shares the same systematic bias with respect to
distance. Even spectroscopic
determinations have generally neglected multiplicity, and the
resulting spectrophotometric distance estimates contributed
to the wide adoption of the 1.6 kpc distance. 

Revising the distance from 1.6 to
2.1 kpc multiplies the luminosity of every star in Table
\ref{OBstars} by 1.7. This factor of ${\sim}2$ is suggestive. 
The majority of O stars in clusters have been observed to be in
near-equal-mass binary systems
\citep{HS08,BM08}. \citet{VH08} found that each component of
the historical O4+O4 binary CEN 1, the most luminous star system in M17, is
itself a spectroscopic binary with approximately equal-luminosity 
components and suggested that several other known O stars in NGC 6618
could also be unresolved binaries. Assuming the stars in Table \ref{OBstars}
are at 2.1 kpc, we find that the majority of O stars in
NGC 6618 are far more luminous in the mid-IR than expected from the
cataloged main-sequence spectral types. Three of these
stars, CEN 43, CEN 2, and CEN 18, each have mid-IR luminosities exceeding an
O3 V star. 
Three others, OI 174, CEN 3, and CEN 61, are unlikely to be single
main-sequence stars because their SED-derived spectral types are
${>2}$ subclasses earlier than their cataloged spectral types (Table \ref{OBstars}).
Because NGC 6618 is very young and even O4 V stars take ${\sim}4$ Myr to
evolve off the main sequence \citep{AB93}, 
it is unlikely that the massive stars in NGC 6618
are giants. We therefore give binary
classifications for the over-luminous stars in Table \ref{OBstars},
with spectral types 
estimated for each component assuming (1) that the components have nearly
equal masses and (2) that their combined light
yields the cataloged or SED-derived spectral type for a single star at
1.6 kpc. We
stress that our spectral typing for each component of the binary
systems is not definitive, but the high fraction of candidate
binary systems is a robust result. 

At 2.1 kpc, one source, OI 352, has mid-IR luminosity equivalent to a
binary O3+O3 V system. If this were the true classification, then OI
352 would compete 
with CEN 1a,b for domination of the \hii\ region, but in
fact OI 352 is located in an unremarkable part of the nebula, outside
the central wind-dominated cavity \citep{T03,paper1}.  We therefore speculate
that this star, like CEN 1, could be a 
multiple system of 3 or more later-type O-dwarfs.
BFT07 report an unusually hard (4.4 keV) thermal component
from their plasma fits to the X-ray spectrum of OI 352. Strong shocks
produced by colliding-wind
binaries are the most promising explanation for O stars with ${>}1$
keV hard X-ray emission. CEN 1a,b, CEN 43, CEN 2, and CEN 18 all
produce hard X-ray emission (BFT07), and all are
identified as candidate binaries by H08 and/or by our SED fitting, with
primaries of type O6 V or earlier. CEN 25, a solitary O6.5 V star,
lacks hard X-ray emission.
In contrast, OI 174, CEN 3, and CEN 61 are likely binary systems with
primaries of type O6 V or later, and these systems have only soft-component
X-ray spectra. Evidentally hard X-ray emission is not a 
ubiquitous property of massive binary systems, but this phenomenon may
be more common in earlier-type binaries.
CEN 16 seems anomalous, for although it boasts a very hard X-ray 
spectrum (3.5 keV), it is among the least luminous candidate O stars,
equivalent to an early B binary system (Table \ref{OBstars}).

\subsubsection{Comparison of X-ray Sources with  GLIMPSE YSOs}
The {\it white} box in the bottom panel of Figure \ref{clusters}
outlines the field observed to date by {\it Chandra} (Townsley et
  al.\ in prep., Broos et al.\ in prep.). BFT07 presented 886 X-ray
  point sources extracted from the first
  {\it Chandra} Advanced CCD Imaging Spectrometer (ACIS) observation
  of M17. 
  By scaling the X-ray luminosity function (XLF) of the Orion
  Trapezium Cluster to match the XLF of the M17
  sources, BFT07 estimated a total young stellar population of
  8,000--10,000 members (to a limiting mass of 0.1 
  \Msun) in the NGC 6618 cluster and its immediate vicinity in
  M17. BFT07 assumed a distance of 1.6 kpc, and scaling their 
  XLF to 2.1 kpc increases the total population to 11,000--14,000
  stars and YSOs.

Of the 195 candidate YSOs in our M17 subsample (Table
\ref{counts}), only 34 fall within the $17\arcmin \times 17\arcmin$
field of the ACIS 
observations. This number would be many times higher if not for
the presence 
of extremely bright diffuse mid-IR emission from the \hii\ region
and PDR that drastically reduces the point-source
sensitivity in all IRAC bands.
The source density in the GLIMPSE Archive within the ACIS
field analyzed by BFT07 is 107,000 deg$^{-2}$. This is only
${\sim}50\%$ 
of the average density of Archive sources in the larger M17 target field
(see Table \ref{counts}), {\it despite} the fact that the ACIS field
contains the dense center of the NGC 6618 cluster.
Because of this incompleteness, we cannot attempt any
meaningful quantification of the YSO population most directly
associated with NGC 6618. X-ray observations, by contrast, do not
suffer from the 
deleterious effects of bright backgrounds, and moreover the 
(near-axis) spatial resolution of ACIS is higher than that of
IRAC. 
Because very different physical mechanisms are responsible for X-ray
versus IR emission from young stars, we are interested in learning whether the
IR-excess sources selected for our YSO sample represent a subsample of
the X-ray-selected population or a distinct population.

\citet{COUP} performed a detailed analysis of the X-ray source
population observed as part of the {\it Chandra} Orion Ultradeep Project
(COUP). They found that the 838.1-ks COUP observation detected 100\%
of the known young F- and G-type young stars and YSOs in Orion but
only ${\sim}50\%$ of A-type stars. \citet{COUP} concluded that
the X-ray emission observed in the low-mass stars in the COUP sample
is consistent with an origin in stellar coronae analogous to the
solar corona, but at a considerably higher level of activity. The
disappearance of solar-type convective envelopes in more massive stars
on the main sequence could explain the drop in the COUP
completeness for intermediate-mass stars.\footnote{Microshocks in OB stellar winds are
  believed to be the primary source of soft X-ray emission from massive
  stars; this very different from the emission mechanism operating in low-mass stars.} 
\citet{COUP} also observed that X-ray emission in
T Tauri stars is partially suppressed by accretion. 
As we have shown, our IR-selected sample of candidate YSOs is highly
sensitive to intermediate- to  
high-mass objects that are still accreting via circumstellar disks
and/or envelopes. We therefore expect that our YSO sample
complements the (much larger) X-ray-selected sample, as
we preferentially select the YSOs that are the most likely to be missed
by {\it Chandra}.

We have correlated our final YSO sample with an updated X-ray source
catalog covering the entire area 
observed to date by {\it Chandra}/ACIS, with 400 ks total integration time
(Broos et al.\ in prep). 
Out of the 34 candidate YSOs in the overlap region, 14 have
X-ray counterparts. 
The 95\% confidence intervals for $M_{\star}$ and
$L_{\rm TOT}$, along with the evolutionary Stage derived from the well-fit
(Equation \ref{best}) YSO models, are given in Table \ref{x}
for each source.
The sources in Table \ref{x}
exhibit no obvious trends in mass,  
luminosity, or evolutionary Stage with which to distinguish candidate
YSOs with X-ray counterparts from those without. The majority of the candidate
YSOs in Table \ref{x} are fit with models of intermediate to high
stellar mass corresponding to spectral types A and B on
the main sequence. There may be a slight preference for less-evolved
sources with higher accretion rates in the X-ray quiet population in
Table \ref{x}, but this is not significant, given the small sample size. 
Intrinsic X-ray variability may play a role in determining which YSOs are
detected. It is also possible that some of the luminous YSOs with ACIS
detections are themselves X-ray quiet but possess binary
companions that are X-ray bright. {\it There is no evidence that the 34
candidate YSOs listed in Table \ref{x} represent a
distinct population from the X-ray-selected sources.}

\begin{figure}
\epsscale{1.2}
\plotone{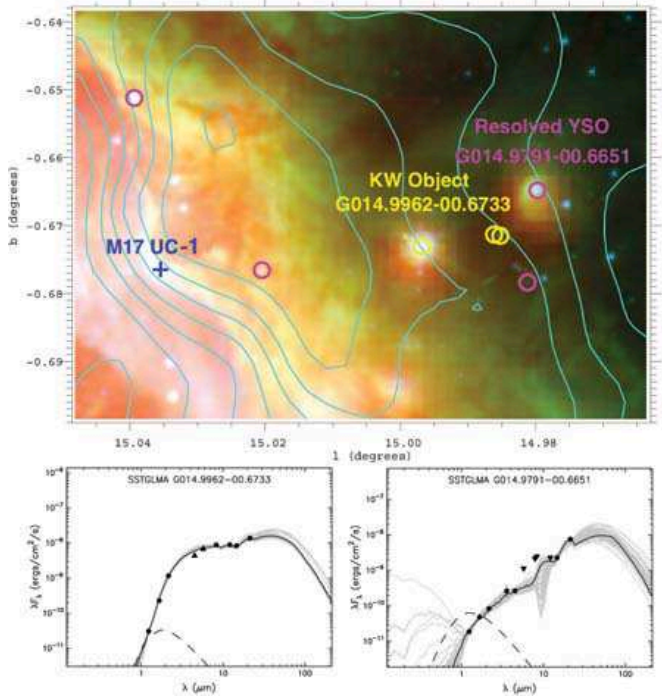}
\caption{\small Region containing the peak molecular gas density in M17, near
  the \hii\ region--PDR interface. The image combines IRAC [4.5] ({\it
    blue}) and [5.8] ({\it 
    green}) with \MSX\ E (21.3 \um; {\it red}). The {\it cyan}
  contours are the HHT $^{13}$CO
  moment 0 map from the {\it top} panel of Fig.\ \ref{CO}. Candidate YSOs are
  overplotted as circles using the color scheme of Fig.\
  \ref{YSOverview}. The ultracompact \hii\ region M17 UC-1 is
  marked. The KW Object ({\it left}) and neighboring resolved YSO ({\it
  right}) are also
  highlighted, and the model fits to their SEDs are plotted as in
  Fig.\ \ref{IRDC}. Fluxes used as upper or lower limits for the model
  fitting are plotted as downward- and upward-pointing triangles,
  respectively. 
\label{seqSF}}
\end{figure}
Two interesting candidate massive YSOs are located near the peak molecular 
gas density of M17 South and the interface with the \hii\
region, shown in
Figure \ref{seqSF}. One, the KW Object, is a binary system. A candidate
Herbig Be star is the more luminous component (KW-1) and dominates the
IR emission 
\citep{ZJ02,RC04a}. The lower-luminosity component (KW-2) is X-ray
bright (BFT07) but unresolved in our IR survey data, hence the KW
Object is considered an X-ray detected source (X4)  
in Table \ref{x}. KW-1 is saturated in the
GLIMPSE and MIPSGAL images, so models were fit to its 2MASS and \MSX\
fluxes only. The model fits indicate a central star
of $M_{\star}\sim 10$ \Msun\ and $T_{\rm eff}\sim 26,000$ K,
consistent with an early B star and in
good agreement with the findings of \citet{RC04a}. The models indicate
that the star may have a substantial circumstellar disk ($0.01$ \Msun$\le
M_{\rm disk} \le 0.3$ \Msun) with a large inner hole of radius ${\sim}40$
AU (${\sim}10$ times the dust sublimation radius). The inner hole is
responsible for the precipitous rise in the near- through mid-IR
SED (Fig.\ \ref{seqSF}) and indicates that the central star has ceased accreting material
from the disk. Our model SEDs produce a broad 9.7 \um\ silicate
feature in emission, 
but ISOCAM spectra presented by \citet{RC04a} show that the ``bumps''
in the SED near 8 and 11 \um\ are actually 
due to PAH emission features excited by the hot, central star. We have
therefore used the {\it MSX} A and C bands as upper limits to the SED
fitting, since these bands contain PAH features that increase their
flux with respect to the continuum provided by the YSO models.
The 1-D envelope model 
employed by \citet{RC04a} predicts the silicate feature in absorption
and does not explicitly include a disk. 
These authors report a total circumstellar mass of 10 \Msun\ within
0.1 pc from the star. The total mass and envelope radius are very
difficult to constrain in the absence of mm/sub-mm data, hence we
cannot confirm these reported values.

 The other prominent candidate YSO highlighted in Figure
\ref{seqSF}, G014.9791-00.6651, is resolved by IRAC. This
source, NX5 in Table \ref{x}, is not detected in X-rays. Like KW-1,
NX5 is massive ($M_{\star}
\sim 10$ \Msun), but it may be somewhat
less evolved. The 
parameters of the 47 well-fit models poorly constrain the
circumstellar environment of NX5, allowing for a wide range of both
$M_{\rm disk}$ and 
$\dot{M}_{\rm env}$, so the evolutionary Stage is ambiguous, split
between 60\% Stage II and 40\% Stage 0/I models. The resolved outer
envelope of this star is large, with radius ${\sim}10^5$ AU, and the SED
exhibits the signatures of a PAH spectrum. As with our modeling of
KW-1, we set the bands most affected
by PAHs as upper limits for the model fitting, and the
majority of well-fit models cluster around $T_{\rm eff}\sim 23,000$ K,
hot enough for the star to produce sufficient UV radiation to excite
PAHs in its circumstellar material. Together with M17 UC-1, the
nearby hypercompact 
\hii\ region that is excited by a B0 star \citep{MS04},
KW-1 and NX5 are 3 very luminous YSOs with very similar stellar masses
but with very different circumstellar environments. This
demonstrates that, in spite of their close proximity to each other, these
3 sources are not precisely coeval.


\subsubsection{Challenges of Identifying Cluster Members and
  Determining the Age of NGC 6618}
The fraction of stars exhibiting IR excess emission due to
circumstellar disks is a
diagnostic of the age of embedded clusters \citep{HLL01b}.
BFT07 find that only ${\sim}10\%$ of the sources in their X-ray
selected sample exhibit IR excesses. This fraction is very low in
comparison to recent results using IR-selected source
samples. \citet{ZJ02} found a $JHK$ excess fraction of 53\% in the
central regions of the NGC 6618 cluster, and
\citet{VH08} reported a 74\% excess fraction from
deeper $JHKL$ observations 
of a similar field. Some variation in excess fraction is
inevitable due to differences in sample selection, and 
excess fractions measured in the
$JHKL$ color plane are expected to be higher than those in the $JHK$
color plane. The low excess fraction reported by BFT07 is partially
explained by selection biases against embedded sources and 
against low-mass sources, both of which are fainter in
X-rays. To gain depth beyond the 2MASS and GLIMPSE catalogs, BFT07 correlated
their X-ray sample with
SIRIUS $JHK_S$ photometry presented by J02, hence the difference
between the excess fractions reported by these two 
works reflects the different 
selection criteria and analysis techniques rather than differences in
IR photometry.

The large discrepancy between reported IR excess fractions
reflects a more basic issue: The M17 \hii\ region is a
difficult place to search for 
YSOs. The diffuse IR emission is bright at all wavelengths
with a steep red spectral index and exhibits complex
morphological structure on multiple spatial scales \citep{paper1}. This makes
reliable mid-IR photometry difficult even for bright sources and impossible for
faint ones. Nebular structures can be mistaken for IR
excess emission associated with stars. 
As
\citet{bowshock} note, two of the high-mass Class I sources identified by
\citet{MN01} are actually mid-IR emission from
stellar-wind bow shocks that could be 
mistaken for circumstellar disks. \citet{CL91} claimed that the
majority of high-mass stars in NGC 6618 display $JHK$ excess, but
this was disputed by \citet{VH08}. 
Among the O stars in NGC 6618 detected by GLIMPSE, all are detected at
wavelengths as long as 4.5 \um\ and many are detected at 8.0 \um, but
we can state with high confidence that none displays an IR excess, with
the possible exception of star No.\ 8 in Table \ref{OBstars}, which is
associated with a third mid-IR bow shock \citep{bowshock}.

Near-IR photometry can be more reliable, because
the diffuse emission is less bright.
But $JHKL$ observations are challenged by the large spatial
variations in interstellar extinction across the M17 cloud (see the
$A_V$ values in Table \ref{OBstars}, for example). While the
extinction due to molecular gas in the M17 North and South cloud
components is generally $A_V>15$ mag, and the dense core of M17 South produces
$A_V> 90$ mag, the average extinction through the central cavity is
$A_V< 15$ mag (Fig.\ \ref{extinction}). 
The average extinction observed toward the
ionizing stars of the M17 \hii\ region is $A_V\sim 8$ mag \citep[Table
\ref{OBstars};][]
{CEN,HHC97,VH08}.  
The foreground
extinction to the cloud is  $A_V\sim 2$ mag \citep{geo,VH08}.
We thus deduce that the column density of the M17 cloud is divided into
approximately equal halves behind ($A_V \sim 7$~mag) and in front ($A_V \sim
6$~mag) of the \hii\ 
region. 
Including foreground extinction, the
 total extinction along a typical line-of-sight through the
cavity is $A_V < 17$~mag, or $A_K <
2$~mag. Hence background stars are detectable even in
relatively shallow near- and
mid-IR observations along the line of sight through the central cavity
and the ionizing cluster. In contrast, adjacent dense regions of the M17 North
and South cloud
components produce sufficient extinction to screen out background
stars against even deep near-IR observations. Because of this complex
extinction morphology, there is no comparable
nearby field that can be used to estimate the background
contamination in observations of NGC 6618. Any attempt to measure accurately the
disk fraction or low-mass IMF in NGC 6618 must overcome this
significant challenge.

A major strength of X-ray observations is the ability to separate
young cluster members, which are 2--3 orders of magnitude brighter in X-rays
than older main-sequence or evolved stars, from the field star
population (BFT07 and references therein). The contamination from
unassociated sources in the X-ray sample of the M17 young stellar population
is therefore very low, ${<}5\%$ (BFT07). \citet{ZJ02} recognized the
difficulty in using their off-cluster field to correct for source
contamination and instead used a Galactic disk population synthesis
model. While we are not in a position to verify their results, the
conservative upper limit of 3 Myr on the cluster age derived by
\citet{ZJ02} is reasonable. 
H08 assumed that all stars with IR excess,
and by extension all sources with reddening equivalent to $A_V \ga
10$~mag, were cluster members. H08 therefore concluded that the M17 molecular
cloud behind NGC 6618 
was opaque. 
Both the assumption and the conclusion were
not valid. The extinction
through the central cavity is relatively low ($A_K < 2$~mag)
and the deep 
observations used by H08 detected stars as faint as $K=20.1$ mag and
$L=15.6$ mag. 
The sample of H08 therefore contains a potentially large fraction of
background stars, casting doubt upon the reported disk fractions and
the correspondingly young age of 0.5 Myr claimed for the M17 stellar population.

It is tempting to conclude that the inclusion of diskless field stars in a
magnitude-limited sample of the M17 young stellar population would
decrease the measured IR excess fraction. 
But stars with disks are not well-separated from reddened stellar
photospheres in the $JHK$ or even in the $JHKL$ color-color diagrams
\citep{BW03b,grid}. 
Color-color spaces or SEDs that include photometry at 4.5 \um\ and
longer wavelengths provide a significantly 
more reliable means for distinguishing circumstellar extinction from
interstellar extinction.
Of the 64
candidate X-ray emitting protostars identified by BFT07 (their Table
7) most of them on the basis of
the $K-[3.6]$ color (analogous
to $K-L$), only 3 were selected as YSOs by our more conservative
multiband SED fitting (see Table \ref{x}), but 16 were well-fit 
by reddened stellar photospheres (the remaining sources are detected
in too few 2MASS/GLIMPSE bands to be fit reliably with SED models). 
Highly-reddened contaminating sources that are intrinsically red, such
as the numerous red giants in the Galactic plane behind the M17 \hii\
region, are the most 
difficult to separate from YSOs on the basis of $JHKL$ colors
alone. If correction for contamination is 
not done, then highly-reddened, background red giants may provide a
significant number of {\it false} IR excess sources. This would
artificially increase the derived disk fraction, leading to an
underestimate of the cluster age.

We are left without a strong constraint on the age of the M17 young
stellar population. There is a population of PMS stars
lacking inner disks (BFT07), and there is also ample evidence for ongoing star
formation (Nielbock et al.\ 2001, J02, this work).
The preponderance of observational evidence indicates that M17 has not
formed all of its stars 
simultaneously, but has experienced continuous or episodic star
formation over a period of ${<}3$ Myr. Sources like the candidate
massive YSOs and M17 UC-1 shown in Figure \ref{seqSF} are almost
certainly younger than 0.5 Myr, the average age of the extended YSO
population (\S4.2 above).
To minimize the impact of the lingering age uncertainty, we adopt a
mean age of ${\sim}1$ Myr for the young stars and YSOs in NGC 6618 and 
the surrounding molecular 
cloud. 

We can use the BFT07 results to
make an estimate of the star formation rate (SFR) that has produced the
ionizing cluster of M17 and its entourage of YSOs. Scaling the stellar
initial mass function (IMF) measured by \citet{Muench} for the Orion
Trapezium Cluster to the 11,000--14,000 stars in M17, we arrive at a
total mass in stars of 8,000--10,000~\Msun. 
Stars have been forming in M17 at an
average rate of 0.008--0.01~\Msun~\peryr.

\section{A History of Propagating Massive Star Formation}

We have shown that the major sites of star formation in the extended
M17 region are located around the rim of the large bubble
M17 EB. We will now examine the structure and possible origins of M17
EB and show that the beginnings of a plausible star formation
history of M17 are found within this bubble.

\subsection{The Origin of M17 EB}

The most likely production mechanisms for a bubble the size of M17 EB
are the expansion of an \hii\ region or a supernova explosion, both of
which are expected to occur in regions of massive star formation. We
favor the \hii\ region interpretation. The morphology of
the mid-IR emission outlining M17 EB is similar to that of other 8
\um\ bubbles outlining the PDRs of \hii\ regions
\citep{bubbles,WP08}. 
 One of the best ways to discriminate between \hii\ regions and
supernova remnants (SNRs) is by the radio continuum
spectral index. \hii\ 
regions are filled with thermal emission from ionized gas and exhibit
a flat spectral index for optically thin radio emission, while SNRs
are often characterized by steep, non-thermal spectral indices from
synchrotron emission \citep[e.g.][]{CB06}.
M17 EB
is included in two single-dish radio Galactic plane surveys of
comparable spatial resolution.
Images at 3 cm from the
Nobeyama 45-m telescope \citep[HPBW=2.7\arcmin;][]{nobe} and at 11 cm from the Effelsberg
100-m telescope \citep[HPBW=4.3\arcmin;][]{11cm}\footnote{See
  http://www.mpifr-bonn.mpg.de/survey.html.} show
a plateau of radio continuum emission that fills the half of M17 EB
farthest from M17 (other half is dominated by emission from
the M17 \hii\ region 
itself) and drops sharply at the location of the 8~\um\
emission outlining M17 EB. Comparison of the 3~cm
and 11~cm surface brightnesses at various points in the interior of
M17 EB shows that the radio continuum flux density is approximately constant (with
a typical value of ${\sim}50$ mJy) over
this wavelength range, consistent with a flat, thermal emission
spectral index. 

One caveat regarding the interpretation of the radio emission within
M17 EB should be kept in mind. The non-thermal emission from SNRs
eventually fades (on the order of a few $10^5$ yr) to the point of becoming
indistinguishable from the background. The residual thermal emission
from the parent \hii\ region lasts much longer. Hence, we cannot rule
out the possibility that a supernova has occurred within M17
EB. But because the observed thermal radio emission 
completely fills the bubble, there is no need to invoke a
supernova explosion to explain its origin.

\begin{figure}
\epsscale{1.2}
\plotone{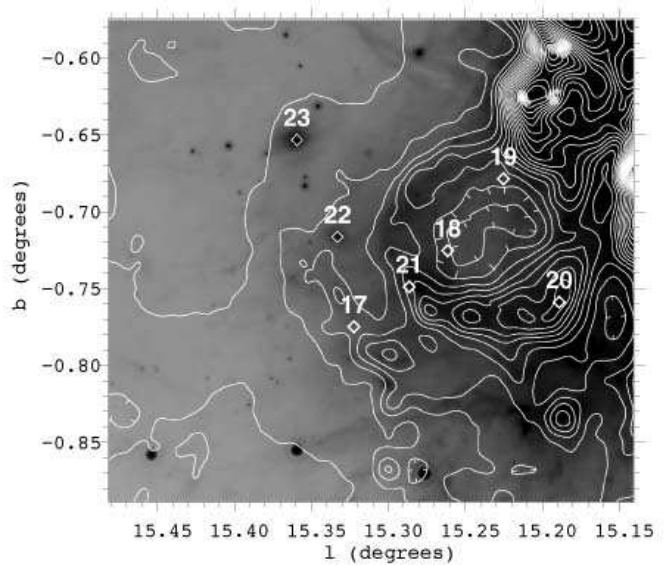}
\caption{\small MIPS 24 \um\ image highlighting the OB stars of NGC 6618PG
  (heavy diamonds) and the
  90 cm radio ring (contours). Two candidate B stars associated with
  24 \um\ emission 
  are indicated by thin diamonds. Source numbers are from
  Table \ref{OBstars}.
  \label{ring}}
\end{figure}

A search of SIMBAD objects with coordinates inside M17 EB reveals a
group of 5
stars with OB spectral classifications. Although a few of these stars
have proper motion estimates \citep{Tycho}, none have parallax
distances. The spectral types of these sources listed in the SIMBAD
database are of unknown provenance and questionable quality, so while
we list the catalogued spectral types for reference in the lower 2 sections of Table
\ref{OBstars}, we trust the classifications only to the extent that they
identify hot stars.
Again following \citet{WP08}, we  fit SEDs generated 
from \citet{Kurucz} stellar atmospheres to available $B$, $V$, 2MASS, and
GLIMPSE fluxes of
these stars, and we find that their IR luminosities are consistent with
very early spectral 
types at either 1.6 kpc or 2.1 kpc (Table \ref{OBstars}). If
  the stars were at a significantly
 different distance, their SEDs would be
inconsistent with early-type stars. We therefore
suggest that these stars are the 5 most massive members of NGC
6618PG, the ionizing cluster driving M17 EB (Figs.\ \ref{ring} and \ref{overview}).  

The OB stars of NGC 6618PG
are significantly less 
reddened ($A_V\le 5$ mag) than the O stars in NGC 6618, as expected
since NGC 6618PG is located in a 
much more diffuse \hii\ region. NGC 6618PG is associated with both H$\alpha$
emission and {\it ROSAT} soft X-ray emission \citep{D03}. The soft
X-ray emission is also apparent in an archival {\it Einstein}
observation taken 13 years earlier than the {\it ROSAT} survey. 
Significant differences in 
morphology between these 2 images show that the X-ray emission is
variable on ${\la}10$-yr timescales. We therefore infer that the emission
is dominated by unresolved X-ray-bright point sources; these are the young
stars of the NGC 6618PG cluster. High-resolution X-ray observations of
the interior of M17 EB with {\it Chandra} or {\it XMM-Newton} are
needed to separate individual cluster members from field stars and to
detect any truly diffuse X-ray emission. 

The NGC 6618PG cluster must be significantly more evolved than NGC
6618 because, even in the absence of bright diffuse mid-IR emission
from the interior of M17 EB, we do not detect {\it any} candidate YSOs
above the background level in the vicinity of the candidate ionizing
stars and X-ray emission. Given that elsewhere in the target field we detect at
least a few solar-mass stars with optically thick disks, the absence
of candidate YSOs means that the age of
NGC 6618PG is ${>}2$ Myr \citep{HLL01}. We therefore consider the
possibility that the OB stars have evolved off the main sequence, and
list possible spectral types of luminosity class III for the most
luminous stars, BD-16 4831 and BD-16 4826 (Nos.\ 17 and 18 in
Table \ref{OBstars}). These stars place an upper limit on the age of
NGC 6618PG. Assuming negligible mass loss, an O6.5 III star
spent ${\sim}4.5$ Myr as an O5 V star followed by ${\la}0.5$ Myr of
post-main-sequence evolution \citep{AB93,MSH05}, for a maximum age of
${\sim}5$ Myr.
Stars 17 and 18 are the best candidates for ionizing the \hii\ region
inside M17 EB, although 
we cannot rule out the possibility that other luminous stars once
contributed to the ionization and have since gone supernova. 

Locations of all 5 candidate ionizing stars in NGC 6618PG
and the surrounding 
interior of M17 EB are marked on a 
MIPSGAL 24 \um\ image in Figure \ref{ring}. Also shown are contours of 90
cm radio continuum from VLA observations \citep{CB06}. An
obvious feature of this image is a ring-shaped
structure, centered at $(l,b)=(15.23\degree,-0.72\degree)$ and
${\sim}6\arcmin$ in diameter, that is apparent in both the IR and radio
continuum. 
While this structure is superficially suggestive of a SNR, careful
analysis of available VLA data at 20 cm and 90 cm 
combined with 3.6 cm observations from the commissioning of the 100-m
Robert C. Byrd Green
Bank Telescope\footnote{See
  http://wiki.gb.nrao.edu/bin/view/Observing/M17publicImage.} reveals
that the radio spectral index is consistent with thermal
emission. There is no significant difference between the spectral index of
the radio ring and the rest of the M17 \hii\ region.
The 24 \um\ morphology is reminiscent of a stellar-wind bubble
produced by a luminous star \citep[like bubble N49 in][]{WP08}, but
there is only a faint hint of a PDR in the GLIMPSE 8.0 \um\ images, and
this structure is not detected in the HHT CO maps. We have fit SEDs to
every Archive source located within the ring, and we find no candidate
early-type star apart from those included in Figure
\ref{ring} and Table \ref{OBstars}. Star 17 is the closest O star to
the center of the ring, 
but it is not central enough to explain the ring
geometry. 
The origin of the ring
is therefore puzzling. 
Recent structural models of the M17 \hii\ region suggest that the
blister has broken to the North \citep{geo,paper1}, and hence could
vent energy into the interior of M17 EB. The ring structure may be
material blown from the side of the M17 \hii\ region
by such venting. This raises the possibility that
NGC 6618 contributes energy to M17 EB. 

\subsection{Constraints on the Expansion Timescale of M17 EB}

M17 EB is a PDR around a faint, diffuse \hii\
region. 
In the classical model for an expanding \hii\ region (neglecting stellar winds), 
the timescale for expansion to a radius $R$ is given by
\begin{equation}\label{phototime}
  t_{\rm phot}[{\rm Myr}] = (3.175 \times 10^{-14}) \times \frac{4R_S}{C_{\rm
      II}}\left[\left(\frac{R}{R_S}\right)^{\frac{7}{4}} - 1\right],
\end{equation}
from Equation 12-20 of \citet{LS78}. 
We assume typical properties of ionized gas at $T_e=8,000$ K:
Isothermal sound speed $C_{\rm II}=10$ \kms\ and recombination
coefficient $\alpha^{(2)}=3.09\times 10^{-13}$ cm$^3$ s$^{-1}$. The
Str\"{o}mgren radius $R_S$ depends upon the ambient ISM gas density
$n_0$. We can make a rough estimate of $n_0$ from the column density
of molecular gas swept up by the expansion of the bubble. Using the
column density of molecular and atomic gas at $v=12$--26 \kms\ derived
from the CO emission 
(see \S4.1), the total gas mass in M17 EB (including M17
North and part of MC G15.9-0.7) is ${\sim}50,000$ \Msun. If we model M17 EB
as a thin spherical shell with diameter $2R= 20$ pc, then redistributing
the gas throughout the shell interior yields a mean ambient density of
$n_0\sim 350$ cm$^{-3}$, consistent with an inhomogeneous molecular
cloud environment. 
Photoionzed expansion driven by stars 17 and 18 
in NGC 6618PG cannot produce an 
\hii\ region the size of M17 EB within the stellar lifetimes ($t_{\rm
  phot} > 5$~Myr). 


The expansion of M17 EB could instead be dominated by stellar winds. 
According to the analytical model of \citet{W77}, the
timescale for expansion of a windblown bubble to radius $R$ is
\begin{equation}
  t_{\rm wind}[{\rm Myr}] = \left[n_0\left(\frac{R({\rm
      pc})}{27}\right)^{5}\left(\frac{L_{\rm wind}}{10^{36}}\right)^{-1}\right]^{\frac{1}{3}}.
\end{equation}
Typical wind luminosities for massive stars are $L_{\rm
  wind}/L_{\rm bol}=2.5\times 10^{-3}$ for dwarfs and $L_{\rm
  wind}/L_{\rm bol}\sim 5\times 10^{-3}$ for giants \citep{VdKL01}. 
For a given $n_0$, wind-driven expansion produces a large bubble much more
quickly than photoionized expansion. The \citet{W77} model
with $n_0 \sim 350$ cm$^{-3}$ and the same 2 driving stars produces
M17 EB in $t_{\rm wind} < 1$~Myr.

The analytical models for both photoionized expansion and wind-driven
expansion are oversimplifications of the physics governing the structure
of \hii\ regions. 
Both models underestimate the expansion timescales because they neglect
radiative cooling by dust and, in the case of expansion driven by gas
pressure (Equation 3), depressurization caused by breaks
in the bubble. 
Given the above caveats, the ideal case of photoionized expansion
gives a lower limit on $t_{\rm phot}$ that exceeds the stellar
lifetimes. In the wind-driven case, unless
radiative cooling increases $t_{\rm wind}$ by a factor ${>}10$,
expansion driven by wind 
momentum (Equation 4) 
from the 2 most
luminous stars in NGC 6618 PG can produce the bubble  
in $t_{\rm wind} < 5$ Myr. In the other limiting
case of $t_{\rm wind} > 2$ Myr (see \S5.1 above), the average 
expansion velocity of the bubble is 4.5 \kms, and the current velocity
is lower because the expansion slows with time. The signature of
such slow expansion would be undetectable in our spatially-integrated CO
line profiles (\S4).

\subsection{Sequence of Events}

We now reconstruct a plausible star formation history for M17 based
upon the wealth of observational data presented in this work.

\begin{figure*}
\epsscale{1.0}
\plotone{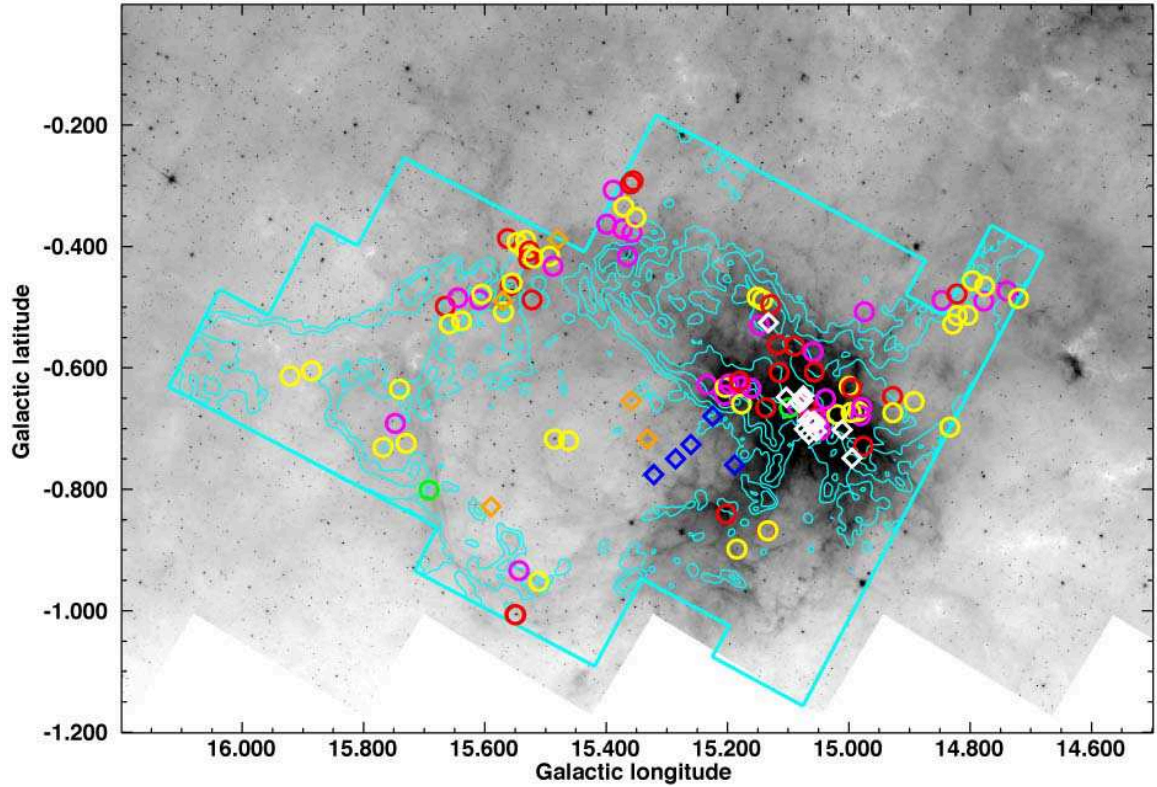}
\caption{\small Overview of the extended massive young stellar population
  associated with M17 described in this work. {\it Cyan} contours of
  $^{12}$CO emission at $v=19$ 
  \kms\ are 
    overlaid on a GLIMPSE 8 \um\ image. Candidate YSOs
  associated with the extended M17 molecular cloud and M17 EB
  are marked by colored circles as in Fig.\ \ref{clusters}. 
Diamonds mark the 
  known O stars of 
  NGC 6618 ({\it white}), the 5 most massive OB stars detected in NGC
  6618 ({\it blue}), and the 5 candidate OB stars identified by
  associated diffuse 24 \um\ emission ({\it orange}).
  \label{overview}}
\end{figure*}
Initially, the area shown within the {\it white}
boundaries in Figure \ref{overview} was spanned by a large GMC
complex that included the gas now observed in the M17 molecular cloud and MC
G15.9-0.7. Between 2 and 5 Myr ago, a massive cloud subcomponent in 
the center of the 
complex underwent gravitational collapse, forming the first cluster of
perhaps 2,000--3,000 stars, NGC 6618PG, the predecessor and possible
progenitor of 
subsequent star formation in the complex. The OB
stars in NGC 6618PG ({\it blue} diamonds in Fig.\ \ref{overview})
ionized an \hii\ region that expanded, sweeping
ambient ISM gas and dust into a bubble, M17 EB. 

The M17 molecular cloud, with initial mass 
${>}7\times 10^4$ \Msun, lay to   
the south of NGC 6618PG, in the path of the expanding M17 EB. What
happened next is debatable. The onset of
star formation 
in M17 may have been triggered when M17 EB collided with the M17
molecular cloud. But
triggered star formation is impossible to prove, because
we cannot prove that the molecular cloud would {\it not} have
collapsed on its own, 
independent of the existence of NGC 6618PG. Indeed, it is curious that
NGC 6618, the more massive cluster, formed later than
NGC 6618PG. Measurements of the
magnetic field strength in the PDR
interface between the \hii\ region and M17 South (see Fig.\ \ref{seqSF})
indicate that M17 South is in approximate dynamic equilibrium, with
thermal, turbulent, and magnetic pressure combined supporting the
cloud \citep{BT01,geo}. These support mechanisms 
might have allowed M17 initially to resist gravitational collapse.

The M17 molecular cloud began to collapse and form stars not more than
2 Myr ago. We know that NGC 6618 is younger than NGC 6618PG,
because: (1) Despite being a more luminous cluster exhibiting many signs of
strong stellar winds \citep{T03,paper1,bowshock}, NGC 6618, unlike NGC 6618PG,
has not yet dispersed 
its natal cloud; and (2) NGC 6618 has numerous highly-embedded intermediate- to
high-mass stars (ongoing massive star formation), while 
NGC 6618PG has no intermediate-mass stars with optically thick disks
(massive star formation has ceased). 
The burst of star formation that created the most massive
members of NGC 6618 was probably very rapid.
The time-averaged SFR in the central
regions of M17 has been
0.008--0.01~\Msun~\peryr. 

The presence of NGC 6618PG  explains some
puzzling features of the classic ``blister'' structure of the M17 \hii\ region. 
Why should a blister \hii\ region form in the first place? 
This question has received its share of theoretical attention in recent
years. Two-dimensional models of molecular clouds with uniform 
density show that gravitational acceleration peaks at the edges of
clouds, sweeping material inward \citep{BH04}. 
\citet{FH08} present 3-dimensional numerical simulations that lead to
the formation 
of massive protostellar cores on the edges of molecular clouds
facilitated by gravitational instabilities that focus large-scale
colliding flows. 
M17 could be a good example of the end-product of such a
process, with its 
cluster of very massive young stars forming on the northern side of
the massive M17 South cloud. 
The CO maps show a 
narrow molecular ridge tracing M17 North that
appears to be pressure-bounded on 2 sides. The ridge lies between NGC
6618 and NGC 6618PG  (contours between NGC 6618
and NGC 6618PG in Fig.\ \ref{overview}).
The source of outward
pressure from the stars ionizing the M17 \hii\ region is well-known,
and we have now identified a potential source of counteracting pressure from the
expansion of M17 EB. This provides a natural explanation for the
morphology of the M17 North cloud and the associated northern
half of the M17 \hii\ region and PDR.

As the M17 \hii\ region pushes outward from M17 South, M17 North
is eroded. Once this barrier is breached, the built-up pressure
in the \hii\ region can be released into the diffuse interior of M17
EB. There are observational clues that this may already have
occurred. The diffuse X-ray emission identified by \citet{T03} traces
hot plasma filling the interior cavity of the M17 \hii\ region. This
plasma appears
to spill out of the \hii\ region to the East and North (Townsley et
al.\ in prep.). The thermal radio/IR ring shown in Figure \ref{ring}
may also be evidence of a blowout from the northern side of the \hii\
region. In this scenario, it is possible that NGC 6618 now pumps
energy into M17 EB, even though the bubble was originally driven by
NGC 6618PG. Additional observations in X-rays and optical recombination
lines are needed to reveal the detailed structure of M17 EB.

The expansion of M17 EB is asymmetric, with NGC 6618PG offset from the 
center of the bubble toward
its southern rim, near M17 itself. This is reasonable given that the presence of the
large molecular mass in M17 indicates a strong density gradient
across the bubble. Any energy input from NGC 6618 into M17 EB would
provide an additional push to expand the bubble preferentially away
from the M17 \hii\ region. 

The northward expansion of M17 EB overtook another
molecular cloud, MC 15.9-0.7.
The central regions of MC G15.9-0.7 lack IR signatures of star formation. 
As Figure \ref{overview} shows, only the side of this cloud that
intersects the 8 \um\ emission 
from M17 EB exhibits
a significant concentration of candidate YSOs after the removal of background
contaminants. Hence 
the circumstantial evidence for triggered star formation is stronger here
than in the case of NGC 6618. 
Outside of M17 itself, the
bubble-cloud interface produces the brightest CO ($J=2\rightarrow 1$) emission at
$v=20$ \kms. 
The current SFR in the molecular
gas swept up by 
M17 EB and in the extended outer regions of M17 is ${>}16\%$ of
the SFR that produced the well-studied young stellar population of the M17
\hii\ region.

\section{Conclusions}

We have used  IR, radio, and X-ray survey data combined with
targeted millimeter observations to show that star formation in the
M17 complex extends ${\sim}0.5\degree$ farther to the north than
previously thought. Our $^{12}$CO and $^{13}$CO line maps cover a 0.72
deg$^2$ area that contains $1.35\times 10^5$ \Msun\ in gas at
$v=12$--26 \kms. The morphology of the CO map (Fig.\ \ref{overview})
is dominated by an extended bubble, M17 EB, occupying the space
between the well-studied M17 molecular cloud and a 
neighboring molecular cloud to the north, MC G15.9-0.7.

We have found 406 candidate YSOs in a $1.5\degree \times 1\degree$
target field centered on M17 EB by fitting model
SEDs to fluxes of sources in the GLIMPSE Archive supplemented by
MIPSGAL and \MSX\ photometry. We characterized the contamination from
unassociated sources in our YSO sample by analyzing a control field
exhibiting no obvious signs of massive star formation.
We were able 
to identify and remove the brightest ${\ga}50\%$ of dust-rich AGB
stars from our sample. We conclude that ${>}80\%$ of sources in the
control field sample are candidate YSOs distributed at unknown
distances along the long sightline through the inner Galactic plane.

We created a map of the extinction produced by the molecular cloud
structures associated with M17. Outside of a few small regions of high gas
column density, across ${\sim}90\%$ of our target field the extinction
is too low to significantly reduce mid-IR  detections of background
sources. The complex, spatially varying extinction in 
the M17 molecular cloud itself creates special challenges for
observations of the young stellar population associated with NGC
6618. The average extinction through the cloud cavity occupied by the central M17
\hii\ region is $A_V< 17$~mag, too low to prevent the detection of a
significant fraction of background stars even for relatively shallow near-IR
observations. Recent results reported by H08, in particular the high
IR excess fraction and young age of 0.5~Myr for the young stellar
population in M17, appear to have been compromised by the use of an
IR-selected sample that was not corrected for background contamination.

Approximately half of the candidate YSOs in our target field are
unassociated contaminants. 
In studies of massive young clusters, IR excess sources are
commonly assumed to be associated by virtue of their apparent youth,
but this assumption is not valid in the case of star-formation regions
near the inner Galactic mid-plane. By selecting candidate YSOs exhibiting a
significant degree of clustering with respect to the control sample,
we successfully removed the predicted level of contamination.

Among 195 candidate YSOs overlapping the area occupied by
the extended M17 complex, 96 have high statistical likelihood of
association with M17 once the contribution from unassociated 
foreground and background sources is removed. Modeling the YSO
population, we find that our sample consists primarily of
intermediate-mass YSOs with disks \citep[Stage II;][]{grid} and highly
embedded (Stage 0/I) YSOs. We model the accretion rates for the Stage
0/I sources and find that their median age is ${\la}0.075$~Myr, leading
us to derive a circumstellar disk lifetime of ${\sim}0.5$ Myr for
sources with $M_\star\ga 3$ \Msun\ (corresponding to B-type stars on
the main sequence). 
Candidate YSOs distributed
around the rim of M17 EB and the extended outer regions of
M17 represent a ${>}16\%$
 addition to the star formation
rate that has produced the young stellar population in
the central 
regions of M17, as observed by {\it Chandra} (BFT07). The
concentration of star
formation activity tracing the interface between M17 EB and MC
G15.9-0.7 appears to have been triggered by the expansion of the
bubble.

We have identified 5 candidate ionizing
stars inside M17 EB. By fitting stellar atmosphere
models to their IR SEDs we constrained the luminosity
distance of these 5 sources, showing that their IR luminosities give
OB spectral types at the M17 distance. An associated lower-mass
population of young stars is suggested 
by X-ray survey observations by the {\it Einstein} and {\it ROSAT}
satellites, leading us to deduce the presence of a young cluster, NGC
6618PG. The two earliest-type stars in NGC 6618PG, BD-16
4831 and BD-16 4826 (stars 17 and 18 in Table \ref{OBstars}), have luminosities equivalent to main-sequence
spectral types O4--O5, but the stars may be evolved. The stellar winds
from these two stars carry sufficient momentum to produce an
\hii\ region the size of M17 EB in ${<}5$ Myr. The current
expansion velocity of M17 EB is low, a few \kms\ maximum.

We also fit stellar atmosphere SEDs to 11 of the principal
ionizing stars in NGC 6618. 
At the newly-revised 2.1~kpc distance (H08), we find that a high
fraction of the O stars in NGC 6618 are too luminous in the mid-IR to
be single stars on the main sequence, yet the stars are too young to have evolved
off the main sequence.
This finding supports the hypothesis of H08 that most of the O stars in M17
are in binary systems with 
approximately equal-mass components.

The wealth of observational data paints a picture of propagating
star formation in this region. Three main waves of star formation can
be identified: (1) The formation of the NGC 6618PG cluster 
(2,000--3,000 stars), 2--5 Myr in the past, followed by the
expansion of an \hii\ region delineated by M17 EB; (2) The rapid
collapse of the M17 molecular cloud (current mass ${>}6\times 10^4$
\Msun) within the last 2 Myr, producing the massive NGC 
6618 cluster and associated massive star formation that continues to
the present day (11,000--14,000 stars); and (3) The  onset of star
formation in MC G15.9-0.7 and the extended outer regions of M17
(${>}1,000$ stars). 
The
possibility that the later waves of star formation were
triggered by the first raises a tantalizing question: 
Can a massive progenitor cluster trigger the formation of an even more
massive daughter cluster?

Because they are beacons observable at large
distances across the local universe, \hii\ regions are commonly used
as SFR tracers in external galaxies, and have been used to
estimate the SFR of the Milky Way as well \citep{SBM78}. The canonical
Galactic SFR of a few \Msun~\peryr\ is equivalent to several hundred
\hii\ regions the size of M17.
Observational tracers tied to \hii\ regions are sensitive only to the
most massive ${<}1\%$ of stars, and the uncertainty involved in
extrapolating over the IMF to the  
low-mass stars that make up the bulk of the stellar mass is
enormous. Star formation distributed at lower densities but over large
spatial volumes is increasingly observed in
prominent Galactic molecular cloud complexes, including
Orion \citep{M05}, W5 \citep{XK08}, and now M17. 
It is not yet clear if low-density star formation outside of
bright \hii\ regions or dense GMCs significantly increases the
galactic SFRs derived from  
observational tracers of ionized or molecular gas.

\acknowledgements
We are indebted to Leisa Townsley for keen insights and suggestions
that helped guide this work and for sharing early results from {\it
  Chandra}. 
We thank the anonymous referee for numerous comments that helped
to substantially improve this work.
Frank D. Ghigo, Ronald J. Maddalena, and Dana S. Balser provided the
GBT 3 cm image used in our analysis.
We thank Steve Bracker for IR color analysis of candidate AGB stars,
David Pooley for finding the {\it Einstein} image, and Cormac
Purcell for inspecting the CORNISH survey images on our behalf. 
Conversations with Debra 
Shepherd, Fabian Heitsch, and Kyle Westfall helped improve this work.
We thank Dr.\ A. R. Kerr of the National Radio Astronomy Observatory for
providing the single-sideband ALMA Band 6 prototype mixers used in
this study. The Heinrich Hertz Telescope is operated by the Arizona
Radio Observatory, a part of Steward Observatory at The University of
Arizona. This work was supported by NASA/JPL contract 1275394 and NSF
grant AST-0303689 to the University of Wisconsin. Additional support
was provided by the 
\emph{Spitzer} Theoretical 
Research Program (1290701 \& 1310231; B. A. W., T. P. R.), the NASA
Theory Program (NNG05GH35G; B. A. W.), and NSF grant AST-0708131 to
The University of Arizona (M. K., 
J. H. B.). M. K. gratefully acknowledges support from a Korea Research
Foundation Grant (KRF-2007-612-C00050). Additional support was
provided by NASA through the {\it Spitzer Space Telecopse} Fellowship
Program (T. P. R.).
This work relies on
observations made the the {\it Spitzer Space Telescope}, operated by
the Jet Propulsion Laboratory under a contract with NASA.
This research has made use of the SIMBAD database,
operated at CDS, Strasbourg, France. This work has also made use of
data obtained from the High Energy 
Astrophysics Science Archive Research Center (HEASARC), provided
by NASA's Goddard Space Flight Center.

{\it Facilities:} \facility{Spitzer}, \facility{HHT}, \facility{MSX}, \facility{GBT}, \facility{VLA}

\end{document}